\documentclass[11pt]{article}
\usepackage{latexsym,amsmath,amsfonts,amssymb}
\usepackage[american]{babel}
\usepackage{color}
\usepackage[unicode]{hyperref}
\usepackage{enumerate}
\usepackage{float}
\usepackage{subcaption}
\usepackage{xcolor}

\usepackage{enumitem}

\usepackage{graphicx}
\usepackage{amsmath,euscript,array,mathrsfs,amsfonts}
\usepackage{hyperref}
\hypersetup{
colorlinks = true,
linkcolor = red,
linktocpage = true,
citecolor = blue
}

\usepackage{verbatim}

\makeatletter\def\l@subsubsection#1#2{}%
\makeatother

\renewcommand\theequation{\arabic{section}.\arabic{equation}} 

\newcommand{\beqs}{\begin{eqnarray}}
\newcommand{\eeqs}{\end{eqnarray}}
\newcommand{\bal}{\begin{aligned}}
\newcommand{\eal}{\end{aligned}}
\makeatletter
\newcommand\setItemnumber[1]{\setcounter{enum\romannumeral\@enumdepth}{\numexpr#1-1\relax}}
\makeatother
\begin{document}
\baselineskip=15.5pt
\pagestyle{plain}
\setcounter{page}{1}



\def\andrea#1{\textcolor{green}{#1}}
\def\aldo#1{\textcolor{blue}{#1}}
\def\fra#1{\textcolor{red}{#1}}
\def\JL#1{\textcolor{violet}{#1}}

\newcommand{\ber}{\begin{eqnarray}}
\newcommand{\eer}{\end{eqnarray}}



\newfont{\namefont}{cmr10}
\newfont{\addfont}{cmti7 scaled 1440}
\newfont{\boldmathfont}{cmbx10}
\newfont{\headfontb}{cmbx10 scaled 1728}

\numberwithin{equation}{section}



%

%
\setcounter{footnote}{0}
\renewcommand{\theequation}{{\rm\thesection.\arabic{equation}}}

\begin{titlepage}

\begin{center}

\vskip .5in 
\noindent

{\Large \bf{Holographic Baryons as Quantum Hall Droplets} }\bigskip\medskip

Francesco Bigazzi$^a$\footnote{bigazzi@fi.infn.it}, Aldo L. Cotrone$^{a,b}$\footnote{cotrone@fi.infn.it}, Andrea Olzi$^{a,b}$\footnote{andrea.olzi@unifi.it} and Jean-Loup Raymond$^{b,c}$\footnote{jean-loup.raymond@ens-paris-saclay.fr}  \\

\bigskip\medskip
{\small 
$^a$ INFN, Sezione di Firenze, Via G. Sansone 1, I-50019 Sesto Fiorentino (Firenze), Italy.
\\
$^b$ Dipartimento di Fisica e Astronomia, Universit\'a di Firenze, Via G. Sansone 1, I-50019 Sesto Fiorentino (Firenze), Italy.
\\
$^c$ ENS Paris-Saclay, 4 Av.~Des Sciences, 91190 Gif-sur-Yvette, France.
}

\vskip .5cm 
\vskip .9cm 
     	{\bf Abstract }\vskip .1in
\end{center}

\noindent
We provide a first-principle construction of baryons as quantum Hall droplets in single-flavor holographic QCD.
The baryons are described as charged D6-branes with a circular boundary on a flavor D8-brane in the Type IIA backgrounds dual to the confining and non-confining phases.
The holographic description allows us to calculate precisely their properties, such as mass and size.
We also consider other objects with baryonic charge, such as vortons, domain walls with holes and ``sandwich vortons'', and discuss the relative (meta)stability of all these configurations.

 \noindent
\vskip .5cm
\vskip .5cm
\vfill
\eject

\end{titlepage}

\setcounter{footnote}{0}

\small{
\tableofcontents}

\normalsize

\newpage
\renewcommand{\theequation}{{\rm\thesection.\arabic{equation}}}
%

\section{Introduction}

In QCD-like theories with gauge group $SU(N)$ and a single massless flavor ($N_f=1$), the low-energy description of baryons as solitonic Skyrmions is not available, since the latter requires at least a $SU(N_f=2)$ flavor group \cite{Skyrme:1961vq}.
More than two decades ago it was suggested that, at least in the planar limit of these theories, baryons at low energy should correspond to some kind of ``brane-like'' objects related to the singular locus of the low-energy $\eta'$ potential defined by $\eta'=\pi$, see e.g.~\cite{Kogan:1993yw,Gabadadze:2000vw,Forbes:2000et,Son:2000fh,Gabadadze:2002ff}.\footnote{In the planar limit the pseudo-scalar $\eta'$ is the lightest particle of these theories. Throughout this paper, we work with $\theta_{YM}=0$.}
Indeed, in the case of a single flavor, the spins of all the quarks in a baryon must be aligned, and hence the total spin of the baryon should be $N/2$. For small enough quark mass and in the large $N$-limit, the constituent quarks become more mobile, and we can no longer neglect the spin-orbit and spin-spin
interactions. Therefore, it is likely that the one-flavored baryon is not spherically symmetric in shape. One thus expects ``pancake-shaped'' particles, close to (2+1)-dimensional defects.
A few years ago, this proposal was improved by the observation that, in analogy with the case of the stable domain walls at $\theta_{YM}=\pi$, on the worldvolume of these objects there should live a $U(1)_N$ Chern-Simons (CS) theory \cite{Komargodski:2018odf}.
In this respect, the baryons would be analogous to the quantum Hall droplets of the fractional Hall system.

Several expected properties of the baryons have been derived from these premises, yet precise quantitative statements are, as usual, difficult to achieve from purely effective descriptions \cite{Kogan:1993yw,Gabadadze:2000vw,Forbes:2000et,Son:2000fh,Gabadadze:2002ff,Komargodski:2018odf,Ma:2019xtx,Karasik:2020pwu,Ma:2020nih,Karasik:2020zyo,Kitano:2020evx,Nastase:2022pts,Lin:2023qya,Rho:2024ihu}.
In this paper, we provide for the first time a first-principle construction of such baryons, in a QCD-like theory admitting a holographic dual.  
To be more precise, in this paper we consider the gluonic core of the baryons, leaving the study of their mesonic shell for the future. 
We also construct related configurations, such as ``sandwich vortons'' and domain walls with holes, study their stability, and (some of) their possible decay channels. 
Other defects of similar nature in the same theory - domain walls, strings, vortons - have been investigated in previous works \cite{Bigazzi:2022luo,Bigazzi:2022ylj,Bigazzi:2024ahm,Bigazzi:2024mge}. 

The theory we consider, the Witten-Sakai-Sugimoto model (WSS)  \cite{Witten:1998zw,Sakai:2004cn,Sakai:2005yt}, is the top-down holographic theory closest to planar QCD, sharing with the latter the vacuum structure and many crucial properties, from confinement to chiral symmetry breaking. As such, we consider the existence of quantum Hall droplet baryons in this theory as a possible indication of their existence in real QCD. The WSS setup has also been used to holographically model hypothetical planar QCD-like dark sectors and (QCD-like) axions \cite{Bigazzi:2019eks}. Within this framework, the baryons and other configurations we construct in this work can be re-interpreted as defects in beyond the Standard Model theories.
 
Let us detail the content of the paper.
In the next section, we recall basic facts about the WSS model. 
An important property is that it allows to disentangle the chiral symmetry breaking transition from the deconfinement one.
As such, we can study baryons and related configurations in the deconfined phase as well.
In both phases, chiral symmetry breaking corresponds to a single (for $N_f=1$) D8-brane supporting flavor degrees of freedom, treated as a probe of a Type IIA background describing the appropriate vacuum of the Yang-Mills sector.

In section \ref{sec:EulerD6}, which constitutes the core of this paper, we set up the study of various configurations, including the baryons, as D6-branes ending on the flavor D8-brane at a circular boundary, in both confined and deconfined phases.
Indeed, the D6-brane is the object realizing the correct monodromy of the field dual to the $\eta'$ \cite{Bigazzi:2022luo}, and after reduction on the four-sphere of the background, it naturally hosts on its worldvolume the $U(1)_N$ CS theory advocated in \cite{Komargodski:2018odf}.
In order to describe a (meta)stable baryon as a quantum Hall droplet, as well as other baryonic configurations, it is necessary to consider charged configurations under the baryonic symmetry. 
This corresponds to switching on an electromagnetic field on the D6-branes \cite{Bigazzi:2024mge}.

We thus describe in detail the equations of motion, variational principle, boundary conditions, and the relations between the baryon charge $n_B$, the quark charge $q_s$, and the spin $J$ associated with the D6-brane.
In particular, we study configurations such that $J=n_B^2 N/2$ (as in the standard Fractional Quantum Hall Effect (FQHE)) - this reduces to the expected relation $J = N/2$ for the elementary $n_B=1$ baryon made by $N$ quarks with aligned spins.
By analyzing the topology of the configuration, we are able to identify the solitonic mode living on the boundary of the droplet and carrying integer baryonic charge, in close analogy to the description of the electron in the FQHE.
Perhaps not surprisingly, the solitonic mode is a would-be gauge mode of the gauge connection on the D6-brane worldvolume.    

Section \ref{sec:baryon} is devoted to the presentation of the (numerical) solution corresponding to the baryon, \textit{i.e.}~the embedding profile and the electromagnetic field of the D6-brane.
The solution allows to compute precisely, from first principles, the baryon properties, such as its mass and radius.
As envisaged in  \cite{Komargodski:2018odf}, while the mass scales linearly with the number of colors $N$, as customary for baryons in the planar limit \cite{Witten:1979kh}, the radius is $N$-independent.
Instead, both these observables depend on the 't Hooft coupling $\lambda$, the chiral symmetry breaking scale $f_{\eta'}$, and the temperature (in the deconfined phase). 
In particular, it turns out that these baryons are small particles in the holographic regime, with size suppressed by the large value of $\lambda$, as in the case of the standard $N_f \geq 2$ baryons in the WSS model \cite{Hata:2007mb,Hashimoto:2008zw} (but with a different power law suppression). 

In section \ref{sec:sandwich} we study two different but related configurations: a domain wall with a hole and a ``sandwich vorton''.
Both correspond to a D6-brane with a circular boundary at the D8-brane tip (as the baryon).
The former is expected to play a role in the decay of infinite domain walls, which develop holes that eventually eat up all their worldvolume.
We concentrate on metastable configurations with baryonic charge, but analogous metastable uncharged solutions and unstable solutions exist as well.

The ``sandwich vorton'' is a charged string loop whose structure depends on two different fields.
A prototype example is the ``axionic-$\eta'$ string'' (or the ``axionic-pionic'' string)  \cite{Gabadadze:2000vw}.
In the holographic model, it corresponds to a cylindrical D6-brane terminating on two D8-branes whose tips are at different holographic radial positions.
In the WSS model, it can carry baryonic charge, which makes it (meta)stable.
 
Section \ref{sec:decays} discusses possible decay channels between various configurations: baryons, ``sandwich vortons'', ordinary vortons, quarks (in the deconfined phase). The discussion is based on energy and charge considerations.
The results are non-trivial ``phase diagrams'' highlighting the most stable configurations depending on the theory parameters (e.g.~$f_{\eta'}$, temperature, $\lambda$). 

Section \ref{sec:conclusions} contains a summary and some directions for future studies.

Finally, Appendix \ref{ap:topo} provides more details on the topology of the solutions, while appendix \ref{app:numerics} describes the numerical integration procedure followed to solve the equations in the main body of the paper. Appendices \ref{app:deconf1} and  \ref{app:deconf2} contain details on the solutions in the deconfined phase.

\section{The WSS model}\label{sec:model}
The Witten-Sakai-Sugimoto (WSS) model~\cite{Witten:1998zw,Sakai:2004cn,Sakai:2005yt} represents the top-down holographic theory closest to large $N$ QCD, and has proven highly effective in capturing features of its strongly coupled regime. It describes a non-supersymmetric Yang-Mills theory in $(3+1)$ dimensions with gauge group $SU(N)$, interacting with $N_f$ quark flavors as well as with (an infinite tower of) massive Kaluza-Klein (KK) modes transforming in the adjoint representation of $SU(N)$. The model emerges in the low-energy limit of a string theory construction based on $N$ D4-branes compactified on a circle $S^1_{x_4}$, where the coordinate $x_4$ satisfies the periodicity condition $x_4 \sim x_4 + 2\pi/M_{KK}$, with $M_{KK}$ being the KK mass scale. The fundamental flavor degrees of freedom are introduced by means of additional $N_f$ D8 and $\overline{\text{D}8}$-branes transverse to the $S^1_{x_4}$ circle. In the present work, we are going to focus on the case with just a single ($N_f = 1$) massless quark.

At zero temperature, the large $N$ strong coupling regime $\lambda\gg1$ of the unflavored model, where $\lambda$ is the 't Hooft coupling at the KK mass scale $M_{KK}$, is holographically described by a type IIA supergravity background sourced by the D4-branes 
\begin{eqnarray}
	&&ds^2=\left(\dfrac{u}{R}\right)^{3/2}\left(dx^\mu dx_\mu +f(u)\,dx_4^2\right)+\left(\dfrac{R}{u}\right)^{3/2}\dfrac{du^2}{f(u)}+R^{3/2}u^{1/2}d\Omega_4^2\,,\nonumber \\
	&&f(u)=1-\dfrac{u_0^3}{u^3}\,,\quad e^\phi=g_s\left(\dfrac{u}{R}\right)^{3/4}\,,\quad F_4 = \dfrac{2\pi N}{V_4}\omega_4\,,\quad R^3=\pi g_s Nl_s^3\,,
	\label{solitonic}
\end{eqnarray}
where $\mu=(0,1,2,3)$, $u\in [u_0,\infty)$ is the radial coordinate holographically related to the renormalization group energy scale of the dual field theory, and $\omega_4$ is the volume form of the unit $S^4$, with volume $V_4= 8\pi^2/3$. The field $\phi$ is the (running) dilaton, and $F_4$ is the Ramond-Ramond four-form field strength.\footnote{Here and in the following we will adopt the same conventions as in \cite{Sakai:2004cn} for what concerns the Ramond-Ramond forms: they are thus scaled with respect to the standard notation as
	$C_{p+1} \rightarrow k_0^2 \tau_{6-p}\pi^{-1}C_{p+1}$, where $k_0^2=2^6\pi^7l_s^8$ and $\tau_p=(2\pi)^{-p}l_s^{-(p+1)}$.} 
The above background holographically accounts for confinement and mass gap formation in the dual $SU(N)$ gauge theory.

The mass scale of the glueballs is set by $M_{KK}$, which thus essentially also accounts for the dynamical scale $\Lambda$ of the dual quantum field theory
\begin{equation}
	\ M_{KK} \hspace{0.5cm} \longleftrightarrow \hspace{0.5cm} \Lambda\,.
\end{equation}
To avoid conical singularities in the $(u, x_4)$ plane, the KK mass scale is related to the background parameters through
\begin{equation}
	M_{KK} = \dfrac{3}{2}\sqrt{\dfrac{u_0}{R^3}}\,.
\end{equation}
Moreover, the 't Hooft coupling $\lambda$ is related to the string theory parameters through\footnote{Adopting the same conventions as in \cite{Sakai:2004cn}, we define $\lambda = N g_{YM}^2 = 2\pi g_s N l_s M_{KK}$.}
\begin{equation}\label{eq:lambda}
	\dfrac{R^3}{l_s^2} = \dfrac{1}{2}\dfrac{\lambda}{M_{KK}}\,.
\end{equation}

Flavor degrees of freedom are introduced by means of two stacks of $N_f$ D8 and $N_f$ $\overline{\text{D}8}$-branes placed at a distance $L$ along the $S^{1}_{x_4}$ circle \cite{Sakai:2004cn}. The $U(N_f)\times U(N_f)$ gauge symmetry on their worldvolume holographically realizes the classical global chiral symmetry of the dual field theory. The flavor branes are usually treated in the probe approximation, which is reliable if $N_f\ll N$,\footnote{More precisely, if $\lambda^2 N_f\ll 3\pi^3 N$, see e.g. \cite{Bigazzi:2014qsa}.} and which amounts to neglecting their backreaction on the background sourced by the D4-branes.  To minimize their energy in that background, they actually form a unique U-shaped configuration on the ($x_4, u$) cigar-like subspace, dynamically realizing the spontaneous breaking of the chiral symmetry. If $L=\pi /M_{KK}$, i.e.~when the D8 and the $\overline{\text{D}8}$-brane stacks are placed at antipodal points on the $S^{1}_{x_4}$ circle, the tip of the U-shaped configuration is at $u=u_0$. Instead, in the non-antipodal case, $L< \pi/M_{KK}$, the U-shaped configuration has a tip at some $u_J>u_0$ which is implicitly determined by the relation \cite{Aharony:2006da}
\begin{equation}
	L = J(b)\dfrac{R^{3/2}}{u_J^{1/2}} = \dfrac{3}{2}\dfrac{J(b)\sqrt{b}}{M_{KK}},
\end{equation}
where 
\begin{equation}
	b \equiv \dfrac{u_0}{u_J}\,,
\end{equation}
and
\begin{equation}
	J(b)= \dfrac{2}{3}\sqrt{1-b^3}\int^1_0dy\dfrac{\sqrt{y}}{\left(1-b^3y\right)\,\sqrt{\left(1-b^3y-\left(1-b^3\right)y^{8/3}\right)}}\,.
\end{equation}

Other D-brane configurations probing the background sourced by the D4-branes and wrapping the transverse $S^4$ play a relevant role in the dual gauge theory: wrapped D4-branes correspond to baryon vertices \cite{Witten:1998xy} and wrapped D6-branes can either correspond to topological defects like global strings and domain walls \cite{Bigazzi:2022ylj,Bigazzi:2024mge} or, as we will also see in the next sections, to baryonic-like objects.

\subsection{High-temperature phase of the holographic model}
The WSS model exhibits a rich phase structure at non-zero temperature. This includes both a first-order transition between the confined and a deconfined phase, and a first-order chiral symmetry restoration transition. These two transitions occur in general at different critical temperatures, depending on the model's parameters. Importantly, in the large $N$ and strong coupling regime $\lambda \gg 1$, these features can be precisely studied using the dual holographic description \cite{Aharony:2006da}.

The confinement/deconfinement phase transition takes place at the critical temperature
\begin{equation}
T_c = \frac{M_{KK}}{2\pi}\,.
\label{criticalT}
\end{equation}
In the dual gravity picture, this corresponds to a Hawking-Page transition between the solitonic background presented in the previous section, dual to the confining phase at $T<T_c$,\footnote{The solitonic background at non-zero temperature $T$ is the same as in (\ref{solitonic}) with a compact Euclidean time direction $t_E\sim t_E + 1/T$.} and a black brane geometry representing the deconfined phase at $T>T_c$. The corresponding background features a (string frame) metric given by
\begin{eqnarray}\label{eq:bkgmetric}
	&&  ds^{2}_T =\left(\frac{u}{R}\right)^{3/2} \left[-f_{T}(u)dt^{2} + dx^i dx^i + d x_4^2\right] +\left(\frac{R}{u}\right)^{3/2}\left[\dfrac{du^{2}}{f_{T}(u)} + u^{2}d\Omega_4^2\right]\,, \nonumber \\
	&&f_{T}(u) = 1 - \dfrac{u_{T}^{3}}{u^{3}}\,,
\end{eqnarray}
where $i=(1,2,3)$ and $u\in [u_T, \infty)$, with $u=u_T$ being the radial position of the event horizon. The latter is related to the black brane temperature $T$ (which corresponds to the field theory temperature) through
\begin{equation}\label{eq:uT}
	9u_T = 16\pi^2 R^3 T^2\,.
\end{equation}
The dilaton and the four-form RR field strength in the black-brane background are the same as in equation (\ref{solitonic}). 

At $T<T_c$, chiral symmetry is always broken, with the confinement and chiral symmetry breaking scales being generically different (as $u_J \ge u_0$, the condition being saturated only when the flavor branes are antipodal). 

\begin{figure}
	\center
	\includegraphics[height = 5 cm]{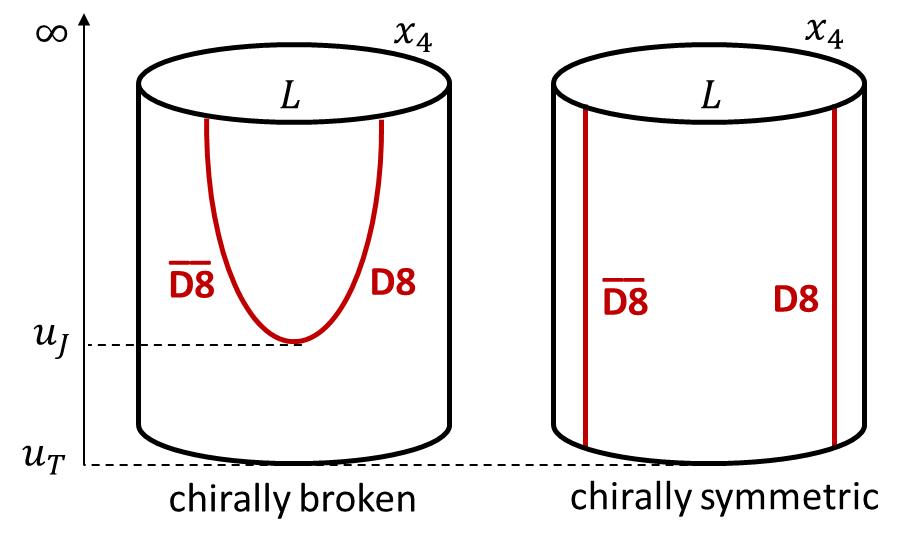}\caption{Schematic picture of the chirally broken (left) and chirally symmetric (right) phases in the deconfined phase of the Witten-Sakai-Sugimoto model. 
		The $x_4$-direction is compactified on a circle with radius $1/M_{KK}$,
		and $u$ is the holographic coordinate with $u =\infty$ being the boundary where the dual field theory lives. If the asymptotic distance $L$ between the flavor branes is sufficiently small, a deconfined, chirally broken phase (left figure) becomes possible. In this phase, the connected flavor branes are embedded non-trivially in the background according to a definite function $x_4(u)$.}
	\label{fig:wssphases}
\end{figure}

When $T>T_c$, the flavor phase structure of the model depends on the value of $L$. If $L M_{KK} \gtrsim 0.97$, chiral symmetry is always restored: correspondingly, the two stacks of D8/$\overline{\text{D}8}$-branes stay disconnected and fall into the black brane horizon. However, for non-antipodal configurations with $L M_{KK} \lesssim 0.97$, an intermediate phase emerges in which, depending on the temperature, the theory is deconfined but chiral symmetry remains broken. This actually happens when $T<T_a$, where
\begin{equation}
	T_a \simeq \frac{0.154}{L}\,,
\label{tadef}	
\end{equation}
is the critical temperature for a first-order phase transition between the connected U-shaped flavor brane configuration at $T<T_a$ and the disconnected one at $T>T_a$, where chiral symmetry is restored. See figure \ref{fig:wssphases} for a pictorial representation of these two phases in the $(x_4,u)$ subspace.

In the chiral symmetry broken phase occurring at $T_c<T<T_a$, the position $u=u_J$ of the tip of the U-shaped flavor brane configuration is implicitly given by 
\begin{align}\label{eq:LRuJ}
	L = \int dx_{4} = 2\int\dfrac{du}{u'} = J_{T}(\tilde{b})\dfrac{R^{3/2}}{u_J^{1/2}}\,,
\end{align}
where 
\begin{equation}
	\tilde{b} \equiv \dfrac{u_T}{u_J}\,,
\end{equation}
and
\begin{equation}
	J_T(\tilde{b})= \dfrac{2}{3}\sqrt{1-\tilde{b}^3}\int^1_0dy\dfrac{\sqrt{y}}{\sqrt{\left(1-\tilde{b}^3y\right)\left(1-\tilde{b}^3y-\left(1-\tilde{b}^3\right)y^{8/3}\right)}}\,.
\end{equation}

Using (\ref{eq:uT}) we see that $u_{J}$ becomes a non trivial function of $T$
\begin{equation}
	LT = \dfrac{3}{4\pi}J_T(\tilde{b})\sqrt{\tilde{b}}\,.
\end{equation}
This relation in turn puts a bound on the maximal allowed value of $\tilde b$ according to the requirement $L T< L T_a\simeq 0.154$, with $L M_{KK} \lesssim 0.97$, namely $\tilde{b}\lesssim 0.75$.
\section{The D6-brane system}\label{sec:EulerD6}
In this section, partly reviewing what has been already shown in the deconfined phase in \cite{Bigazzi:2024mge}, we present a detailed analysis of the Euler-Lagrange equations and the related variational problem for a D6-brane wrapped on the background $S^4$, extended along the holographic coordinate $u$ with a circular boundary at the tip of a U-shaped D8-brane. We find a zoo of solutions associated with different D6-brane profiles. We present the analysis in both the confined and deconfined phases of the WSS model.\footnote{Let us remind that the physics of fractional quantum Hall effects has been studied widely in the literature. See, for example \cite{Fujita_2009,Hikida:2009tp} and related works.}

The worldvolume action for the D6-brane is composed of two parts: the Dirac-Born-Infeld (DBI) action and the Chern-Simons one. The former reads
\begin{equation}
	S^\text{DBI}_\text{D6} = -T_6 \int d^7x\, e^{-\phi} \sqrt{-\det\left(g + \left(2\pi l_s^2\right) f + B\right)}\,,
	\label{eqn:DBIconf}
\end{equation}
where $T_6 = (2\pi)^{-6}l_s^{-7}$ is the D6-brane tension and $f$ is the field strength associated to the $U(1)$ gauge connection $a$ living on the D6-brane worldvolume. Locally, we can write $f=da$. The Kalb-Ramond field $B$ appearing in the DBI action will be set to zero throughout this work. Integrating out the four-sphere, the above action reduces effectively to the DBI action of a membrane with a three-dimensional worldvolume $M_3$. The worldvolume coordinates of the manifold $M_3$ are identified with $t,\psi,u$, where $\psi$ is the polar angular variable of a circle in Minkowski, whose radius $\rho(u)$ (which we take to be $(t,\psi)$-independent) has to be determined by solving the Euler-Lagrange equations. 
The manifold $M_3$ has a boundary $\partial M_3$, which has in general a non-trivial topology. It has one connected circular component in $u_J$, and depending on the nature of the solution for $\rho(u)$, it might have a second connected component at the position where the D6-brane ends, which we denote by $u_*$. In the following, we use the convention $\epsilon^{t\psi u} = +1$ which is equivalent to $dt\wedge d\psi\wedge du = dt\,d\psi\,du$. Note that this induces an orientation on $\partial M_3$, spanned by $\psi,t$.

The Chern-Simons action for the D6-brane is given by
\begin{align}\nonumber
	S_\text{D6}^\text{CS} &= \frac{1}{8\pi^2} \int C_3 \wedge f \wedge f =\\
    &= \frac{N}{4\pi} \int_{M_3} a \wedge f -\dfrac{1}{8\pi^2}\int_{\partial\text{D6}} C_3\wedge a\wedge f\,,
	\label{eqn:CSconf}
\end{align}
where in the last equality, we have integrated by parts and used $\int_{S^4}F_4 = 2\pi N$. We have also used that $d^2a=0$, and we denoted with $\partial$D6 the boundary of the seven-dimensional D6-brane worldvolume. The effective theory on the D6-brane is thus described by a three-dimensional DBI-CS action with $U(1)_N$ CS term \cite{Acharya:2001dz,Argurio:2018uup}. In the case of a D6-brane with a boundary $\partial \text{D}6 = \partial M_3\times S^4$, there is also a six-dimensional boundary term.

Let us comment on this result. The authors of \cite{Sakai:2004cn}, following the results presented in the seminal paper \cite{Green:1996dd}, observe that the standard Chern-Simons term, involving a RR form $C$, for a wrapped brane in the presence of a non-trivial flux on the compact manifold, has to be traded (integrating by parts) with the term involving the field strength of $C$. This is the case whenever $C$ is not single-valued. Looking at \eqref{eqn:CSconf}, the first line has to be traded with the first term of the second line. However, our configuration is more complicated than the one described in \cite{Green:1996dd,Sakai:2004cn}, since we have a physical boundary. This leads to the presence of a boundary term in the second line of \eqref{eqn:CSconf}, explicitly involving $C_3$, which could also be problematic. We will circumvent 
this problem by requiring that the curvature of the gauge connection, and therefore the 
boundary term, vanishes at the boundary of the D6-brane worldvolume $\partial \text{D}6$:\footnote{This condition preserves the gauge invariance of the  RR form $C_3$ in the seven-dimensional D6-brane worldvolume with a boundary. This is not the only choice that preserves this gauge invariance.}
\begin{equation}
	f \underset{\partial \text{D}6}{=} 0\,.
	\label{curvboundaryD6}
\end{equation}
This way, the CS theory on the D6-brane worldvolume reduces effectively to three-dimensional $\mathrm{U(1)}_N$ CS theory, ubiquitous in this work. 

Let us comment now on the gauge invariance of the theory. Consider the first line of \eqref{eqn:CSconf}: since the action depends only on the gauge curvature $f$, it is manifestly gauge invariant. However, after the integration by parts, performed in the second line of (\ref{eqn:CSconf}), the two terms are not independently gauge invariant. In particular, the $\mathrm{U(1)}_N$ CS theory alone is not gauge invariant if the manifold $M_3$ has a boundary (see {\it e.g.} \cite{Elitzur:1989nr}). We study this issue by allowing for a general variation of the gauge connection $a\to a+\delta a$, and then by specializing this variation to a gauge transformation. The corresponding variation of the CS action to first order in the fluctuations reads
\begin{equation}
	\delta S_{CS} = \frac{N}{4 \pi} \int_{M_3} d( \delta a \wedge a) + 2\delta a \wedge da\,.
	\label{eq:var}
\end{equation}
In the specific case of a pure gauge transformation, we have $\delta a = -i g^{-1} d g = d\varphi$, where $g = e^{i \varphi}$ is an element of $U(1)$. With these definitions, the phase $\varphi$ is a periodic scalar field with period $2\pi$ (see \cite{Elitzur:1989nr} for more details). We can use Stokes's theorem to rewrite \eqref{eq:var} for pure gauge transformations:
\begin{equation}
	\delta S_{CS} = \frac{N}{4\pi} \int_{M_3} d\varphi \wedge f =  \frac{N}{4\pi} \int_{\partial M_3} \varphi \,f\,.
	\label{gauge}
\end{equation} 
Therefore, the condition for gauge invariance is that the curvature of the connection $a$ identically vanishes at the boundary:
\begin{equation}
	f \underset{\partial M_3}{=} 0\,.
	\label{curvboundary}
\end{equation}
One can realize that this condition is nothing but the reduction on $M_3$, of the condition \eqref{curvboundaryD6} that we imposed to get rid of the six-dimensional boundary term in the Chern-Simons brane action. Note that we are considering a gauge connection $a$ which does not depend on the four-sphere coordinates, and with null components along $S^4$. Therefore, imposing \eqref{curvboundaryD6}, we get for free the gauge invariance of the effective theory after the integration by parts.

A simple way to guarantee that \eqref{curvboundary} vanishes is to make an ansatz for the gauge connection corresponding to time-independent axisymmetric solutions, \textit{i.e.}~such that its components depend on the holographic coordinate $u$ alone.\footnote{Note that, with our choice of coordinates and orientation, the equation \eqref{curvboundary} can be written in components and reads $f_{t\psi} \underset{\partial M_3}{=} 0$. By choosing a time-independent axisymmetric configuration, this is automatically satisfied.} In addition, to simplify our study, in the following we will work in the radial gauge $a_u = 0$. Therefore, our gauge connection configuration will be given by 
\begin{equation}
	a_t(u)\,,\quad a_\psi(u)\,.
\end{equation}

Let us now derive the equations of motion both in the confined and deconfined phases for the D6-brane embedding and its gauge connection. We do this assuming to have a well-posed variational problem, and then we will discuss how, imposing particular boundary conditions, this analysis is consistent.

\subsubsection*{Confining phase}
We start with the confining background (\ref{solitonic}). The induced metric on the D6-brane is
\begin{equation}
	ds^2_\text{D6} = \left(\dfrac{u}{R}\right)^{3/2} \left( -dt^2 + \rho^2(u)d\psi^2 + \left(\rho'(u)^2  + \left(\frac{R}{u}\right)^3 \frac{1}{f(u)} \right)du^2\right) + R^{3/2}u^{1/2}d\Omega_4^2\,.
	\label{eqn:metricd6conf}
\end{equation}
After some algebra, the full action can be written as
\begin{align}\label{eq:D6actionconf}
	S_\text{D6}= -\dfrac{N}{2\pi} \int dt\,d\psi \int_{u_*}^{u_J} du\, \bigg\{  \dfrac{(a_t\, \partial_ua_\psi - a_\psi \,\partial_ua_t)}{2}+ \dfrac{u\,\rho(u)D(u)}{3(2\pi l_s^2)^2}\bigg\}\,,
\end{align}
where we have introduced the function
\begin{equation}\label{eq:Dconf}
	D(u) = \sqrt{\rho'(u)^2 \left(\frac{u}{R}\right)^3 + \frac{1}{f(u)} + (2\pi l_s^2)^2\left( \dfrac{(\partial_ua_\psi)^2}{\rho^2} - (\partial_ua_t)^2 \right)}\,\,.
\end{equation}

Let us now write the Euler-Lagrange equations for $\rho(u)$, $a_t(u)$, and $a_\psi(u)$, defining a Lagrangian density $\mathcal{L}_\text{D6}$ from the action by 
\begin{equation}
    S= \int dt\,d\psi\, du \,\mathcal{L}_\text{D6}\,.
\end{equation}
The equations of motion read:
\begin{itemize}
	\item Equation of motion for $\rho(u)$
	\begin{align}\label{eq:eqrhoconf}
		\partial_u \left(\frac{u^4\rho'(u) \rho(u)}{D(u)}\right) = \dfrac{R^3u}{D(u)}  \left(\rho'(u)^2\left(\frac{u}{R}\right)^3 + \frac{1}{f(u)} - (2\pi l_s^2)^2(\partial_ua_t)^2\right).
	\end{align}
	\item Equation of motion for $a_t(u)$
	\begin{align}\label{eq:eqatconf}
		3\, \partial_ua_\psi + \partial_u\left( \dfrac{u\,\rho(u) \partial_ua_t}{D(u)}   \right) = 0\,.
	\end{align}
	\item Equation of motion for $a_\psi(u)$
	\begin{align}\label{eq:eqapsiconf}
		3 \,\partial_ua_t + \partial_u\left( \dfrac{u\,\partial_ua_\psi}{\rho(u) D(u)}\right) = 0\,.
	\end{align}
\end{itemize}

We can rescale the gauge connection by a factor $(2\pi l_s^2)$ in such a way that there is no more explicit dependence on $l_s$ in the equations:
\begin{equation}\label{eq:arescale}
	\bar{a}=(2\pi l_s^2)a\,.
\end{equation}
The equations of motion \eqref{eq:eqatconf} and \eqref{eq:eqapsiconf} for the gauge connection components are total derivatives, so from them we get
\begin{align}\label{eq:defktkpsiconf}
	&\dfrac{u\,\rho(u)}{D(u)}\partial_u\bar{a}_{t}+3\bar{a}_\psi = \bar{\mathbf{k}}_t\,,\\\label{eq:defG}
	&\dfrac{u}{\rho(u)D(u)}\partial_u\bar{a}_{\psi}+3\bar{a}_t = \bar{\mathbf{k}}_\psi\,,  
\end{align}
where $\bar{\mathbf{k}}_t$ and $\bar{\mathbf{k}}_\psi$ are two constants of motion to be fixed. As for the gauge connection, the barred constants are related to unbarred ones through $\bar{\mathbf{k}}_{t,\psi} \equiv 2\pi l_s^2\,\mathbf{k}_{t,\psi}$. Defining
\begin{align}\label{eq:ashift}
	\tilde{a}_\psi &= \bar{a}_\psi - \dfrac{\bar{\mathbf{k}}_t}{3}\,, &   \tilde{a}_t &= \bar{a}_t - \dfrac{\bar{\mathbf{k}}_\psi}{3}\,,
\end{align}
the equations of motion can be rewritten as
\begin{align}\label{eq:eomsconf}
	\dfrac{\partial}{\partial u} \left( \dfrac{u^4\rho'(u)\rho(u)}{D(u)} \right)  &- R^3 u\, D(u)\left(1- \frac{9}{u^2}\tilde{a}_t^2 \right) = 0\,, \nonumber \\
	& \partial_u\tilde{a}_\psi + \frac{3}{u} \rho(u)D(u)\,\tilde{a}_t = 0\,, \nonumber \\
	& \partial_u\tilde{a}_t + \frac{3}{u} \dfrac{D(u)}{\rho(u)}\,\tilde{a}_\psi = 0\,.
\end{align}
Using the latter we can write $D(u)$ as 
\begin{equation}
	D(u) = \sqrt{\dfrac{\rho'(u)^2 \left(\dfrac{u}{R}\right)^3 + \dfrac{1}{f(u)}}{1+ \dfrac{9}{u^2} \left(\dfrac{\tilde{a}_\psi^2}{\rho^2(u)}-  \tilde{a}_t^2  \right)}}\,.
\end{equation}
This on-shell expression for $D(u)$ will be helpful in the numerical resolution of the differential equations.

\subsubsection*{Deconfined phase}
In the deconfined phase, where the background metric is given by (\ref{eq:bkgmetric}), a similar approach can be followed. The D6-brane induced metric reads
\begin{equation}
	ds^2_{T, \text{ D6}} = \left(\frac{u}{R}\right)^{3/2} \left( -f_T(u) dt^2 + \rho^2(u) d\psi^2 + \left(\rho'(u)^2  + \left(\frac{R}{u}\right)^3 \frac{1}{f_T(u)} \right)du^2\right) + R^{3/2}u^{1/2}d\Omega_4^2\,.
	\label{eqn:metricd6deconf}
\end{equation}
The full D6-brane action is
\begin{align}
	S_\text{D6} &= -\dfrac{N}{2\pi} \int dt\,d\psi \int_{u_*}^{u_J} du\, \bigg\{\dfrac{(a_t \,\partial_ua_\psi - a_\psi \,\partial_ua_t)}{2}+ \dfrac{u\,\rho(u) D_T(u)}{3(2\pi l_s^2)^2}\bigg\}\,,
	\label{Ddeconf}
\end{align}
where we have defined $D_T(u)$ as
\begin{align} \label{eq:Ddeconf}
	D_T(u) &= \sqrt{1+\rho'(u)^2 \left(\frac{u}{R}\right)^3 f_T(u) +(2\pi l_s^2)^2\left(\frac{f_T(u) (\partial_ua_\psi)^2}{\rho^2(u)}- (\partial_ua_t)^2  \right)}\,\,.
\end{align}
Using the rescaling \eqref{eq:arescale} the equations of motion read
\begin{eqnarray}
	&& \partial_u\left(\dfrac{u\,\rho(u)}{D_T(u)}\left(\dfrac{u}{R}\right)^3f_T(u)\rho'(u)\right)- u\,D_T(u) + \dfrac{u\,f_T(u)}{D_T(u)}\dfrac{(\partial_u\bar{a}_{\psi})^2}{\rho^2(u)}=0\,,\\
	&&\dfrac{u\,\rho(u)}{D_T(u)}\partial_u\bar{a}_{t}+3\bar{a}_\psi = \bar{\mathbf{k}}_{t}\,,\\
	&&\dfrac{u}{D_T(u)}\dfrac{f_T(u)}{\rho(u)}\partial_u\bar{a}_{\psi}+3\bar{a}_t = \bar{\mathbf{k}}_{\psi}\,,  
\end{eqnarray}
where $\bar{\mathbf{k}}_{t}$ and $\bar{\mathbf{k}}_{\psi}$ are two constants of motion which, despite we adopt the same notations, are {\it a priori} different from their counterparts in the confined phase.

Using the analogous redefinition as in \eqref{eq:ashift}, the complete set of equations of motion in the deconfined phase can be written as
\begin{align}\label{eq:eomsdeconf} \nonumber
	\partial_u \left( \dfrac{u^4\rho'(u) \rho(u) f_T(u)}{D_T(u)} \right)  &- R^3u\, D_T(u)\left(1- \dfrac{9\tilde{a}_t^2}{u^2 f_T(u)} \right) = 0\,, &&\\ \nonumber
	& \partial_u\tilde{a}_\psi + \frac{3}{u\,f_T(u)} \rho(u)\,D_T(u)\,\tilde{a}_t = 0\,,\\
	& \partial_u\tilde{a}_t + \dfrac{3}{u} \dfrac{D_T(u)}{\rho(u)}\,\tilde{a}_\psi= 0\,.
\end{align}
Using the latter we can rewrite $D_T(u)$ as 
\begin{equation}
	D_T(u) = \sqrt{\dfrac{1+\rho'(u)^2 \left(\dfrac{u}{R}\right)^3 f_T(u)}{1+ \dfrac{9}{u^2} \left(\dfrac{\tilde{a}_\psi^2}{\rho^2(u)}-  \dfrac{\tilde{a}_t^2}{f_T(u)}  \right)}}\,.
\end{equation}

\subsection*{Baryon number, fundamental string charge and angular momentum}
In the WSS model, the baryon vertex is associated with a D4-brane wrapped on the background four-sphere $S^4$. The presence of $n_B$ D4-branes sources the $C_5$ Ramond-Ramond field through the D-brane Chern-Simons term
\begin{equation}
	n_B\int_{S^4 \times \mathbb{R}_t} C_5\,.
\end{equation}
If the gauge connection on the wrapped D6-brane is turned on, the D6-brane can carry a wrapped (on $S^4$) D4-brane charge, hence a baryon charge $n_B$. The D6-brane can also carry a fundamental string charge, which corresponds in the field theory picture to the number of quarks building up the baryonic configuration. These features recall those of a charged cylindrical D2-brane (eventually ending on a D4-brane) in flat spacetime, known as \textit{supertube}, which supports both fundamental string charge and D0-brane charge \cite{Mateos:2001qs,Kruczenski:2002mn}.

The induced baryon charge can be read from the D6-brane CS coupling. Let us write the D6-brane worldvolume as $M_3\times S^4 = \mathbb{R}_t\times\Sigma\times S^4$, where $\Sigma$ is a surface extended along $\psi$ and $\rho(u)$, or equivalently along $\psi$ and $u$ with $u\in [u_*,u_J]$. Therefore, the D6-brane CS Lagrangian includes the term
\begin{equation}\label{C5coupling}
	\frac{1}{2\pi}\int_\text{D6} C_5\wedge f = \frac{1}{2\pi}\int_{\Sigma} f\,\cdot\int_{S^4\times \mathbb{R}_t} C_5\,,
\end{equation}
so that
\begin{equation}
	n_B = \frac{1}{2\pi}\int_{\Sigma}\, f = -\frac{1}{2\pi}\int_{\Sigma}du\, d\psi\, \partial_u a_{\psi}=  a_\psi(u_*) - a_\psi(u_J)\,,
	\label{eq:nbD6}
\end{equation}
where in the last equality, we used the axisymmetry of the gauge connection. Notably, $n_B$ appears as an integrated magnetic flux for the $U(1)$ gauge connection on the D6-brane, and it is related to the boundary values of the connection. We will return to the topological interpretation of the baryon number in section \ref{vortex}. If we define the shifted gauge connection as
\begin{align}
	\hat{a}_\psi &= a_\psi -\dfrac{\mathbf{k}_{t}}{3},&  \hat{a}_t &= a_t -\dfrac{\mathbf{k}_{\psi}}{3},
\label{shiftedgf}
\end{align}
we can rewrite (\ref{eq:nbD6}) as 
\begin{equation}
	n_B = \hat{a}_\psi(u_*) - \hat{a}_\psi(u_J)\,.
	\label{eq:baryonnumber}
\end{equation}

Fixing the baryon number thus gives a boundary condition on $a$ or, equivalently, on $\hat{a}$. As we will see shortly, the solutions to the equations of motion have to satisfy appropriate boundary conditions to describe metastable objects with baryon number.

The induced fundamental string charge on the D6-brane can be computed by performing the variation of the total action with respect to $f$\footnote{The identification of the string charge $q_s$ with the variation with respect to $f$ comes from considering that the latter appears in the fundamental string action as
	\begin{equation}\label{stringac}
		S_\text{f-string} \supset -\dfrac{q_s}{2\pi l_s^2}\int\left(B+2\pi l_s^2 f\right)\,.\nonumber
	\end{equation}
	Then, we can write
	\begin{equation}
		q_s = -2\pi l_s^2\frac{\delta S_\text{f-string}}{\delta B} = -\frac{\delta S_\text{f-string}}{\delta f}.\nonumber
		\label{qspt}
	\end{equation}
}
\begin{equation}
	q_s =\dfrac{\partial}{\partial(\partial_u a_t)}\,\int d\psi\,\mathcal{L}_\text{D6}= N\,\dfrac{\mathbf{k}_{t}}{3}\,,
	\label{eq:qscharge}    
\end{equation}
where in the last identity, we used the equations of motion. We stress that the above formula holds both in the confined and in the deconfined phases since the equation of motion for $a_t$ has the same expression in both phases (up to replacing $D$ by $D_T$).

As we have already shown in equation (\ref{eqn:CSconf}), on the D6-brane worldvolume, there is also the term
\begin{equation}\label{C3coupling}
	\frac{N}{4\pi} \int_{M_3} a\wedge f \,,
\end{equation}
which in the presence of the non-trivial magnetic flux in (\ref{eq:nbD6}) becomes
\begin{equation}\label{C3coupling2}
	N\, n_B \int_{\mathbb R_t} a_t(t)dt \quad \Rightarrow \quad q_s = N n_B\,,
\end{equation}
where $a_t(t)$ is now a gauge connection fluctuation.\footnote{It can be shown that our time-independent axisymmetric gauge connection can be gauge transformed into this new gauge configuration.} The result $q_s = N\,n_B$ comes from the identification of the D6-brane CS Lagrangian \eqref{C3coupling2} with the fundamental string Lagrangian; it is physically expected to hold in the presence of baryonic objects.\footnote{This relation holds only for confined baryonic objects, and it has to be modified in the presence of free quarks.} From this result and the definition (\ref{shiftedgf}), it also follows that
\begin{equation}\label{eq:ashiftnb}
    \hat{a}_\psi = a_\psi -n_B\,.
\end{equation}
Because of the matching condition we enforced gave $q_s = N \,\mathbf{k}_t/3 = Nn_B$, we have to impose $\mathbf{k}_t/3= n_B$ when we solve the equations of motion for the gauge connection.

Another property of the charged D6-brane configurations we are focusing on is that, as already discussed in \cite{Bigazzi:2024mge}, they carry angular momentum.
Following Noether's theorem, the angular momentum along the $u$ axis is
\begin{equation}
	J = \int_{\Sigma\times S^4} d^6x \,\sqrt{-g}\,g_{\psi\psi}\, T^{t\psi}.
	\label{definJ}
\end{equation}
The Chern-Simons term is topological, and hence its contribution to the stress-energy tensor vanishes. We are thus left with the stress-energy tensor contribution from the DBI action
\begin{equation}\label{eq:stress-energy}
	T_{MN} = -\frac{2}{\sqrt{-g}} \frac{\delta \mathcal{L}_\text{D6}}{\delta g^{MN}} = T_6\, e^{-\phi}\, \left(-g_{MN} \, \sqrt{1+ \frac{1}{2}g^{AB} g^{CD} \bar{f}_{AC} \bar{f}_{BD}} + \dfrac{ g^{AB}\,\bar{f}_{MA}\bar{f}_{NB}} {\sqrt{1+ \dfrac{1}{2}g^{AB} g^{CD} \bar{f}_{AC} \bar{f}_{BD}}}\right)\,,
\end{equation}
where $g$ is the induced D6-brane metric and the indices $M,N,A,B,C,D$ run over the D6-brane worldvolume coordinates. In particular 
\begin{equation}
	T^{t\psi} =\dfrac{ T_6\,e^{-\phi}}{\sqrt{-g_{tt}g_{\psi\psi}g_{uu}}}\dfrac{\partial_u\bar{a}_t\partial_u\bar{a}_\psi}{\sqrt{-[g_{tt}(g_{\psi\psi}g_{uu}+(\partial_u\bar{a}_{\psi})^2)+g_{\psi\psi}(\partial_u\bar{a}_{t})^2]}}\,.
\end{equation}
Using the above expression and the equations of motion, we get, from (\ref{definJ})
\begin{align}\label{eq:JD6}
	J = \dfrac{N}{2}\int du\,\partial_u\left[\left(a_\psi -\dfrac{\mathbf{k}_t}{3}\right)^2\right] 
	.
\end{align}
This formula holds for every D6-brane embedding and in both the confined and the deconfined phases. We can integrate it in $u$ between the generic endpoint in $u_*$ and $u_J$ to get
\begin{align}\nonumber
	J &= \dfrac{N}{2}\bigg[\left(a_\psi(u_J) -\dfrac{\mathbf{k}_t}{3}\right)^2-\left(a_\psi(u_*) -\dfrac{\mathbf{k}_t}{3}\right)^2\bigg]=\\
	&=\dfrac{N}{2}\left[ \left(\hat{a}_\psi(u_J)\right)^2 - \left(\hat{a}_\psi(u_*)\right)^2\right] = -\dfrac{N}{2}\left( \hat{a}_\psi(u_*) + \hat{a}_\psi(u_J)\right)n_B. \label{eq:spin}
\end{align}

\subsection{Zoology of (meta)stable solutions}\label{sec:zoology}

For a given value of $\mathbf{k}_t$ and $\mathbf{k}_\psi$, the space of solutions of the system of differential equations derived in (\ref{eq:eomsconf}) and (\ref{eq:eomsdeconf}) is a four-parameter space whose general structure is challenging to determine. 

We focus on the zoology of time-independent axisymmetric solutions, which can be interpreted from the field theory point of view:
\begin{itemize}
	\item \textbf{Baryons}. They correspond to D6-brane configurations that shrink to zero size at some value of the holographic coordinate $u_*  = u_E$ so that $\rho(u_E) = 0$. See figure \ref{fig:holobaryon}.
    \begin{figure}[H]
	\center
	\includegraphics[height = 4cm]{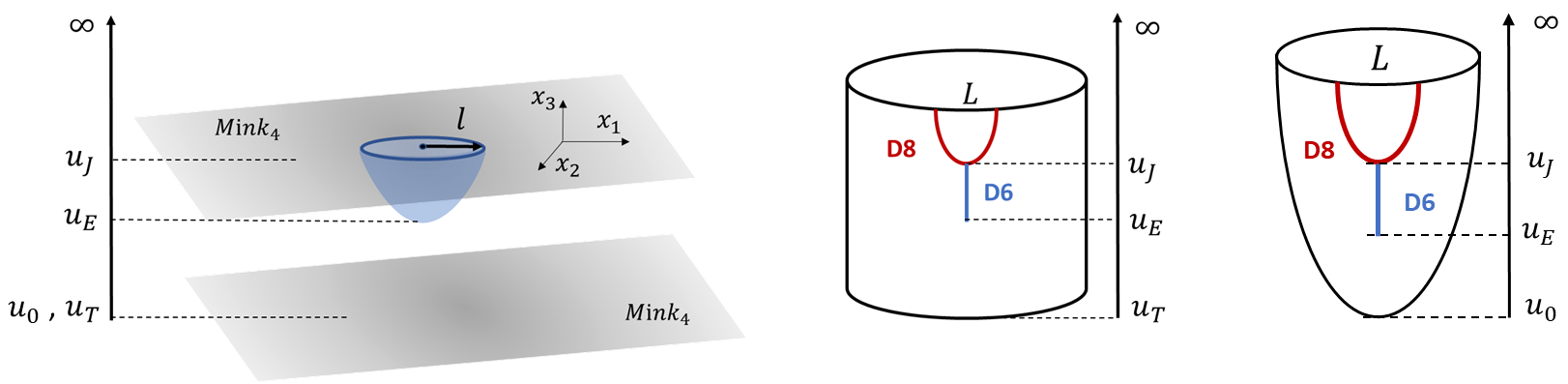}
	\caption{Pictorial representation of the D6-brane embedding describing holographically the Hall droplet baryon. The D6-brane has a radius $l$ at $u = u_J$ where it is attached to the D8-brane, and it shrinks to zero size at $u=u_E$. We present the embedding both in the deconfined (cylinder) phase and the confined (cigar) phase of the WSS model.}
	\label{fig:holobaryon}
\end{figure}

	\item \textbf{Vortons}. They are solutions in the deconfined phase for which $\rho(u)\neq 0$ for all $u$, and they extend from $u_J$ to the horizon $u_* = u_T$. See figure \ref{fig:holovorton}.

\begin{figure}[H]
	\center
	\includegraphics[height = 4cm]{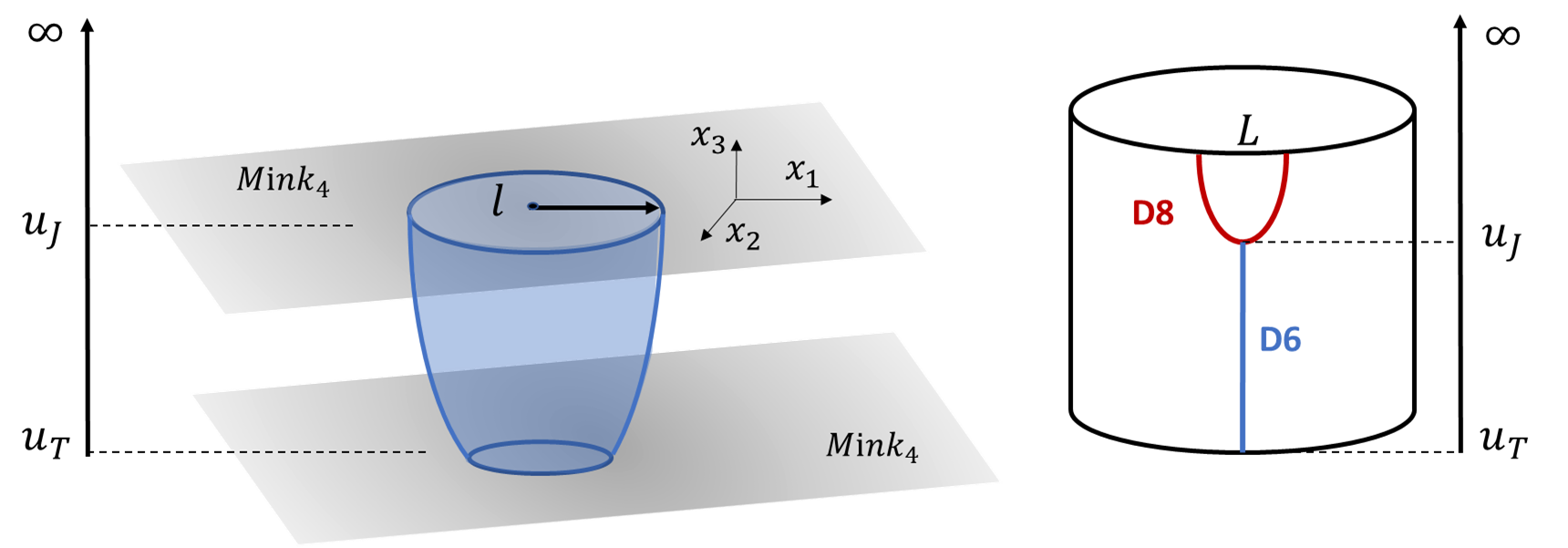}
	\caption{ Pictorial representation of the D6-brane embedding describing holographically the vorton. The D6-brane has a radius $l$ at $u = u_J$ where it is attached to the D8-brane, while it ends on the horizon with a non-trivial profile.}
	\label{fig:holovorton}
\end{figure}

	\item \textbf{\textit{Sandwich} vortons}. These are solutions such that $\rho(u)\neq 0$ for all $u$, and they extend from $u_J$ to $u_* = u_K$, which corresponds to the tip of another U-shaped flavor D8-brane.\footnote{Thus, since $u_K \neq u_J$, the two flavors condense at different scales. For simplicity, in this paper we only consider the two flavor branes with the tips at the same position along $x_4$.} These topological defects can be interpreted as excitations of two
different flavors leading to a ``sandwich'' structure. We will return to the interpretation of these objects in section \ref{sec:sandwich}.\footnote{We can also think of \textit{sandwich} defects with a straight boundary on the two D8-branes. In this case, the composite objects are straight global strings associated with two different flavors.} See figure \ref{fig:holosandwich}.

\begin{figure}[H]
	\center
	\includegraphics[height = 7cm]{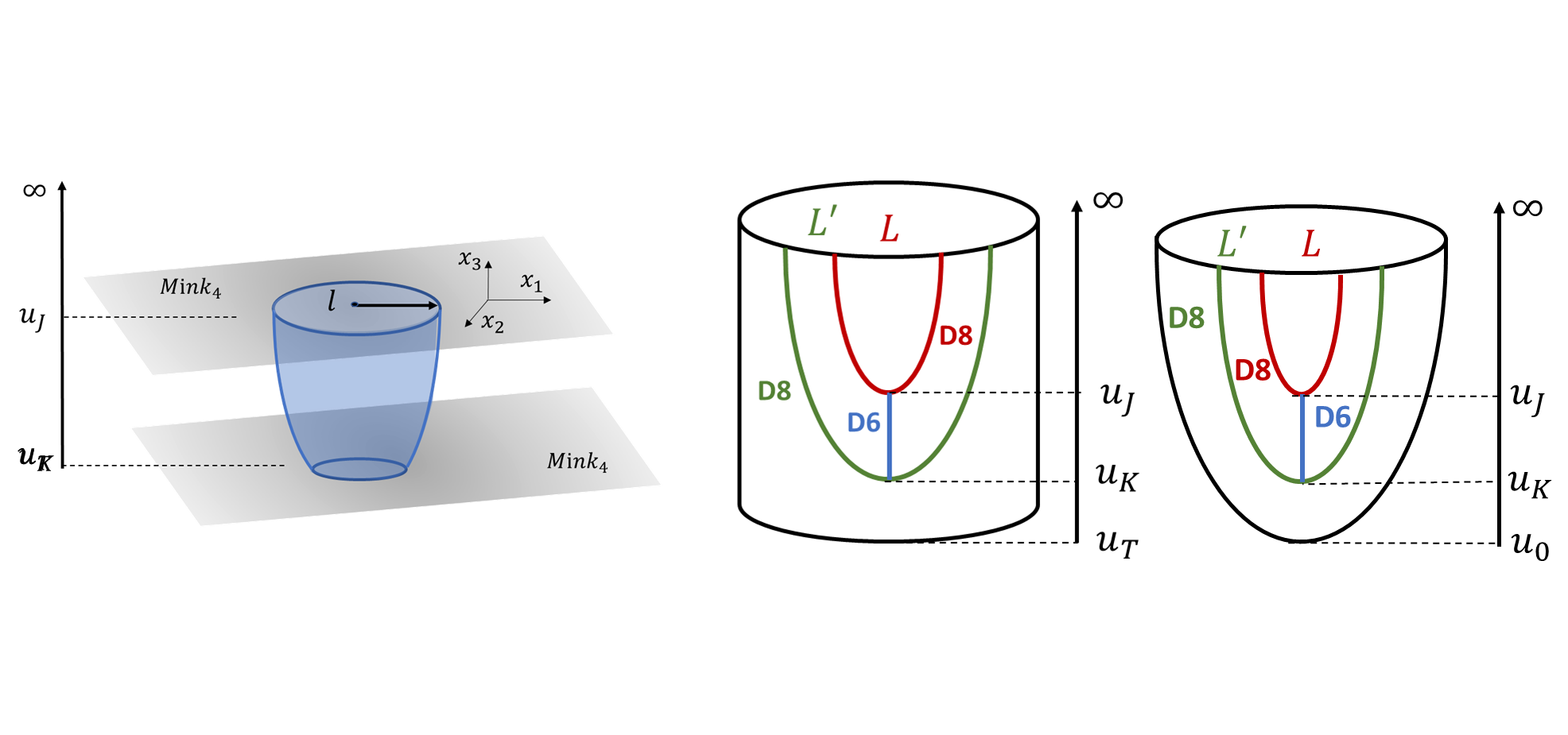}
	\caption{Pictorial representation of the D6-brane embedding describing holographically the \textit{sandwich} vorton. The D6-brane has a radius $l$ at $u = u_J$ where it is attached to the first D8-brane, while it ends on a second D8-brane at $u=u_K$. The embedding is presented in the deconfined (cylinder) and the confined (cigar) phase of the WSS model.}
	\label{fig:holosandwich}
\end{figure}

    \item \textbf{Punctured domain walls}. They are domain wall solutions attached to a circle of radius $l$ at $u_J$ that extend to $\rho \to \infty$ at some $u_*\geqslant u_0$ in the confined phase and $u_*\geqslant u_T$ in the deconfined phase. See figure \ref{fig:holopunctured}.

\begin{figure}[H]
	\center
	\includegraphics[height = 4cm]{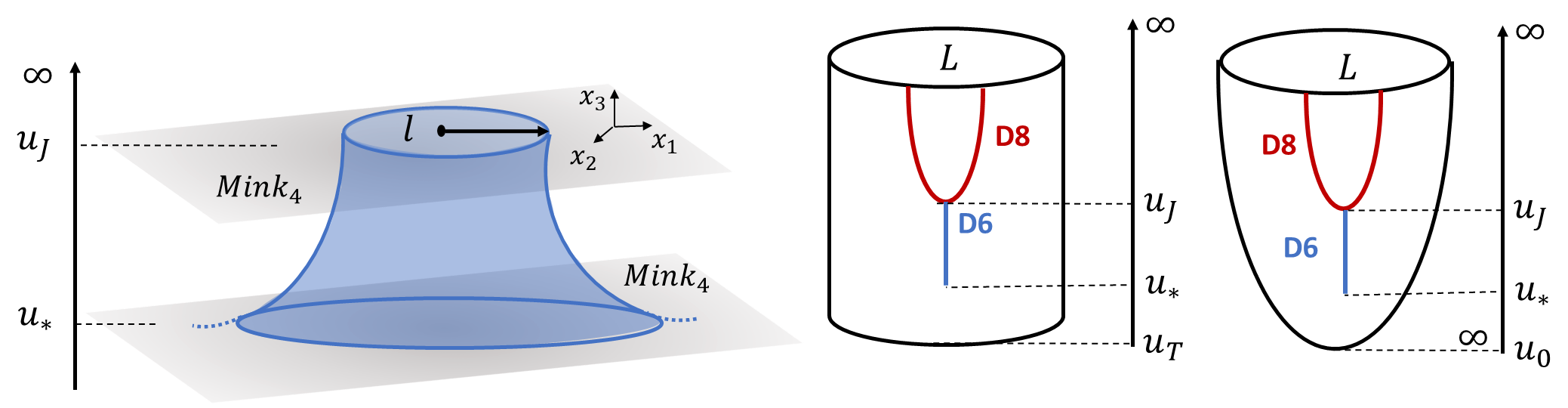}
	\caption{Pictorial representation of the D6-brane embedding describing holographically the punctured domain wall. The D6-brane has a radius $l$ at $u = u_J$ where it is attached to the D8-brane. It extends to $\rho\to\infty$ for a finite $u=u_*$. The embedding is presented in the deconfined (cylinder) phase and the confined (cigar) phase of the WSS model.}
	\label{fig:holopunctured}
\end{figure}
\end{itemize}

These configurations are not necessarily metastable. Time-independent stable or metastable solutions are generally a very restricted subclass of solutions for which we need to impose other constraints.

The first necessary criterion for (meta)stability, as already discussed in \cite{Bigazzi:2024mge}, comes from a no-force condition for intersecting branes, 
which is achieved when the intersection is orthogonal. 
This, in our case, will result in a first boundary condition
\begin{equation}
	\rho'(u_J) = 0\,.
\end{equation}
For D6-brane embeddings associated with baryon solutions, we also have to require that the embedding is a smooth manifold. For example, a conical shape at $u_*$ 
would signal some unbalanced tension,
causing the brane to 
change shape.
As a result, in the baryon case, we must have
\begin{equation}
	\rho'(u_E) \rightarrow\infty\,,
\end{equation}
where $u_*=u_E$ is the value of the holographic coordinate such that $\rho(u_E) = 0$. 

The \textit{sandwich} vortons have to be supported also with the orthogonality condition at $u_K$
\begin{equation}
	\rho'(u_K) = 0\,.
\end{equation}
For vortons in the deconfined phase, the orthogonality requirement at the horizon is discussed in \cite{Bigazzi:2024mge}. Expanding the metric in the near-horizon region, it is possible to show that orthogonality is guaranteed if and only if the embedding $\rho(u)$ admits a (regular) Taylor expansion in $u_T$.

\subsection{Variational principle for a DBI-CS action}\label{sec:variationalprinciple}
In this section, we address the variational problem for a DBI-CS action in the presence of a boundary. 
Let us recall that $M_3=\Sigma\times\mathbb{R}_t$ (and $\partial M_3= \partial \Sigma\times\mathbb{R}_t$) with $\mathbb{R}_t$ the time axis and $\Sigma$ a surface extended along $\psi, \rho(u)$, or equivalently along $\psi, u$, with $u\in [u_*,u_J]$ and with some profile $\rho(u)$. 

In our setup, both the DBI action and CS action produce a boundary variation under $a\to a+\delta a$. As a consequence of the presence of the DBI action, unlike in the pure CS case \cite{Elitzur:1989nr}, the variables $\delta a_t$ and $\delta a_\psi$ are not conjugate to each other, and we can safely choose Dirichlet boundary conditions for both of them to work out the variational problem. See \textit{e.g.} \cite{Tong:2022gpg} for a recent discussion on this topic. Therefore, we set
\begin{align}
    &\delta a_t\bigg|^{u_J}_{u_*}= 0\,,& &\delta a_\psi\bigg|^{u_J}_{u_*} = 0\,,&
\end{align}
and the variational problem is automatically well-posed. Let us recall that $u = u_J$, and $u=u_*$ represent the location of the boundary.

\subsubsection*{Boundary conditions for the gauge connection}
\label{sec:bc}
We are now ready to discuss the boundary conditions, consistent with the variational principle, that we will impose to solve the equations of motion. In the baryon setup, and in both phases, assuming that the electric field is finite at the tip of the D6-brane, the equation of motion for $a_t$ ((\ref{eq:defktkpsiconf}) in the confined phase or the one in (\ref{eq:eomsdeconf}) in the deconfined phase) implies that, at the tip where $\rho(u_E) = 0$ and $\rho' (u_E)\rightarrow\infty$, we have $a_\psi(u_E) = n_B$. This, together with the baryon number definition $n_B= a_\psi(u_E)-a_\psi(u_J)$, implies $a_\psi( u_J) = 0$ for the baryon.

In both phases and for all the other solutions with baryon number $n_B$, we will impose the same conditions, $a_\psi(u_J) = 0$ and $a_\psi(u_*) = n_B$. As a consequence for all solutions we have $\hat{a}_\psi(u_*) = 0$ and $\hat{a}_\psi(u_J) = -n_B$ using \eqref{eq:ashiftnb}. Let us summarize the boundary conditions of each configuration presented in section \ref{sec:zoology} separately:

\begin{align}
    &\textbf{Baryons:} & a_\psi(u_E)&=n_B  & a_t(u_E)&=0 & a_\psi(u_J)&=0\,.\label{bcfinalbar} \\   
    &\textbf{Vortons:} & a_\psi(u_T)&=n_B  & a_t(u_T)&=0 & a_\psi(u_J)&=0\,. \\
    &\textbf{Sandwich vortons:} & a_\psi(u_K)&=n_B  & a_t(u_K)&=0 & a_\psi(u_J)&=0\,.\\
    &\textbf{Punctured domain walls:} & a_\psi(u_*)&=n_B  & a_t(u_*)&=0 & a_\psi(u_J)&=0\,. 
\end{align}

Imposing the above conditions and recalling equation (\ref{eq:spin}), we find
\begin{equation}\label{eq:Jnbrelation}
J = \frac{N}{2}n_B^2\,,
\end{equation}
for every configuration. We insist that in the baryon case this is derived, while it is a consequence of our choice of boundary conditions for the other solutions. This quadratic dependence of the angular momentum on the charge is a typical feature of anyonic systems, and it is consistent with the observation that the charged D6-brane configurations we are focusing on are excitations of a topological phase of a $U(1)_N$ CS theory.\footnote{As already observed in \cite{Bigazzi:2024mge}, 
the state with opposite spin $J = -(N/2)n_B^2$ can be obtained by considering a $\overline{\text{D6}}$-brane for which the CS term changes sign.
} The above relation was already found and discussed in \cite{Bigazzi:2024mge} for the case of vortons. Here we see how a proper treatment of the boundary conditions allows us to extend it to other D6-brane configurations.
\subsection{Topological interpretation of the baryon number}
\label{vortex}
In this section, we classify the principal $U(1)$-bundles that can be defined on the D6-brane worldvolume. We show that these bundles are precisely classified by the baryon number. We identify the topological data at the intersection of the D6-brane and the D8-brane, we connect this boundary data to the baryon number, and show the similarities with the well-known Yang-Mills instanton description of baryons. Then, we provide a physical interpretation of the shifted gauge connection $\hat{a}$, which, as we shall see, describes in the baryon case a vortex living on the D6-brane. Finally, we outline some similarities and differences between the well-known Nielsen-Olesen vortex \cite{Nielsen:1973cs} and the one we study throughout this work, which we will refer to as \textit{DBI-CS vortex}.

\subsubsection*{Topological properties of the D6-brane}
In this section, we do not consider the $S^4$ on which the D6-brane and the D8-brane are wrapped. Moreover, for our purpose, the time axis will be neglected; therefore, the manifold of interest is the above-defined $\Sigma$. All the D6-brane configurations considered in this work have an intersection with the D8-brane in $u=u_J$, which has the topology of a circle $S^1$. This circle is a connected component of $\partial \Sigma$. For baryon solutions, it is {\em a priori} the only one, but the other solutions may have other connected components. Hence, we discuss separately the solutions for which $\rho(u)\neq0$ for all $u$ and the baryon.

\smallbreak
In the $\rho(u)\neq0$ case, which corresponds to the \textit{sandwich} vorton, the vorton and the punctured domain wall cases presented in section \ref{sec:zoology}, the boundary has two circular connected components. One is the circle in $u_J$, and the other is in $u_*$. This last circle is either the intersection with another U-shaped D8-brane (for \textit{sandwich} vortons), the intersection with the horizon (for vortons), or a circle at infinity (for the punctured domain walls). In all these cases, $\Sigma$ has the global topology of a cylinder between two circles, which we will denote $S^1_{J,*}$.

\smallbreak
In the baryon case, $\Sigma$ has a priori the topology of a disk. However, as we have seen in equation (\ref{bcfinalbar}), $a$ has a {\em vortex-like} singularity in $u_E$. We will interpret this $a_\psi(u_E) \neq 0$ as a topological defect at the tip of the D6-brane. A common workaround procedure to study these defects is to consider instead the manifold $\Sigma$ with a small neighborhood of the defect excised. This has multiple topological and physical consequences. Firstly, it adds another circular boundary component in $u_*=u_E$, which allows for a non-trivial value of the gauge connection at the tip without a singularity. Moreover, this implies that the discussion of baryons and the other D6-brane configurations can be conducted simultaneously, as both can be treated by considering a cylinder between two circles.

\subsubsection*{Classification of the principal $U(1)$-bundles}
\begin{figure}[H]
    \centering
    \begin{subfigure}{0.45\textwidth}
        \centering
        \includegraphics[width=\textwidth]{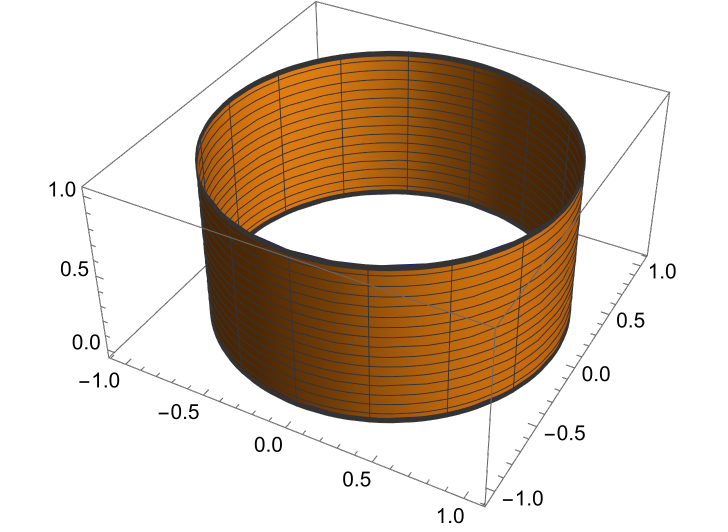}
        \caption{}
    \end{subfigure}
    \hfill
    \begin{subfigure}{0.45\textwidth}
        \centering
        \includegraphics[width=\textwidth]{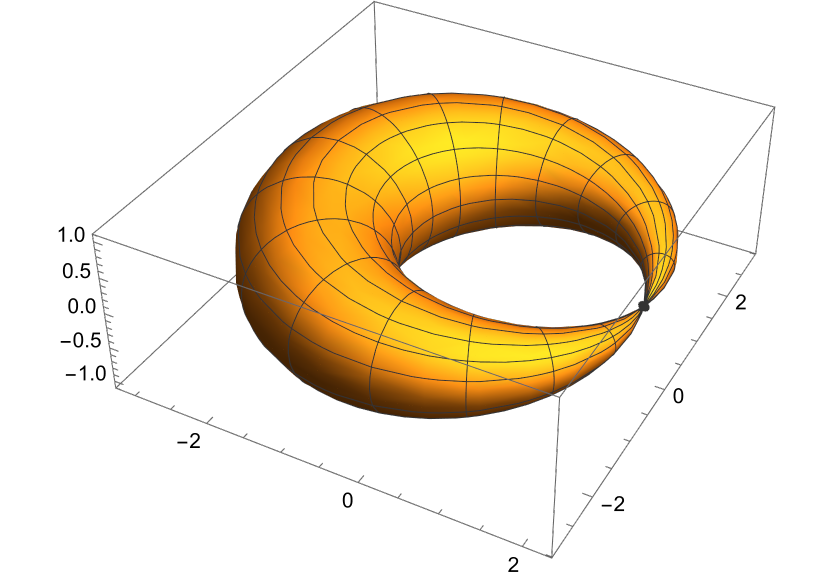}
        \caption{}
    \end{subfigure}
    \caption{Cylinder configuration (a) and its one-point compactification $\mathbb{PT}$ (b). The boundary is indicated in blue.}
\label{drawing}
\end{figure}

\smallbreak
We now want to work out a classification of the principal $U(1)$-bundles defined on the manifold $\Sigma$, which, as we have shown, can be reduced to the study of a cylinder. Note that, because of (\ref{curvboundary}), the curvature $f$ has to vanish on both bounding circles of the cylinder. Therefore, $f$ restricts to a map on the one-point compactification of $\Sigma  \setminus \partial \Sigma$. This corresponds to identifying the entire boundary $\partial \Sigma$ as a single point. Doing so, we obtain a {\em pinched} torus, noted $\mathbb{PT}$, illustrated in figure \ref{drawing}.

It has been shown that $U(1)$-principal bundles are classified up to gauge transformations by the second cohomology group of this manifold, with integer coefficients (see appendix \ref{ap:topo} for a more detailed explanation of this statement). This is a crucial difference with the baryons constructed usually with more flavors, which rely on non-trivial $SU(N_f)$-principal bundles, classified by the fourth cohomology group. The cohomology of the pinched torus is
\begin{equation}
    H^*(\mathbb{PT}; \mathbb{Z}) = (H^0(\mathbb{PT}; \mathbb{Z}),H^1(\mathbb{PT}; \mathbb{Z}), H^2(\mathbb{PT}; \mathbb{Z}), 0\dots)  = (\mathbb{Z}, \mathbb{Z}, \mathbb{Z}, 0\dots),
    \label{cohomopt}
\end{equation}
and all the higher cohomology groups vanish because of $\dim (\mathbb{PT})=2$. $H^0 = \mathbb{Z}$ is a direct consequence of the connectedness of $\mathbb{PT}$. The coefficient of $H^1$ describes a winding number.\footnote{Though in this case it comes from the ``gluing" operation we performed and is not physical.} The last integer classifies $U(1)$-principal bundles up to gauge transformations. It can be extracted by computing the first Chern number over the pinched torus:
\begin{equation}
    \frac{1}{2\pi}\int_{\mathbb{PT}} f \in \mathbb{Z}\,.
\end{equation}
This quantity is a quantized magnetic flux over $\mathbb{PT}$. The compactification we used mapped $\Sigma$ to $\mathbb{PT}$ and therefore the above integral is exactly the baryon number previously defined on $\Sigma$ in equation (\ref{eq:nbD6}). Therefore, the baryon number classifies the $U(1)$-principal bundles on $M_3$ up to gauge transformations.
\smallbreak
It is useful to consider the link between this construction and $SU(2)$ instantons in four-dimensional Euclidean space, which share many similarities. The mathematical objects we used and the analogous ones appearing in the instanton study are reported in table \ref{tab:instan}.

\begin{table}[H]
    \centering
    \begin{tabular}{|c|c|c|c|c|c|}
        \hline
          & Manifold & Boundary & Compactification & Group & Classification \\
        \hline
        Baryon & $S^1 \times [u_*, u_J]$ & $S^1_* \cup S^1_J$ & $\mathbb{PT}$ & $U(1)$ & $H^2(\mathbb{PT}; \mathbb{Z})$ \\ 
        \hline
        Instanton & $\overline{\mathbb{R}}\,^4$ & $S^3_{\infty}$ & $S^4$ & $SU(2)$ & $H^4(S^4; \mathbb{Z})$ \\ 
        \hline
    \end{tabular}
    \caption{Relevant groups and manifolds to the topology of the baryon and instanton case.}
    \label{tab:instan}
\end{table}
\smallbreak 
Though there are many similarities between these constructions, there is an additional element in our setup. The structure of the large gauge transformations on $\Sigma$ is non-trivial. The entire cylinder has a non-trivial first cohomology group, equal to $\mathbb{Z}$. Moreover, as is well known \cite{Elitzur:1989nr}, the allowed holonomies around circles for an Abelian Chern-Simons theory at level $N$ like the one we consider are given in units of $1/N$. Therefore, for every integer (divided by $N$), we can define a unique large gauge transformation\footnote{A large gauge transformation is a gauge transformation that is not homotopic to the identity. As we will see, it acts non-trivially on the boundary and has to be viewed as a global symmetry of the theory. The argued uniqueness is assuming it does not spoil the axisymmetry and time invariance of our ansatz.} over the manifold. We denote the fraction associated with such a transformation as $n_G$. As an example, in the particular case of an integer $n_G = n_B$, we can define a scalar $\chi$ as
\begin{align}
      \chi= \dfrac{\mathbf{k}_\psi}{3}\,t+\dfrac{\mathbf{k}_t}{3}\,\psi = \dfrac{\mathbf{k}_\psi}{3}\,t + n_B\, \psi\,,
    \label{eq:defchi}
\end{align}
where we used the result $\mathbf{k}_t = 3 n_B$ coming from \eqref{C3coupling2}, such that we can send $a$ to the shifted gauge connection $\hat{a}$ through the following large gauge transformation
    \begin{align}
    a &= \hat{a} + d\chi\,.
    \end{align}
\smallbreak
Importantly, after this transformation, $\hat{a}$ is always a regular field, \textit{i.e.}~$\hat{a}(u_E) = 0$ in the baryon case. This was already known from the boundary conditions (\ref{bcfinalbar}), though in the baryon case, connecting them by a large gauge transformation was only possible because of this excision of the defect.

We will now connect the bulk data (the baryon number) to the topological data of the boundary. As we have seen, the latter consists of two disconnected circles on which $f=0$, therefore, locally $a=d\varphi$ but not globally, as $a$ can wind around circles. This winding is classified by $\pi_1(U(1)) = \mathbb{Z}$. The possible pure gauge configurations are therefore split into topological sectors, labeled by their holonomies. We define two fractional topological numbers from the holonomy of $\hat{a}$ (the regularized gauge connection) over the bounding circle. 
The orientation we chose for $M_3$, $dt\wedge d\psi\wedge du = +dt\,d\psi\,du$, induces an orientation on the bounding circles, which results in an opposite sign for the $u_J$ circle and the $u_*$ one:\footnote{Subtleties might arise when $u_*=u_T$ is the event horizon of the spacetime.}
\begin{equation}
    n_J = -\frac{1}{2\pi} \int_{S^1_{J}} \hat{a}(u_J)\,, \qquad   n_* = \frac{1}{2\pi} \int_{S^1_*} \hat{a}(u_*)\,.
    \label{eq:windingflavor}
\end{equation}
Recall that for baryon configurations $u_*=u_E$ and therefore $n_*=n_E=0$, while for sandwich solutions $u_*=u_K$ and hence $n_*=n_K$. These numbers are connected to the winding numbers $n^{(W)}_{J,*}$ by the following relations:
\begin{equation}
    n^{(W)}_{J} \equiv -\frac{N}{2\pi} \int_{S^1_{J}} \hat{a}(u_J) = Nn_J\,, \qquad   n^{(W)}_{*}  \equiv \frac{N}{2\pi} \int_{S^1_*} \hat{a}(u_*) = Nn_*\,.
\end{equation}

The gauge connection $\hat{a}$ can then be interpreted as a topological soliton interpolating between two topological sectors, one at $u=u_J$ characterized by $n_J$ and the other $u=u_*$ by $n_*$. Using Stokes' theorem, we can then rewrite the baryon number $n_B$ as
\begin{equation}
    n_B = n_* + n_J\,,
    \label{windingchern}
\end{equation}
where we stress that, even though $n_*$ and $n_J$ are fractional, their sum has to be an integer in virtue of (\ref{cohomopt}). Importantly, these two fractional numbers are not invariant under a global action $n_G$, which can, for the same reason, take fractional values $ n_G = n^{(W)}_G/N,$ with $n^{(W)}_G \in \mathbb{Z}$. One can check that the large gauge transformation $n_G$ transforms $n_J \to n_J - n_G$ and $n_* \to n_* + n_G$, which confirms that even though these numbers are not invariant, their sum $n_B$ is. Moreover, equation (\ref{windingchern}) implies that for a D6-brane solution to have a non-zero baryon number, it has to have at least one non-trivial winding over one of the circles at the boundary. With the choice of conventions that we made, the winding number on an intersection with a U-shaped flavor D8-brane counts exactly the net number of fundamental strings coming out from the aforementioned flavor brane. This number corresponds in the field theory to the number of quarks associated with the corresponding flavor.

We can rewrite the formula for the angular momentum \eqref{eq:spin} in terms of the holonomies of $a$ \eqref{eq:windingflavor}
\begin{align}\label{eq:Jholonomies}
    J = \dfrac{N}{2}(n_J^2-n_*^2)=\dfrac{N\,n_B}{2}(n_J-n_*)\,.
\end{align}
Then we can change the topological sectors defined by the two holonomies by $n_J \to n_J - n_G$ and $n_* \to n_* + n_G$. As we have seen before, this transformation associated with $n_G$ does not change the baryon number $n_B$, however, it affects the total spin of the D6-brane:
\begin{align}
    J \to J - N\,n_B\,n_G = J -n_B\,n^{(W)}_G.
\end{align}
Therefore, the spin is shifted by an integer number $-n_B\,n^{(W)}_G$.\footnote{We want to outline a connection between formulas \eqref{eq:baryonnumber}, \eqref{eq:spin} or equivalently \eqref{windingchern}, \eqref{eq:Jholonomies} and the results presented in \cite{Jackiw:1990pr} where they study an Abelian Chern-Simons theory coupled with a charged scalar field in (2+1)-dimensions. They find the very same formula for the magnetic flux, and the anyonic-like angular momentum for solitons in this theory. See also \cite{Bolognesi:2007ez} for more recent investigation of Chern-Simons solitons.}

\begin{figure}[H]
    \centering
    \begin{subfigure}{0.45\textwidth}
        \centering
        \includegraphics[width=\textwidth]{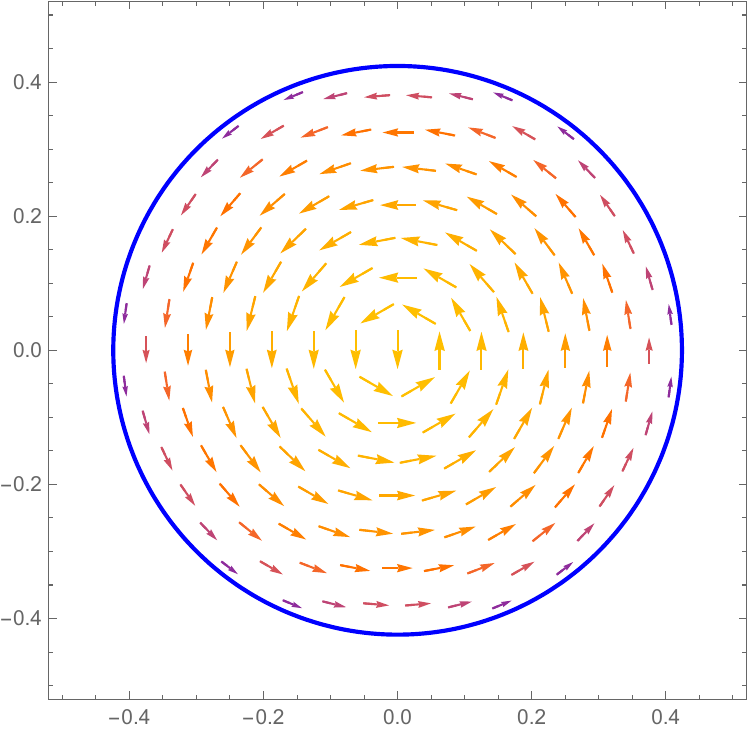}
        \caption{Singular connection $a_\psi$.}
    \end{subfigure}
    \hfill
    \begin{subfigure}{0.45\textwidth}
        \centering
        \includegraphics[width=\textwidth]{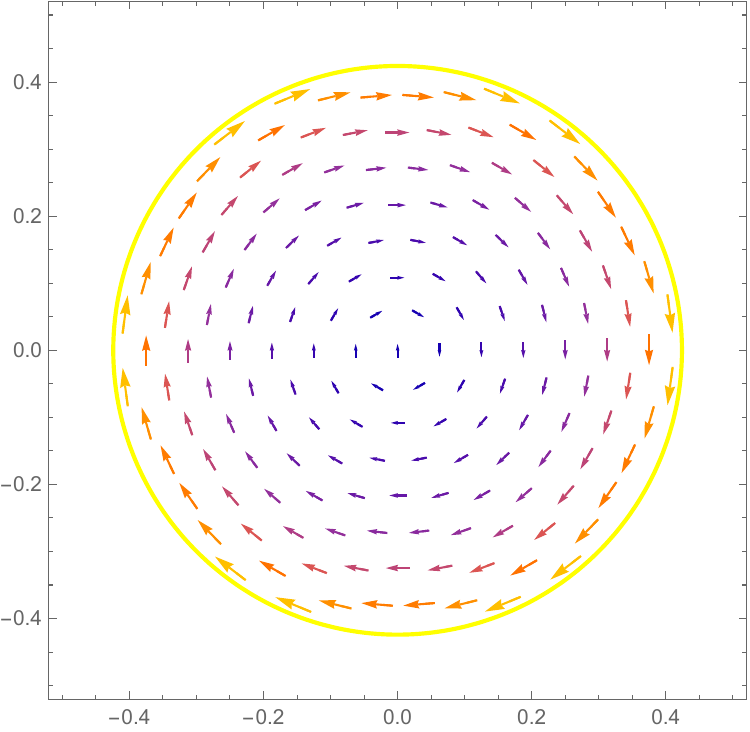}
        \caption{Regular connection $\hat{a}_\psi$.}
    \end{subfigure}
    \caption{Projection of the revolution plot of $a_\psi$ (left panel) and $\hat{a}_\psi$ (right panel). This solution is computed for $\lambda = 100$ and $b=0.4$ in the confined phase.}
\label{fig:vortex}
\end{figure}

It is instructive to visualize the gauge connection component $a_\psi$ for a baryon solution to illustrate the singularity we discussed. We have not yet illustrated the strategy to solve the equations of motion, but for the purposes of this section, we can present directly the result for the gauge connection component $a_\psi$, and its shifted version $\hat{a}_\psi$. Figure \ref{fig:vortex} shows the revolution plot around $u$ of $a_\psi(u)$ (left panel) and $\hat{a}_\psi(u)$ (right panel) projected into a plane of fixed $u$. The solid circumference corresponds to the value of the field at $u=u_J$, while the center corresponds to $u=u_E$. The color represents the strength of the field, going from 0 in blue to $|n_B|$ in yellow. As explained previously, the gauge connection $a_\psi$ is singular at the D6-brane tip while it is zero in $u=u_J$, as imposed by the boundary conditions discussed in the previous section. The shifted gauge connection $\hat{a}_\psi$ is regularized, but becomes nonzero at the boundary $u=u_J$ due to the action of the large gauge transformation.

Summarizing the discussion of this section, in all the cases under study, we can define a global baryon number, which is shown to be given by the first Chern number, and classifies the possible $U(1)$-principal bundles up to gauge transformations. Up to removing (in the baryon case) from the manifold a small circle around the point where the defect is located, one can define a global topological number $n_G$ around the D6-brane profile. $\chi$, a large gauge transformation connecting $a$ and $\hat{a}$, has a non-trivial $n_G=n_B$. For the baryon, $\hat{a}$ appears as a regularized connection, in the sense that the singularity is removed from the tip of the D6-brane in $u_E$. This regular field is then used to define two boundary holonomies, $n_*$ and $n_J$, that are directly related to the baryon number and count the number of field theory quarks (divided by $N$) that form the baryonic state.

\subsubsection*{The DBI-CS vortex}
We shall now point out an interesting interpretation of $\hat{a}$ in the baryon case, which has a single non-zero integer holonomy in $u_J$.
The baryon solution we are studying shares many similarities with the Nielsen-Olesen vortex \cite{Nielsen:1973cs} (see also \cite{Tong:2005un} for a review). Most importantly, both of them arise from a similar topological construction: as we discussed, the shifted gauge connection $\hat{a}$ is regular in $u_E$, so that the non-trivial topological number is given just by $n_J$ (since $n_E=0$, this is an integer), which is associated with the homotopy group
\begin{equation}
    \pi_1(U(1)) = \mathbb{Z}\,,
\end{equation}
as for the Nielsen-Olesen vortex. Additionally, both configurations carry a magnetic flux, quantized for topological reasons. However, there are two significant differences with respect to the Nielsen-Olesen case. Nielsen-Olesen vortex solutions are constructed by starting with a non-trivial singular solution with non-zero winding at the boundary at infinity, and then allowing it to depend on a radial coordinate to regularize it at the origin. Instead, we first obtained the gauge connection $a$, which did not have winding around the boundary $u_J$, and it was singular in $u_E$ (see left panel of figure \ref{fig:vortex}). Then, the singularity is resolved by applying a non-trivial gauge transformation $\chi$, which gives it a winding at the boundary in $u_J$. 

The second difference is that Nielsen-Olesen vortices are often introduced for theories with another scalar field charged under the $U(1)$, typically a Higgs field, which is absent here. In these Maxwell-Higgs models, the scalar vacuum expectation value connects $a_t$ and $\partial_ua_\psi$ as well as $a_\psi$ and $\partial_ua_t$, see \cite{Horvathy:2008hd}. In our system, by looking at the equations of motion in the confined and the deconfined phase \eqref{eq:eomsconf} and \eqref{eq:eomsdeconf}, we observe that the quantity that connects them comes from the DBI Lagrangian and it is mediated by the quantity $D_{(T)}(u)$. We can interpret this observation as if the role of the scalar field is played by geometrical degrees of freedom in our system.


\subsubsection*{Comments on the interpretation of baryons as quantum Hall droplets}\label{sec:edge}
The D6-brane configurations presented in this work correspond to non-trivial excitations of a topological phase of matter given by the $U(1)_N$ CS theory living on their worldvolume (after the reduction on the four-sphere). Such excitations are also commonly studied from the boundary point of view, after adding a dynamical term living on the boundary for phenomenological reasons. For comprehensive reviews on the topic, see \textit{e.g.}~\cite{Wen:2004ym,Tong:2016kpv,Arouca:2022psl}. This famously known edge dynamics is then described by the Floreanini-Jackiw action \cite{Floreanini:1987as}, and is the one governing the boundary physics of quantum Hall states. In our setup, we do not have access to the standard Floreanini-Jackiw action, since there is no explicit boundary term in the effective three-dimensional theory. 
However, we can identify the solitonic mode of the gauge configuration $a$ with $\chi$, being $a = d\chi$.

With this observation in mind, we can discuss the topological sectors of the edge theory, which on the field theory side allow us to define the ``electron operators" of integer charge in the FQHE. They are used to define the baryon charge and the corresponding charged operator. See, for example, \cite{Komargodski:2018odf}. In our holographic model, such an operator can be formulated as
\begin{equation}
    \mathcal{O}_B = e^{i N \chi}\,.
\end{equation}
This is the minimal local operator that carries a baryon charge.

For the field $\chi$, it is easy to verify that
\begin{equation}
    n_B = \frac{1}{2\pi}\int_{u_J} d\chi\,.
\end{equation}
Therefore, as in the standard FQHE, the solitonic part of the edge mode gives the charge of the droplet configuration.

\subsection{Analysis of the D6-brane energy}
\label{sec:energy}
In this section, we derive the formula for the energy of the D6-brane configuration.
The total D6-brane energy can also be derived from Noether's theorem by means of the stress-energy tensor \eqref{eq:stress-energy}\footnote{Just as the CS term, the boundary action does not contribute to the stress-energy tensor.}
\begin{equation}
    E = - \int du \,d\psi \,d\Omega_4 \sqrt{-g}\,\, T_t\,^t\,,
\end{equation}
with
\begin{equation}
    T_t\,^t = g^{tt}T_{tt} = - \left( \frac{R}{u}\right)^{3/4} \frac{1}{g_s (2\pi)^6 l_s^7} \left(u^{7/4} R^{9/4}\rho(u) D(u) + \frac{u^{7/4} R^{9/4}\rho(u) (\partial_u\tilde{a}_t)^2}{D(u)} \right)\,.
\end{equation}

After some algebra, we get
\begin{equation}
    E = \frac{N}{3(2\pi l_s^2)^2} \int^{u_J}_{u_*} du \,u\,\rho(u) \left( D(u)  +  \frac{(\partial_u\tilde{a}_t)^2}{D(u)}\right)\,.
\label{eq:EE}    
\end{equation}
Using the equations of motion for $\tilde{a}_\psi$ in (\ref{eq:eomsconf}) and the definitions in (\ref{eq:ashift}) we get
\begin{align}\nonumber
    E &= \frac{N}{3(2\pi l_s^2)^2} \int^{u_J}_{u_*} du\left( u\,\rho(u) D(u) - 3 \tilde{a}_\psi\partial_u\tilde{a}_t\right)\\
    &= \frac{N}{3(2\pi l_s^2)^2} \int^{u_J}_{u_*} du\left( u\,\rho(u) D(u) - 3 \bar{a}_\psi\partial_u\bar{a}_t\right) + N\,n_B\,\mu\,,
\end{align}
where we have introduced a would-be chemical potential $\mu$ conjugate to $n_B$
\begin{equation}
    \mu\equiv\int^{u_J}_{u_*}du\,\partial_u a_t = a_t(u_J)\,.
\end{equation}

In the deconfined phase, the result looks formally the same
\begin{equation}
     E  = \frac{N}{3(2\pi l_s^2)^2} \int^{u_J}_{u_*} du\left( u\,\rho(u) D_T(u) - 3 \bar{a}_\psi\partial_u\bar{a}_t\right) + N\,n_B\,\mu\,.
\end{equation}

Let us observe that the energy is equal to the Helmholtz free energy $F$, calculated from the on-shell value of the action $\tilde S = \int dt \tilde L = -\int dt F$, where $\tilde L$ is the Legendre transformed Lagrangian in the u-coordinate (see \textit{e.g.} \cite{Bigazzi:2011it}). We also point out that the final form of the energy is the non-supersymmetric version of the BPS result present in \cite{Mateos:2001qs,Emparan:2001ux,Mateos:2001pi,Kruczenski:2002mn} where the sum of the charges (with appropriate coefficients) gives the energy.

The energy of the D6-brane configurations we are focusing on, is bounded from below by a \textit{Bogomol'nyi bound} that arises because of the topological nature of the solutions we are considering. The derivation of the bound is very similar in both phases, though the result differs. We show the derivation in the confined phase, and we indicate the result in the deconfined phase at the end. Let us start from the total energy of the configuration
\begin{equation}
    E = \frac{N}{3(2\pi l_s^2)^2} \int^{u_J}_{u_E} du\, \frac{u\,\rho(u)}{D(u)} \left(\frac{1}{f(u)}+\rho'(u)^2 \left(\frac{u}{R}\right)^3 +  \frac{(\partial_u\tilde{a}_\psi)^2}{\rho^2(u)}  \right).
    \label{energy2}
\end{equation}
The first two terms can be combined to give the Lagrangian in the uncharged ($a=0$) case, and we can use the usual Bogomol'nyi trick of completing the square to get
\begin{align}\nonumber
    E = &\frac{N}{3(2\pi l_s^2)^2} \int^{u_J}_{u_E} \,\bigg\{du\frac{u\,\rho(u)}{D(u)}  \left(\sqrt{\frac{1}{f(u)}+\rho'(u)^2 \left(\frac{u}{R}\right)^3 } +  \frac{\partial_u\tilde{a}_\psi}{\rho(u)} \right)^2 \\
    &-  2 \partial_u\tilde{a}_\psi \sqrt{u^2 + 9 \left(\frac{\tilde{a}_\psi^2}{\rho^2(u)}- \tilde{a}_t^2\right)}\bigg\}\,.
\end{align}

Now, we can easily derive the Bogomol'nyi bound
\begin{equation}
    E \geqslant  E_{B}^{\text{conf.}}\equiv-\frac{2N}{3(2\pi l_s^2)^2} \int^{u_J}_{u_E} \, du \,u\,\partial_u \tilde{a}_\psi \sqrt{1 + \frac{9}{u^2} \left(\frac{\tilde{a}_\psi^2}{\rho^2(u)}- \tilde{a}_t^2  \right)} \,,
    \label{eq:Bogolboundconf}
\end{equation}
with the associated Bogomol'nyi equation
\begin{equation}
    \partial_u\tilde{a}_\psi = - \rho(u) \,\sqrt{\rho'(u)^2 \left(\frac{u}{R}\right)^3 + \frac{1}{f(u)}}\,,
    \label{eq:Bogoleqconf}
\end{equation}
which has to hold if the bound is saturated. 

In the deconfined phase, the Bogomol'nyi bound reads
\begin{equation}
    E \geqslant  E_{B}^\text{deconf.}\equiv-\frac{2N}{3(2\pi l_s^2)^2} \int^{u_J}_{u_*} \, du \,u\,\partial_u \tilde{a}_\psi\sqrt{f_T(u)} \sqrt{1 + \frac{9}{u^2} \left(\frac{\tilde{a}_\psi^2}{\rho^2(u)}- \frac{\tilde{a}_t^2}{f_T(u)}  \right)} \,,
    \label{eq:Bogolbounddeconf}
\end{equation}
and the associated Bogomol'nyi equation
\begin{equation}
    \partial_u\tilde{a}_\psi = - \rho(u)\, \sqrt{\rho'(u)^2 \left(\frac{u}{R}\right)^3 + \frac{1}{f_T(u)}}\,.
    \label{eq:Bogoleqdeconf}
\end{equation}
We do not expect a solution to the full problem satisfying \eqref{eq:Bogoleqdeconf} or \eqref{eq:Bogoleqconf} to exist. However, if it were possible to saturate the bound, we would get an energy formally given by the same expression in the two phases
\begin{equation}
E = E_{B}^\text{deconf.}= E_{B}^\text{conf.} = - \frac{2N}{3(2\pi l_s^2)^2} \int^{u_J}_{u_*} du\,\tilde{a}_t\partial_u\tilde{a}_\psi \geqslant C_{\lambda, b}|n_B|\,,
\label{BogoboundE}
\end{equation}
where $C_{\lambda, b}$ is a strictly positive constant, depending on the value of $\lambda$ and $b$.

\section{The Hall droplet baryon}\label{sec:baryon}
In this section, with the aim of providing a D6-brane realization, within the WSS model, of the Hall droplet baryon discussed in \cite{Komargodski:2018odf}, we present the numerical solutions of the Euler-Lagrange equations for the baryon configuration. We remind that the baryon solution corresponds to a charged D6-brane configuration, which is attached at the D8-branes' tip at $u=u_J$, and it smoothly shrinks to zero size at $u=u_E< u_J$. The details on the numerical methods used to solve the equations of motion can be found in the appendix \ref{app:numerics}. We will present the results separately for the confined and deconfined phases of the WSS model. For both phases, we have checked analytically that the boundary conditions for the fields $\rho(u)$, $\tilde{a}_t(u)$, and $\tilde{a}_\psi(u)$ are consistent with the equations of motion by employing a series expansion around $u_E$ and $u_J$.

\subsection{Excursus: rigid rotor estimates}
	
Before we present the numerical results, we approximate the D6-brane as a disk-like rigid rotor in order to extract analytic estimates for the stability radius and the energy. We will then compare these estimates with the numerical results and show that the rigid rotor is a good approximation for the baryonic D6-brane's size.

For a rigid rotor with mass $M$, moment of inertia $I$, spinning with angular momentum of modulus $J$, the total energy in the non-relativistic limit can be approximated as
\begin{equation}
    E = M + \frac{J^2}{2 I}\,.
    \label{eq:Erotor}
\end{equation}
For a disk-like rigid rotor of radius $l$, $I = \frac{1}{2}M \,l^2$. Approximating the D6-brane baryon with such an object is justified since, as we will show, the extension of the D6-brane along the holographic coordinate $u$, compared to its radius on the D8-brane, is found to be parametrically small in 
$\lambda$. This means that the D6-brane is squeezed at $u = u_J$ for which $l=\rho(u_J)\sim$ constant, making the disk approximation consistent. 

Let us consider solutions that satisfy the relation $J=N\,n_B^2/2$. The mass $M$ can be roughly estimated from the D6-brane DBI action by setting the gauge connection $a$ to zero. Let us proceed in the confining phase for which $M$, defined as $S_\text{D6}=-\int dt\,M$, reads
 \begin{equation}
     M = \frac{N}{3(2\pi l_s^2)^2} \int du\,u\,\rho(u)D(u)_{a=0}\,.
 \end{equation}
 
In the disk approximation, the first derivative of the embedding, $\rho'(u)$, tends to infinity throughout the D6-brane extension, and hence $D$ (defined in equation (\ref{eq:Dconf})) at $a=0$ can be approximated as
\begin{equation}
    D(u)_{a=0} = \sqrt{\left(\dfrac{u}{R}\right)^3\rho'(u)^2+\dfrac{1}{f(u)}}= \left(\frac{u}{R}\right)^{3/2} \rho'(u)+...
\end{equation}
Therefore, after some algebra, the mass $M$ can be written as
\begin{equation}
    M = \frac{2N \lambda^2}{3^6\pi^2\,b^{5/2}}\,M_{KK}^3\,l^2\,.
\end{equation}
We can replace this value of $M$ in the equation for the total energy \eqref{eq:Erotor} and minimize it as a function of $l=\rho(u_J)$. As a result, we get an analytical, approximate estimate of  the corresponding stability radius $l_\text{stable}$, given by
\begin{equation}\label{eq:lstablerigidconf}
    l_\text{stable}\, M_{KK} = \frac{3^2 \pi^{2/3} \,b^{5/6}}{\sqrt{2}}\,\left(\dfrac{n_B}{\lambda}\right)^{2/3}\,.
\end{equation}

Therefore, the stability radius in the confined phase scales with the dimensional parameter $M_{KK}\leftrightarrow \Lambda$ as $l_\text{stable} \sim \Lambda^{-1}$, and is independent of $N$, in agreement with field theory expectations \cite{Komargodski:2018odf}. Moreover, $l_\text{stable}\sim(n_B/\lambda)^{2/3}$ and therefore if $n_B=1$ we find $l_\text{stable}\sim\lambda^{-2/3}$. We can introduce the parameter 
\begin{equation}\label{eq:xiconf}
    \xi = \dfrac{n_B}{\lambda},
\end{equation}
to write the stability radius as
\begin{equation}
     l_\text{stable}\, M_{KK} = \frac{3^2 \pi^{2/3} \,b^{5/6}}{\sqrt{2}}\,\xi^{2/3}\,.
\end{equation}
The parameter $\xi$ is parametrically small whenever we consider a baryon with $n_B\sim\mathcal{O}(\lambda^0)$ while it becomes of order one when $n_B\sim\lambda$. We can also derive an analytical estimate for the on-shell energy of the stable configuration as
\begin{equation}\label{energyrigidconf}
    E = \dfrac{N\,n_B^{4/3}\lambda^{2/3}}{6\pi^{2/3}b^{5/6}}\,M_{KK} = \dfrac{N\,\lambda^{2}}{6\pi^{2/3}b^{5/6}}\,\xi^{4/3}M_{KK}\,.
\end{equation}
Therefore, if we consider a solution with baryon number $n_B\sim\mathcal{O}(\lambda^0)$, the energy scales with $\lambda$ as $E\sim\lambda^{2/3}$.

In the deconfined phase, similar formulas can be worked out, starting from the $a=0$ limit of eqns. (\ref{Ddeconf}) and (\ref{eq:Ddeconf}). The mass of the rigid rotor now reads
\begin{equation}
   M = \frac{2N \lambda^2_{T}}{3^6\pi^2\,\tilde{b}^{5/2}}\left(1-\tilde{b}^3\right)\left(\dfrac{T}{T_c}\right)^3\,M_{KK}^3\,l^2\,,
\end{equation}
where we have defined the temperature-dependent 't Hooft coupling
\begin{equation}
    \lambda_T  =\dfrac{T}{T_c}\,\lambda\,,
\end{equation}
with $T_c$ being the critical temperature for deconfinement (\ref{criticalT}). The stability radius and the corresponding energy now read
\begin{align}\label{eq:lstablerigiddeconf}
&l_\text{stable}\,M_{KK}=\dfrac{3^2\pi^{2/3}\,\tilde{b}^{5/6}}{\left(1-\tilde{b}^3\right)^{1/3}\sqrt{2}}\,\dfrac{T_c}{T}\left(\dfrac{n_B}{\lambda_T}\right)^{2/3}=\dfrac{3^2\pi^{2/3}\,\tilde{b}^{5/6}}{\left(1-\tilde{b}^3\right)^{1/3}\sqrt{2}}\,\dfrac{T_c}{T}\,\xi_T^{2/3},\\ \label{eq:energyrigiddeconf}
    &E = \dfrac{N\,n_B^{4/3}\lambda_T^{2/3}}{6\pi^{2/3}\tilde{b}^{5/6}}\,\left(1-\tilde{b}^3\right)^{1/3}\dfrac{T}{T_c}M_{KK} = \dfrac{N\,\lambda^{2}_T}{6\pi^{2/3}\tilde{b}^{5/6}}\,\left(1-\tilde{b}^3\right)^{1/3}\xi^{4/3}_T\,\dfrac{T}{T_c}M_{KK}\,,
\end{align}
where 
\begin{equation}
    \xi_T = \dfrac{n_B}{\lambda_T}\,.
\end{equation}

In the following, we will present the actual numerical results for the D6-brane baryon solutions in both the confined and the deconfined phases and compare them with the above analytical estimates. We will see that the rigid rotor approximation correctly captures the behavior of the stability radius as a function of $b$ (or $\tilde{b}$) and $\lambda$ (or $\lambda_T$) while the energy fails to be reproduced by this approximation.

\subsection{Confined phase}
\begin{figure}[H]
    \centering
    \begin{subfigure}{0.45\textwidth}
        \centering
        \includegraphics[width=\textwidth]{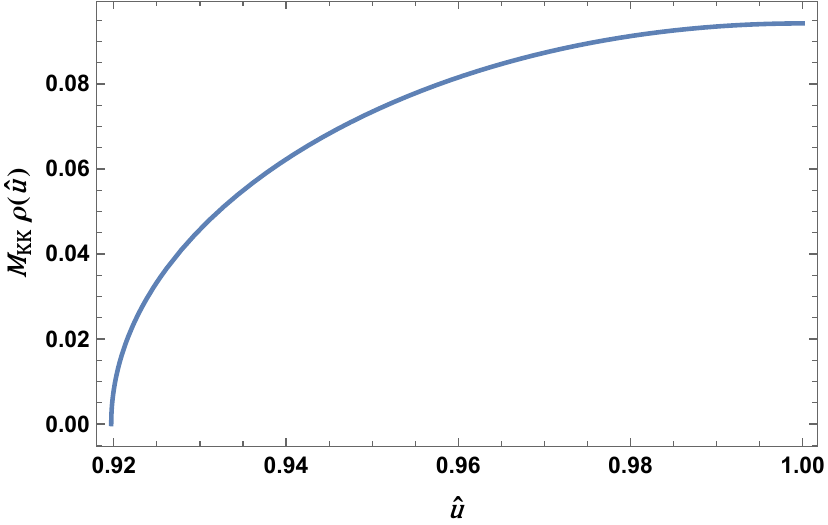}
    \end{subfigure}
    \hfill
    \begin{subfigure}{0.45\textwidth}
        \centering
       \includegraphics[width=\textwidth]{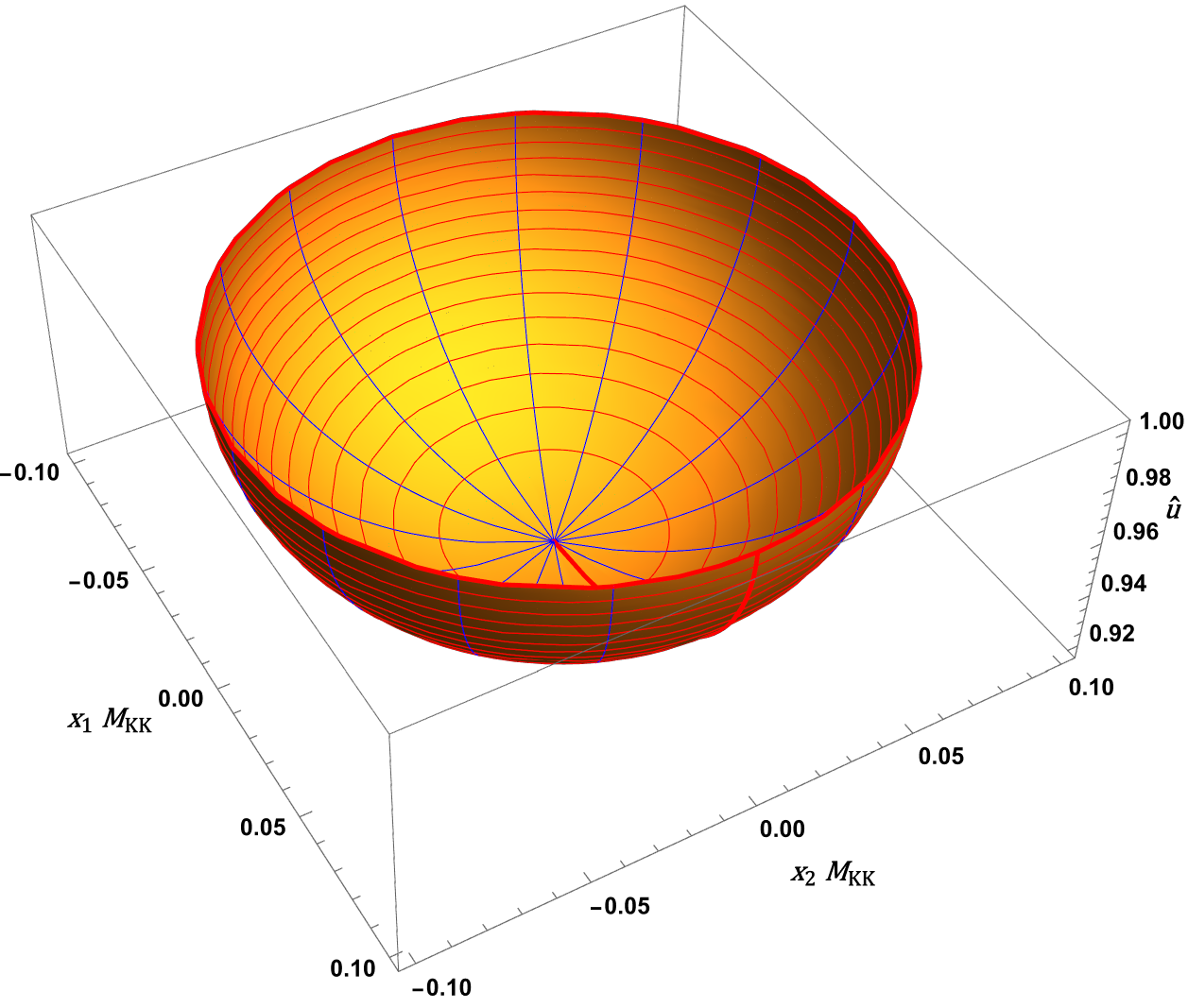}
    \end{subfigure}
\caption{Numerical solution of $\rho$ (times $M_{KK}$) for $b=0.1$ and $\lambda=100$ as a function of the rescaled holographic coordinate $\hat{u}=u/u_J \in [0.92,1]$, describing a stable baryon with baryon number $n_B=1$. The profile's derivative at $\hat{u} = 1$ is zero, indicating the (meta)stability of this embedding, while it goes to $\infty$ at $\hat{u}=u_E/u_J$, consistently with the smoothness of the manifold.}
\label{fig:rhobaryonconf}
\end{figure}

\begin{figure}[H]
    \centering
    \begin{subfigure}{0.45\textwidth}
        \centering
        \includegraphics[width=\textwidth]{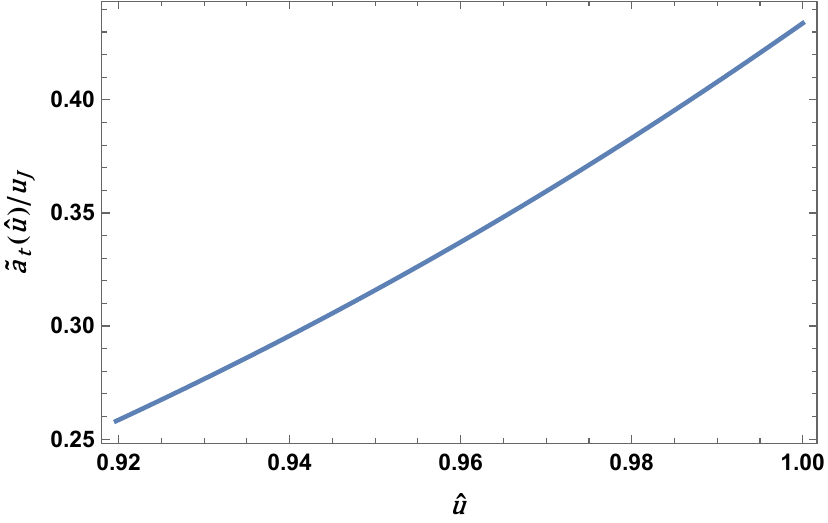}
    \end{subfigure}
    \hfill
    \begin{subfigure}{0.45\textwidth}
        \centering
       \includegraphics[width=\textwidth]{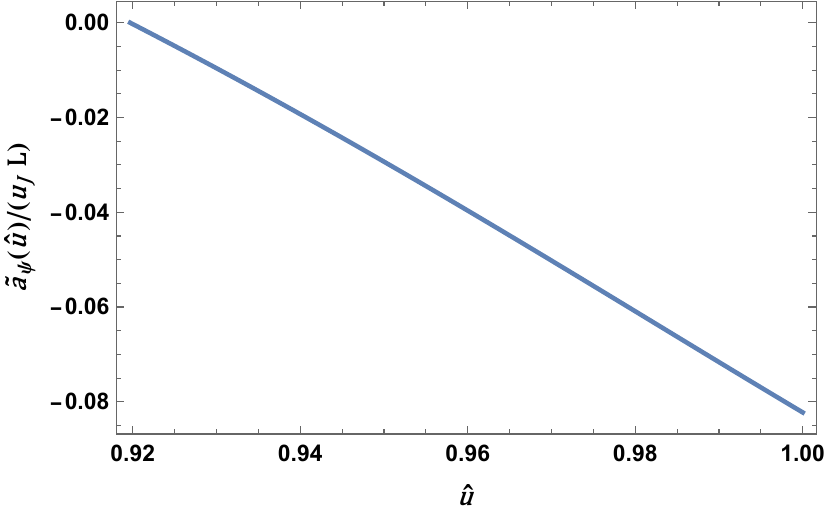}
    \end{subfigure}
    \caption{Numerical solutions for $u_J^{-1}\tilde{a}_t$ (left panel) and $(u_J L)^{-1}\tilde{a}_\psi$ (right panel) as a function of the rescaled holographic coordinate $\hat{u}$ for $b=0.1$ and $\lambda=100$.}
\label{fig:abaryonconf}
\end{figure}

We present the baryon solution with $n_B=1$ in the confined phase in figures \ref{fig:rhobaryonconf} and \ref{fig:abaryonconf} for $\lambda=100$ and $b=0.1$. The integration of the differential equations stops when $\rho'(u)\to \infty$, and this happens at $u=u_E$, which for the present solution is numerically found to be $u_E\approx 0.92\,u_J$. The details of the numerical methods used to solve the set of differential equations can be found in appendix \ref{app:numerics}; let us simply mention that to simplify the numerics, we performed the following coordinate and field redefinitions
\begin{align}
    u&\to u_J\, \hat{u}  ,& \rho &\to L\,\check\rho, & a_\psi &\to L\, u_J \,\check a_\psi, & a_t &\to u_J \,\check a_t\,.
    \label{eq:redefinitionsmain}
\end{align}
We will specify whenever we are dealing with rescaled fields in the following. The plots of the solutions to the equations of motion are shown as a function of the rescaled coordinate $\hat{u}$.

\begin{figure}[H]
    \centering
    \begin{subfigure}{0.45\textwidth}
        \centering
        \includegraphics[width=\textwidth]{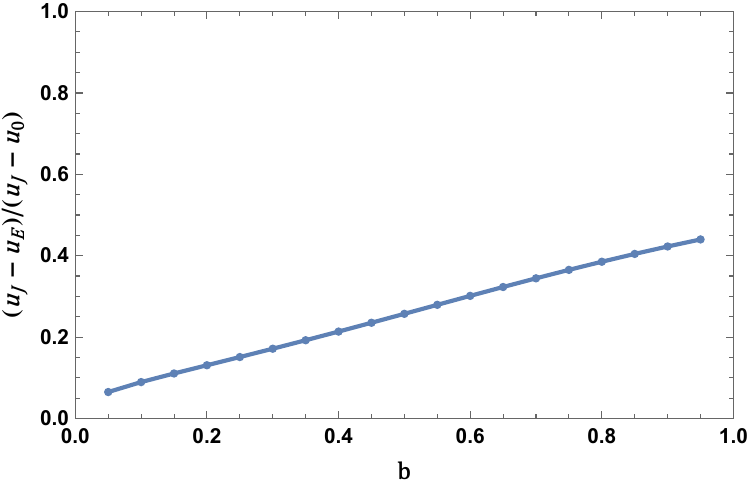}
    \end{subfigure}
    \hfill
    \begin{subfigure}{0.45\textwidth}
        \centering
       \includegraphics[width=\textwidth]{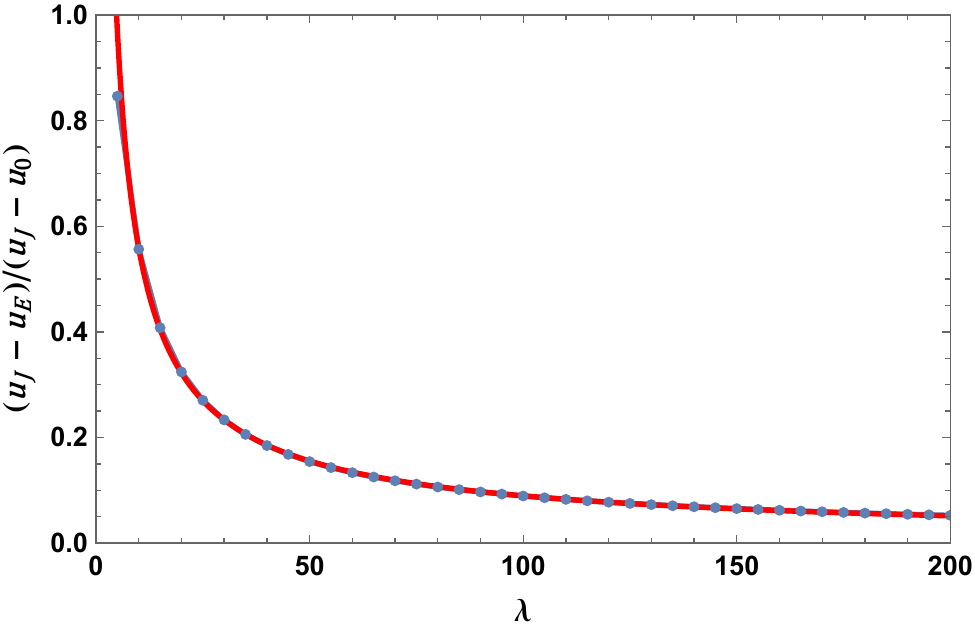}
    \end{subfigure}
   \caption{In these plots, we show the behavior of the extension of the D6-brane along the $u$ coordinate $(u_J-u_E)/(u_J-u_0)$. In the left panel, we fix $\lambda=100$ and we show the dependence on $b$. In the right panel, we fix $b = 0.1$ and we show the behavior of $(u_J-u_E)/(u_J-u_0)$ as a function of $\lambda$.}
\label{fig:u}
\end{figure}

We can compute the extension along $u$ of the D6-brane embedding with baryon number $n_B=1$. This is shown in figure \ref{fig:u}. The left panel shows the behavior of the relative extension $(u_J-u_E)/(u_J-u_0)$ for fixed $\lambda=100$ as a function of $b$. The right one shows $(u_J-u_E)/(u_J-u_0)$ for fixed $b=0.1$ as a function of $\lambda$. Moreover, in red, we present a fit for a fixed $b=0.1$ (the result does not depend on the specific value of $b$), for which we have
\begin{align}
    &\dfrac{u_J-u_E}{u_J-u_0}\bigg|_{b\,\text{fixed}}\sim\,\lambda^{-\gamma}& &\text{with}& &\gamma\approx 0.8\,.&
\end{align}
Since $\lambda\gg 1$ for the reliability of the holographic model at hand, the extension of the D6-brane along $u$ is parametrically small in $\lambda$ for every value of $b$. 

\begin{figure}[H]
    \centering
    \begin{subfigure}{0.45\textwidth}
        \centering
        \includegraphics[width=\textwidth]{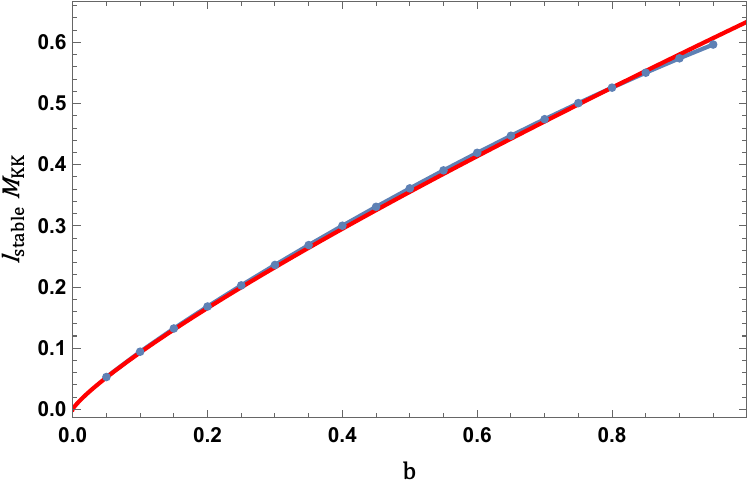}
    \end{subfigure}
    \hfill
    \begin{subfigure}{0.45\textwidth}
        \centering
        \includegraphics[width=\textwidth]{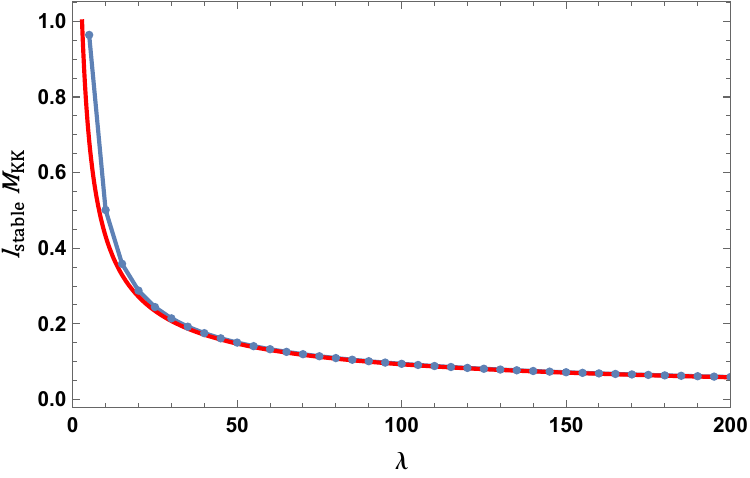}
    \end{subfigure}
   \caption{The plot of $l_\text{stable}M_{KK}$ is shown (in blue) for fixed $\lambda=100$ as a function of $b$ (left panel) and for fixed $b=0.1$ as a function of $\lambda$ (right panel). The red lines are the analytical estimates coming from the rigid rotor.}
\label{fig:lstable}
\end{figure}

The rigid rotor approximation perfectly fits the behavior of the stability radius $l_\text{stable}$, computed as the on-shell value of $\rho(u_J)$, as a function of (large) $\lambda$. In the left panel of figure \ref{fig:lstable}, the plot of $l_\text{stable}$ as a function of $b$ for fixed $\lambda=100$ is shown in blue, together with the analytical result in red coming from the rigid rotor \eqref{eq:lstablerigidconf}. The right panel shows the plot of $l_\text{stable}$ for $b=0.1$ as a function of $\lambda$ (in blue) together with the analytic prediction at fixed $b=0.1$ coming from the rigid rotor (in red). The analytic prediction is very reliable for large $\lambda$, while it starts to fail at very small $\lambda$, where in any case the holographic approximation is not expected to be reliable.

We can observe that the approximation of the D6-brane as a rotating disk seems reasonable since the extension along $u$ of the D6-brane ($\sim\lambda^{-0.8}$) is parametrically smaller than the radius of the D6-brane ($\sim\lambda^{-2/3}$).

\begin{figure}[H]
\center
\includegraphics[height = 4cm]{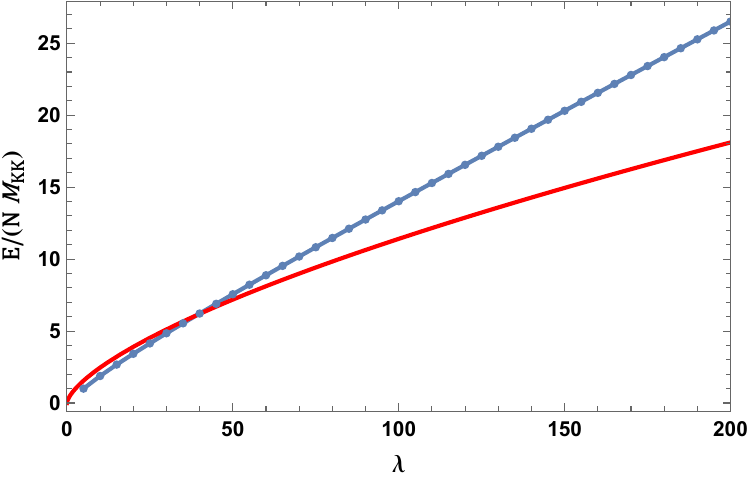}
\caption{The plot shows the energy of the baryon solution with $n_B=1$ in blue for $b=0.1$ as a function of $\lambda\in[5,200]$. The red curve is the corresponding analytical prediction from the rigid rotor.}
\label{fig:energyfit}
\end{figure}

Then, we can compute the energy numerically from the following formula (see equation (\ref{eq:EE}))
\begin{equation}\label{eq:energybaryonconf}
    E = \dfrac{N\,\lambda^2}{2\cdot 3^4\pi^2}\dfrac{J(b)}{b^{3/2}}\,M_{KK}\int^1_{u_E/u_J}d\hat{u}\,\hat{u}\,\rho(\hat{u})\left(D(\hat{u})+\dfrac{(\partial_u\tilde{a}_t)^2}{D(\hat{u})}\right),
\end{equation}
where $\rho(\hat{u})$, and $D(\hat{u})$ are computed for the rescaled fields. In figure \ref{fig:energyfit} we plot the on-shell energy for a fixed value of $b =0.1$ as a function of $\lambda$ and show that for this observable, the rigid rotor approximation \eqref{energyrigidconf} is not reliable. We suspect that dropping the gauge connection in the mass term of the rigid rotor energy might be too drastic, and it spoils the $\lambda$-dependence of the energy.

\begin{figure}[H]
\center
\includegraphics[height = 4cm]{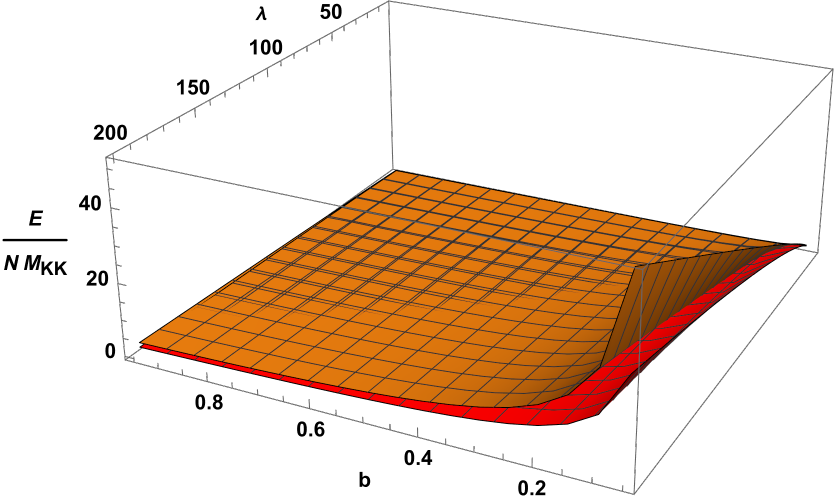}
\caption{The figure shows the plot the energy for a D6-brane solution with $n_B=1$ in orange as a function of $b$ and $\lambda\in[5,200]$. We plot it together with the associated Bogomol'nyi bound \eqref{BogoboundE} in red.}
\label{fig:energybog}
\end{figure}

In figure \ref{fig:energybog}, we plot the energy of a D6-brane solution with $n_B=1$ in orange as a function of $b$ and $\lambda$. We plot it together with the associated Bogomol'nyi bound \eqref{BogoboundE} in red.

\begin{figure}[H]
\center
\includegraphics[height = 4.5cm]{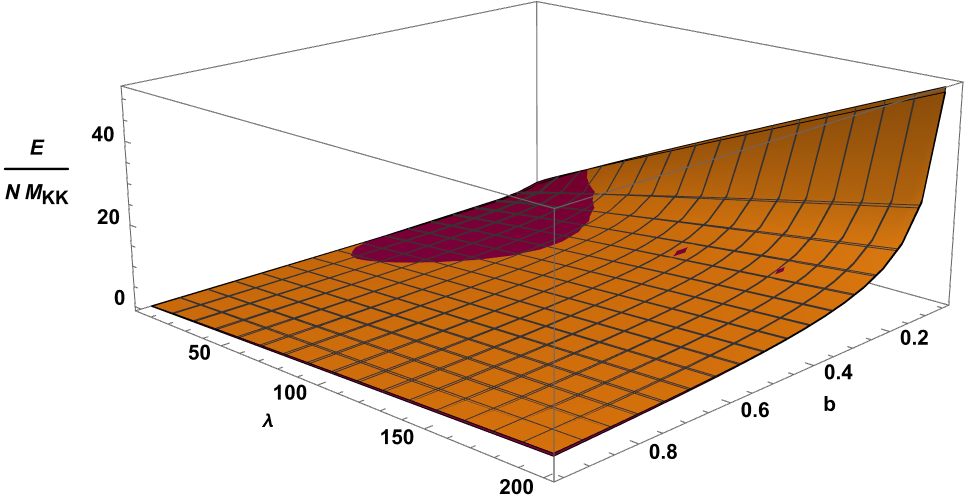}
\caption{The plot shows the energy of the D6-brane baryon solution in violet and the energy of the D4-brane vertex plus the energy of $N$ fundamental strings in orange. The range of parameters is $\lambda\in[5,200]$ and $b\in[0,1]$.}
\label{fig:energyD4strings}
\end{figure}

It is also interesting to shed light on the picture of the D6-brane as composed of a D4-brane vertex and a bunch of fundamental strings. This interpretation has been made precise in the supersymmetric case \cite{Mateos:2001pi} also at the level of the energy. We would like to see the emergence of this structure even in our non-supersymmetric case. Therefore, we can compare the energy of the D6-brane baryon solution with the energy of a wrapped D4-brane placed at $u_E$ plus the energy of $N$ fundamental strings hanging from $u_J$ to $u_E$ giving 
\begin{equation}
    E_\text{D4+f-strings} = \dfrac{N}{6\pi l_s^2} u_E + \dfrac{N}{2\pi l_s^2}(u_J-u_E) .
\label{sommaE}
\end{equation}
As it can be seen from figure \ref{fig:energyD4strings}, the actual energy of the D6-brane baryon solution (in violet) and that given in equation (\ref{sommaE}) (in orange) are, remarkably, almost equal, especially in the large-$\lambda$ limit, reinforcing the reliability of the above-mentioned picture. 

We remind that the equations of motion together with the associated boundary conditions depend on the ratio $\xi = n_B/\lambda$ and not on $\lambda$ alone. We also observe that if $n_B\sim\lambda^0$, the large-$\lambda$ limit corresponds to the $\xi\to 0$ limit. Moreover, we can have $\xi$ of order one in the large-$\lambda$ limit if $n_B\sim\lambda$.\footnote{Note that $\xi$ can be of order one also if $n_B\sim\lambda^0$ and $\lambda\sim\mathcal{O}(1)$, but in this case the holographic model is not reliable.} In the confined phase, baryons are found for every value of $b$ in the regime where $\xi$ is less than some critical value $\xi_\text{cr.}$. In the regime where $\xi$ is close to $\xi_\text{cr.}$, the solutions are typically parametrically larger, and tend to occupy a larger fraction of the space between $u_0$ and $u_J$ until they become almost space-filling, and eventually, we do not find any solution.

\subsection{Deconfined phase}
\begin{figure}[H]
	\center
	\includegraphics[height = 8cm]{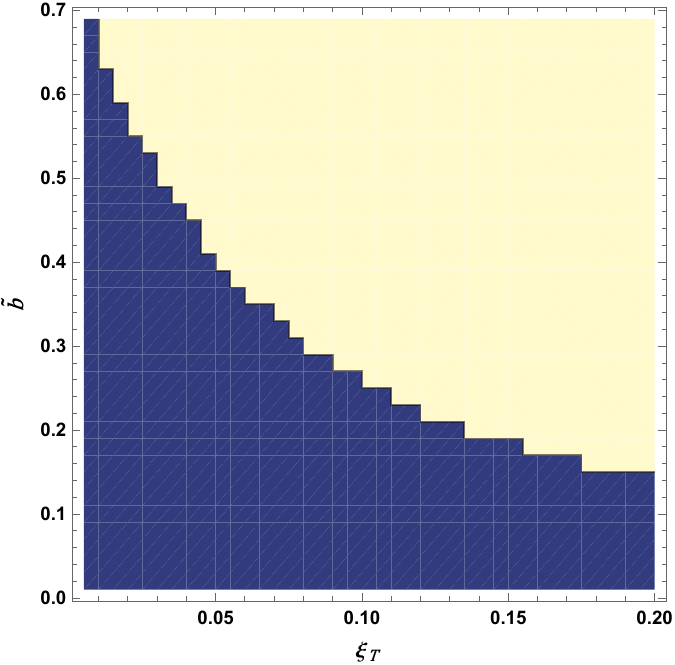}
	\caption{In blue the region, in the parameter space $0<\tilde{b}<0.7$ and $\xi_T\in[0.005, 0.2]$, where D6 baryons exist in the deconfined phase.}
	\label{fig:regiondec}
\end{figure}

The behavior of the solutions in the deconfined phase is analogous to the one in the confined phase, modulo obvious substitutions such as $b \rightarrow \tilde b$, $\lambda \rightarrow \lambda_T$, $\xi \rightarrow \xi_T$. 
We report the detailed discussion in appendix \ref{app:deconf1}.
The only relevant difference w.r.t.~the confined case is the following.
For small enough $\xi_T(\lesssim 0.01)$, we find baryon solutions in the deconfined phase for every value of $\tilde{b}$. However, for small $\xi_T$ such that the holographic model is still reliable in the $n_B\sim\lambda^0$ case, baryon solutions exist only for $\tilde{b}$ less than a critical value $\tilde{b}(\xi_T)$, which decreases as a function of $\xi_T$. 
This means that for sufficiently large temperatures, the baryons cease to exist, since they melt in the plasma.
The existence region of baryons in parameter space is shown in figure \ref{fig:regiondec}.

\subsection{Summary of the baryon's properties}
We conclude the analysis of the D6-brane baryon solution by collecting the results about its physical properties, namely its mass (\textit{i.e.}~the on-shell energy $E$) and its radius. We list them for both the confined and the deconfined phases. In table \ref{tab:baryonstring}, they are shown as functions of the string theory parameters (neglecting multiplicative constants) while in table \ref{tab:baryonproperties} we give their full expressions in terms of ``phenomenological" parameters, like the $\eta'$ decay constant $f_{\eta'}$. The definition of the latter in the confined phase, for generic non-antipodal branes, reads
\begin{align}
    &f_{\eta'}^2 = \dfrac{N\,\lambda}{54\pi^3}\dfrac{M_{KK}^2}{I(b)b^{3/2}}& &\text{with}& &I(b)=\int^1_0dy\dfrac{y^{-1/2}}{\sqrt{1-b^3y-(1-b^3)y^{8/3}}}\,. &
\end{align}
The decay constant at finite temperature is 
\begin{align}
    &f_{T,\eta'}^2 = \dfrac{N\,\lambda_T (2\pi T)^2}{54\pi^3 I_T(\tilde{b}) \tilde{b}^{3/2} }& &\text{with}& &I_T(\tilde{b})=\int^1_0dy\dfrac{y^{-1/2}\sqrt{1-\tilde{b}^3y}}{\sqrt{1-\tilde{b}^3y-(1-\tilde{b}^3)y^{8/3}}}\,.&
\end{align}
See e.g.~\cite{Bigazzi:2019eks} for a review on the derivation of the above expressions. 
Note that they have a non-trivial $N$-dependence.
In the following, we will use the approximations $I(b) \approx I(0) = I=2.48$, and $I_T(\tilde{b}) \approx I_T(0) =  I =2.48$.\footnote{This is a fairly good approximation since $I(b)$ and $I(\tilde b)$ are monotonic functions, with $I(b=1)=\pi$, $I(\tilde b=1)=2$.}. 

The results are given in the large-$\lambda$ limit for a baryon with $n_B=1$:
\begin{table}[H]
    \centering
    \begin{tabular}{|c|c|c|}
        \hline
         & Confined phase & Deconfined phase \\
        \hline
        Radius & $l_\text{stable} \sim  \lambda^{-2/3} \Lambda^{-1} b^{5/6} $  & $l_\text{stable} \sim \lambda_T^{-2/3} T^{-1} \tilde{b}^{5/6}(1-\tilde{b}^3)^{-1/3} $\\ 
        \hline
        Mass & $E\sim N \lambda \Lambda  b^{-1}  $ &  $E\sim N \lambda_T T \tilde{b}^{-1}   $ \\
        \hline
    \end{tabular}
    \caption{Summary of parametric scaling properties of the baryon solution's radius and mass. For the reader's convenience, we used the dynamical scale notation $\Lambda$ instead of $M_{KK}$.}
    \label{tab:baryonstring}
\end{table}

\begin{table}[H]
    \centering
    \begin{tabular}{|c|c|c|}
        \hline
         & Confined phase & Deconfined phase \\
        \hline
        Radius & $l_\text{stable}\approx C_l\,  N^{5/9}  (\Lambda/\lambda)^{1/9} f_{\eta'}^{-10/9}  $  & $l_\text{stable} \approx C_l\, N^{5/9}  (\Lambda/\lambda)^{1/9} f_{T,\eta'}^{-10/9}$\\ 
        \hline
        Mass & $E\approx C_E\, N^{1/3} (\lambda/\Lambda)^{1/3} f_{\eta'}^{4/3}  $ &  $E\approx C_E \, N^{1/3} (\lambda/\Lambda)^{1/3} f_{T,\eta'}^{4/3}  $ \\
        \hline
    \end{tabular}
    \caption{Summary of the main properties of the baryon.}
    \label{tab:baryonproperties}
\end{table}
To obtain the radius estimation, we can use the rigid rotor values, which give
\begin{equation}
    C_l = 2^{-19/18} 3^{1/3}\pi^{-1}I^{-5/9}\approx 0.221\,.
\end{equation}
For the mass estimation, we use a power law fit for $b$ and $\lambda$. $C_E$ is a constant obtained by a fit of the energy scaling law obtained numerically. Notably, the same value is obtained in both phases
\begin{equation}
    C_E \approx 0.129 \,I^{2/3}\approx 0.236\,.
\end{equation}

\section{Punctured domain walls and \textit{sandwich} defects}\label{sec:sandwich}
In this section, we present more exotic solutions to the D6-brane equations of motion, which are not baryons but have non-zero baryon number. In a recent paper \cite{Bigazzi:2024mge}, charged string loop solutions (generally referred to as \textit{vortons}) in the deconfined phase of the WSS model have been discussed in detail. Here we just report that such solutions exist and are metastable only if they have a very large charge $n_B\sim \lambda$.

Other solutions can be found:  
sandwich defects (see figure \ref{fig:holosandwich}) and punctured domain walls (see figure \ref{fig:holopunctured}). 
Before we present the associated numerical solutions of the equations of motion, let us describe in more detail the dual interpretation of these D6-brane configurations.

Punctured domain walls are infinitely extended domain walls with a hole inside. They could form either from the chiral symmetry breaking phase transition, or by tunneling transitions which open up a hole in an infinite DW.
The latter is actually a standard decay channel of infinite DWs (see e.g.~\cite{Forbes:2000et}): holes can form on the worldvolume, then they start to expand, and eventually eat up all of the defect; collisions of the hole boundaries, \textit{i.e.}~global strings, can produce gravitational waves \cite{Hamada:2024dan}.

An amusing feature of these configurations is that the D6-brane can end orthogonally to the D8-brane, signaling a force balance, even in the uncharged case. Indeed, there is a specific value of the radius $\rho(u_J)$ for which we get the necessary metastability condition $\rho'(u_J)=0$. 
We can interpret this behavior by commenting on what happens if we vary the hole radius. If we increase the radius of the hole, the energy of the DW is reduced, as we are decreasing its worldvolume.\footnote{Note that this behavior is opposite w.r.t.~all the other configurations we consider in this paper.}
At the same time, the energy of the bounding string is increased, since its length increases.\footnote{Holographically, this is seen as an increase of the area of the ``vertical portion'' (in $u$) of the D6-brane.}
Thus, there can be a critical radius for which the two opposite forces are exactly balanced.

In this work, we can also consider the charged version of the punctured domain walls under the baryonic symmetry. Holographically, these are charged D6-brane solutions. 
The charge just adds a force component to the picture described above, with the only result that the critical radius of the hole for which there is orthogonality is different w.r.t.~the uncharged case.
In the following, we are going to show only the results for the charged critical (orthogonal) case.

Sandwich defects are even more exotic objects. From the holographic point of view, these are described by D6-branes ending on two different flavor branes.  These objects can be studied in the confined and deconfined phases. The two flavor branes can be interpreted as both dark (for the Standard Model interactions) flavors in the dual QFT, or one of the two flavor branes (the one that condenses at a lower scale), or both, can be associated with QCD flavors. We have such a possibility because the asymptotic separation between a D8-brane and a $\overline{\text{D8}}$-brane $L$, of a given flavor, is a free parameter in the model and hence we can tune $L$ to get the desired $T_a$ (see equation (\ref{tadef})) at which the chiral symmetry is broken. In the framework of $SU(N)$ Yang-Mills theories with flavors, the interplay between a dark and the $\eta'$ meson (or any other dark meson) leads to the formation of rich topological structures. These arise due to a sequence of symmetry-breaking patterns, each introducing different types of defects, which can ultimately combine into composite structures carrying both axion (for example) and $\eta'$ excitations (see e.g.~\cite{Gabadadze:2000vw,Forbes:2000et}). The formation of these defects is a natural consequence of the non-trivial vacuum structure of the theory, and their evolution depends on the dynamics of spontaneous and explicit symmetry breaking. The presence of both $\eta'$ and dark topological defects (or two dark topological defects) in the same system means that, in certain regions, these structures can overlap or interact. 
This is where the idea of \textit{sandwich} structures emerges: a layered configuration where the presence of one field influences the stability and interactions of the other. We can think of these objects both at zero and at finite temperature. In a realistic cosmological scenario at finite temperature, the evolution of these defects depends on their stability and interactions with the surrounding plasma.

Sandwich solutions, if not charged, are not stable and will decay. They can be stabilized, like the other kinds of configurations, by turning on the gauge connection on the D6-brane worldvolume. In this case, these objects will have a baryon number given by (see equation (\ref{eq:nbD6}))
\begin{equation}
    n_B = -\int^{u_J}_{u_K} du \,\partial_u a_\psi= n_J + n_K,
\end{equation}
where $u_K$ and $u_J$ (with $u_K<u_J$ without loss of generality) are the positions of the tip, respectively, of the less and more energetic flavor branes. Importantly, in the deconfined phase, the value of $u_K$ (as the value of $u_J$) is bounded from below by the quantity $\sim 0.75\,u_T$, to ensure that also the second set of branes is in the chirally broken phase. As discussed in section \ref{sec:EulerD6}, $n_J$ and $n_K$ are physically associated with two different flavor degrees of freedom, counting the number of quarks of each flavor (divided by $N$). These sandwiches are therefore prototypes of multi-flavored configurations that support a baryon number, and we expect them to be enclosed in a more general theory that describes $N_f \geq 2$ Hall droplet baryons.\footnote{Let us mention that these sandwich configurations resemble the construction presented in \cite{Callan:1985hy}.}

In the following, we will present the numerical solution for the punctured domain walls and for the sandwich defects with non-zero baryon number $n_B$ and angular momentum $J = N\,n_B^2/2$, obtained from $n_K = 0$ and $n_J = 1$. We first check if the expansion around $u=u_K$ allows for $\tilde{a}_\psi(u_K)=0$ and regularity ($\rho'(u_K)=0$). We demand that the fields have a regular expansion around $u_K$
\begin{align}
    &\rho(u) = \sum\limits_{n=0}^{\infty}c_n(u-u_K)^n,& &\tilde{a}_t(u) = \sum\limits_{n=0}^{\infty}b_n(u-u_K)^n,& &\tilde{a}_\psi(u) = \sum\limits_{n=0}^{\infty}d_n(u-u_K)^n,&
\end{align}
and we impose $c_1= 0$ for orthogonality and $d_0= 0$ to enforce the relation between the angular momentum and the baryon number. We checked that both in the confined and the deconfined phase we get a consistent series expansion in terms of $c_0 = \rho(u_K)$ and $b_0$. An analogous consistency check for the boundary condition with the equations of motion has been done for the punctured domain wall configuration.

\subsubsection*{Rotor estimates}
We end this introduction by giving a numerical estimate of the stability radius and the energy of the sandwich solution, which can be approximated by a rotating circle. For a rotating circle with mass $M$, radius $l$, and angular momentum $J$, the total energy in the non-relativistic limit can be approximated by
\begin{equation}
    E = M + \dfrac{J^2}{2I}\,,
\end{equation}
where $I = Ml^2$ is the moment of inertia. We will then set $J=N\,n_B^2/2$. For this type of solution, we can approximate $\rho(u)$ as a constant $\rho(u)\approx l$ throughout the whole $u$-range and therefore $\rho'(u) = 0$. Let us start from the confined phase. The mass $M$ of the rotating circle is obtained, with the above assumption, from the DBI action in (\ref{eq:D6actionconf}), setting the gauge field $a=0$,
\begin{align}
    &M = \dfrac{N\,\lambda^2}{3^5\pi^2b^2}\alpha\,M_{KK}^2\,l\,,
\end{align}
with 
\begin{equation}
    \alpha=\frac{1}{2} \left(\sqrt{\frac{u_K}{u_J}} \sqrt{\frac{u_K^3}{u_J^3}-b^3} \left(_2F_1\left(\frac{2}{3},1;\frac{7}{6};\frac{u_K^3}{u_J^3\,b^3}\right)-1\right)-\sqrt{1-b^3} \left(_2F_1\left(\frac{2}{3},1;\frac{7}{6};\frac{1}{b^3}\right)-1\right)\right)\,.
\end{equation}
Analogously, in the deconfined phase, setting $a=0$ in (\ref{Ddeconf}), the mass of the rotating circle is approximated by
\begin{align}
    &M = \dfrac{N\,\lambda_T^2}{3^5\pi^2\,\tilde{b}^2}\,\alpha_T\,\left(\dfrac{T}{T_c}\right)^2\,M_{KK}^2\,l\,,& &\text{with}& &\alpha_T = \dfrac{1}{2}\left(1-\dfrac{u_K^2}{u_J^2}\right)\,.&
\end{align}

Minimizing the total energy of the rotor w.r.t. the radius, we find the following results:
\begin{itemize}
    \item Confined phase:\footnote{Remember that $\xi_{(T)}=\frac{n_B}{\lambda_{(T)}}$.}
    \begin{align}
        &M_{KK}\,l_\text{stable}=\dfrac{3^{11/4}\pi\,b}{2^{3/4}\alpha^{1/2}}\,\xi\,,\\
        &E=\dfrac{2^{5/4}N\,\lambda^2}{3^{13/4}\pi\,\,b}\,\alpha^{1/2}M_{KK}\,\xi\,.
    \end{align}
    \item Deconfined phase:
    \begin{align}
         &M_{KK}\,l_\text{stable}=\dfrac{3^{11/4}\,\pi\,\tilde{b}}{2^{3/4}\,\alpha_T^{1/2}}\left(\dfrac{T_c}{T}\right)\xi_T\,,\\
        &E=\dfrac{2^{5/4}N\,\lambda_T^2}{3^{13/4}\pi\,\,\tilde{b}}\,\alpha_T^{1/2}\left(\dfrac{T}{T_c}\right)M_{KK}\,\xi_T\,.
    \end{align}
    \end{itemize}
In both phases, the stability radius and the on-shell energy scale linearly with $\xi_{(T)}$.

We checked that these analytical estimates correctly reproduce the sandwich stability radius and energy in the large-$\lambda$ limit, up to numerical factors.\footnote{This marks a difference with the baryon case, where the energy scaling with $\lambda_{(T)}$ is not well reconstructed. This can be attributed to the less stringent approximation that we did in the rigid rotor analysis for the sandwiches, for which the integration region stays finite.}


\subsection{Confined phase}
\begin{figure}[H]
    \centering
    \begin{subfigure}{0.4\textwidth}
        \centering
        \includegraphics[width=\textwidth]{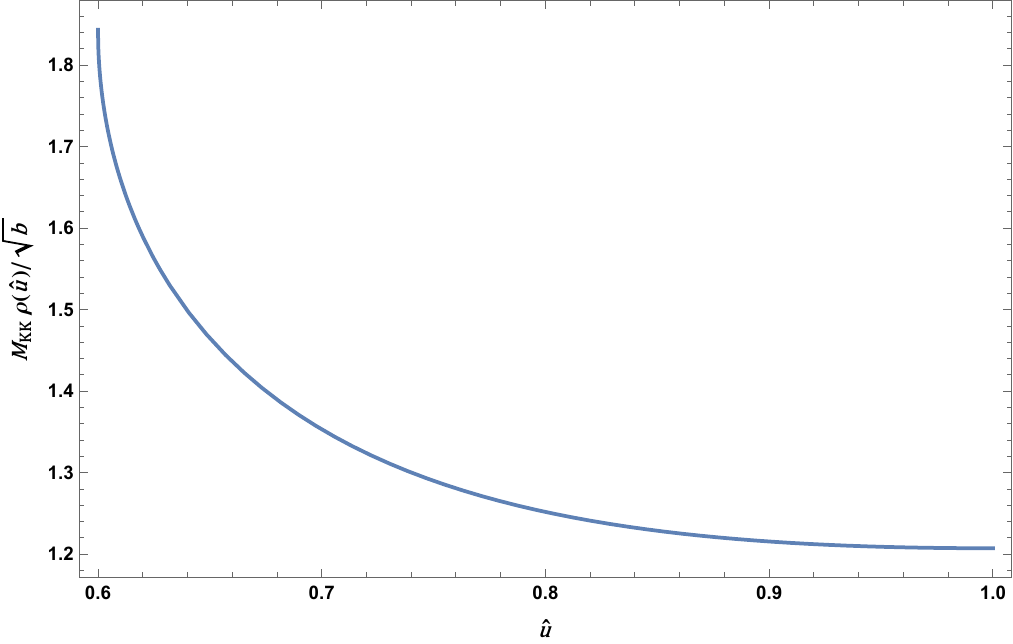}
    \end{subfigure}
    \hfill
    \begin{subfigure}{0.55\textwidth}
        \centering
       \includegraphics[width=\textwidth]{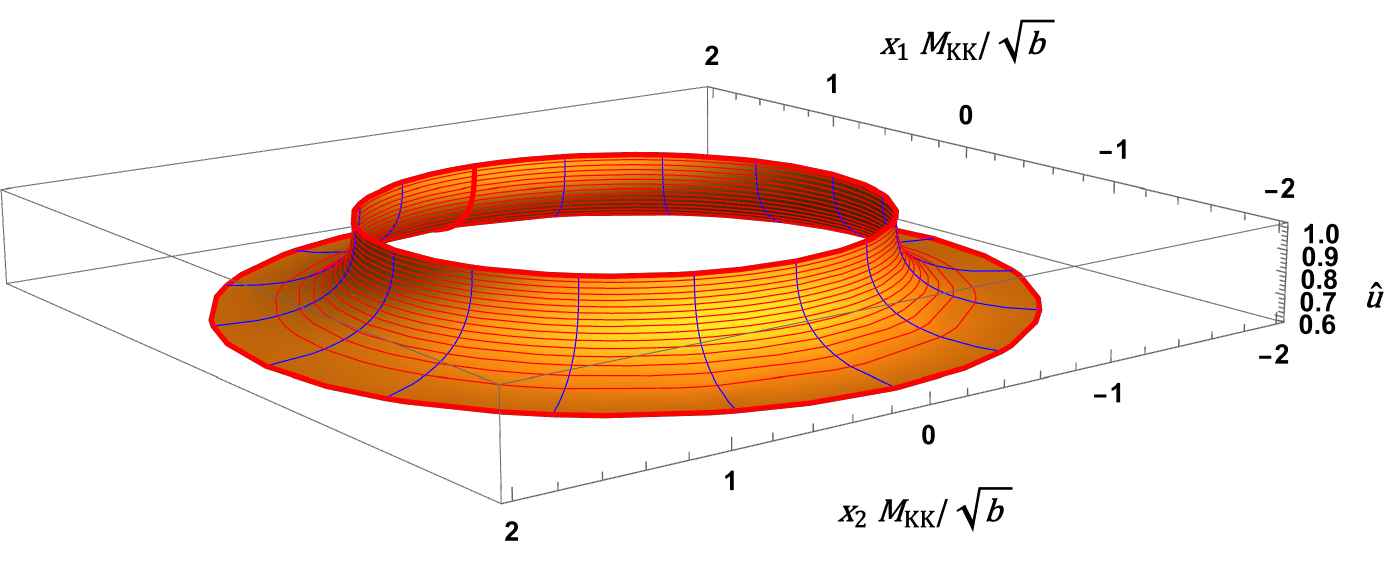}
    \end{subfigure}
\caption{Numerical solution of $\rho$ (times $M_{KK}/\sqrt{b}$) for $b=0.1$ and $\lambda=100$ as a function of the rescaled holographic coordinate $\hat{u}=u/u_J \in [0.6,1]$, describing a stable punctured domain wall with baryon number $n_B=1$. The profile's derivative at $\hat{u} = 1$ is zero, indicating the (meta)stability of this embedding. At $\hat{u}=u_*/u_J$, the profile's radius goes to infinity while in the plot we cut it for clarity at some finite value.}
\label{fig:rhodwconf}
\end{figure}

\begin{figure}[H]
    \centering
    \begin{subfigure}{0.45\textwidth}
        \centering
        \includegraphics[width=\textwidth]{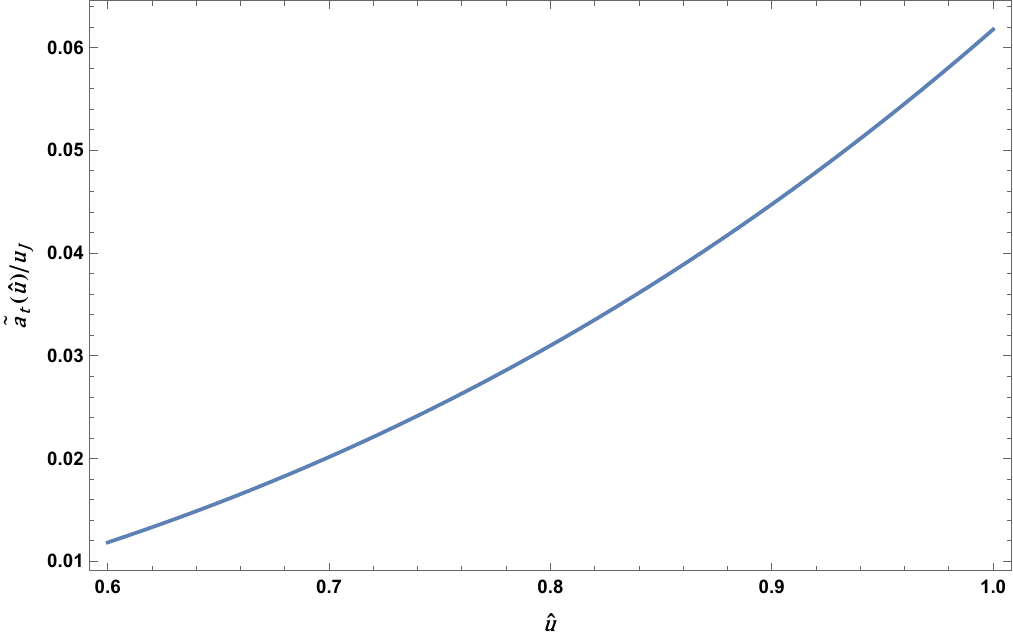}
    \end{subfigure}
    \hfill
    \begin{subfigure}{0.45\textwidth}
        \centering
       \includegraphics[width=\textwidth]{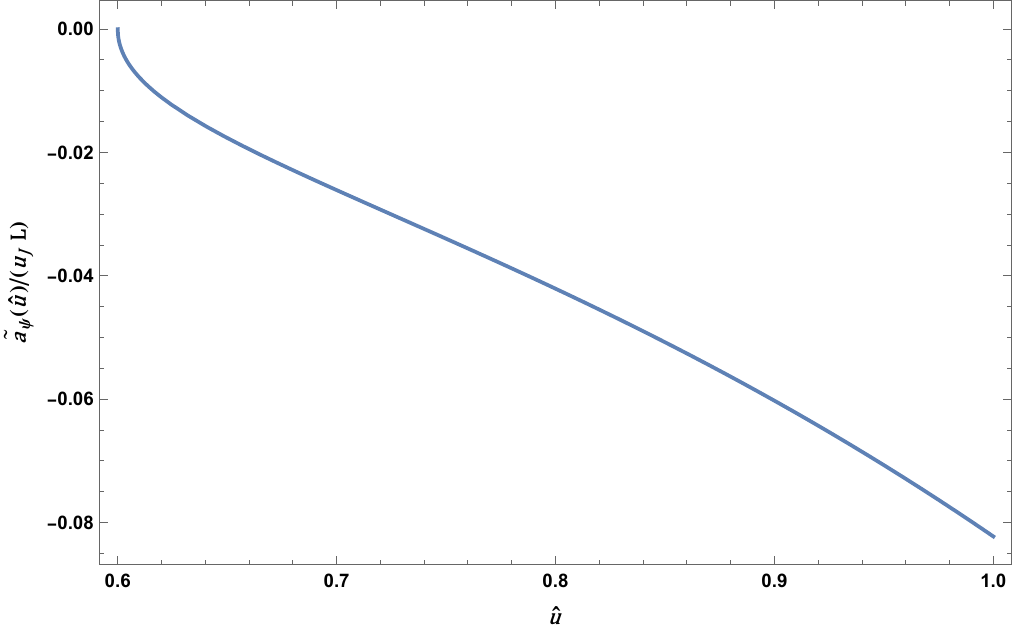}
    \end{subfigure}
\caption{Numerical solutions for $u_J^{-1}\tilde{a}_t$ (left panel) and $(u_J L)^{-1}\tilde{a}_\psi$ (right panel) as a function of the rescaled holographic coordinate $\hat{u}$ for $b=0.1$ and $\lambda=100$.}
\label{fig:adwconf}
\end{figure}

We start by presenting the punctured domain wall solutions. These are infinitely energetic configurations with non-zero baryon number. 
We are free to choose the value of the holographic coordinate $u_*$ at which the D6-brane profile has infinite radius. 
This value fixes the stability radius of the hole.
In figures \ref{fig:rhodwconf} and \ref{fig:adwconf} we show the solutions to the equations of motion for $n_B=1$, $b=0.1$, $\lambda=100$, and for which we have chosen $u_*/u_J=0.6$.

\begin{figure}[H]
    \centering
    \begin{subfigure}{0.4\textwidth}
        \centering
        \includegraphics[width=\textwidth]{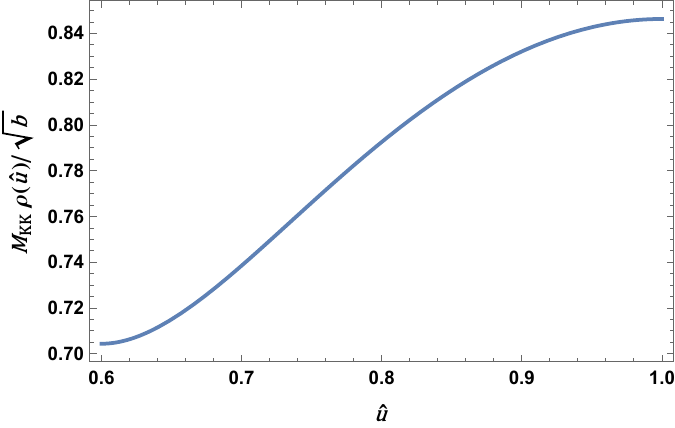}
    \end{subfigure}
    \hfill
    \begin{subfigure}{0.55\textwidth}
        \centering
       \includegraphics[width=\textwidth]{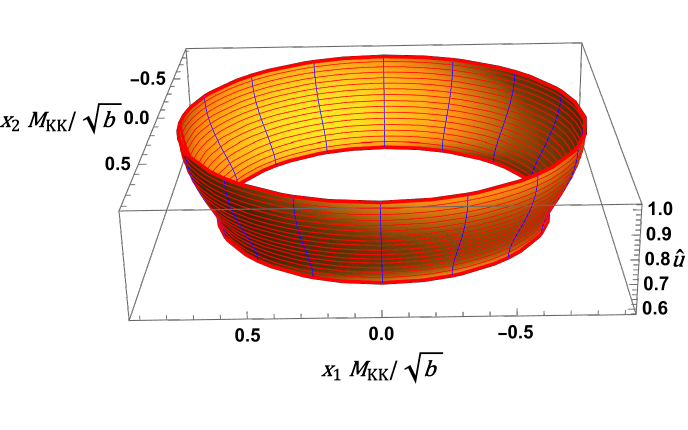}
    \end{subfigure}
\caption{Numerical solution of $\rho$ (times $M_{KK}/\sqrt{b}$) for $\lambda=20$, $b=0.1$ and $\hat{u}_K =u_K/u_J= 0.6$, as a function of the rescaled holographic coordinate $\hat{u}=u/u_J \in [0.6,1]$, describing a stable sandwich defect with baryon number $n_B=1$. The profile's derivative at both $\hat{u}_K$ and $\hat{u} = 1$ is zero, indicating the (meta)stability of this embedding.}
\label{fig:rhoswconf}
\end{figure}

\begin{figure}[H]
    \centering
    \begin{subfigure}{0.45\textwidth}
        \centering
        \includegraphics[width=\textwidth]{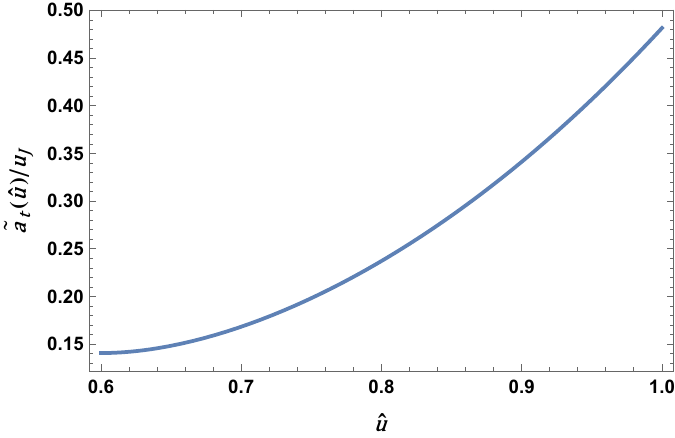}
    \end{subfigure}
    \hfill
    \begin{subfigure}{0.45\textwidth}
        \centering
       \includegraphics[width=\textwidth]{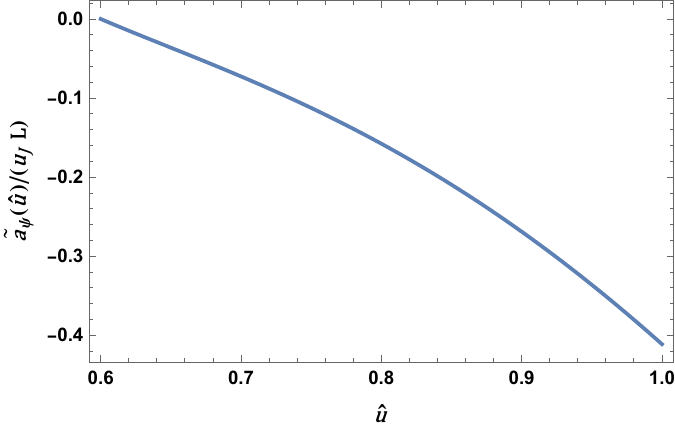}
    \end{subfigure}
\caption{Numerical solutions for $u_J^{-1}\tilde{a}_t$ (left panel) and $(u_J L)^{-1}\tilde{a}_\psi$ (right panel) as a function of the rescaled holographic coordinate $\hat{u}$ for $\lambda=20$, $b=0.1$ and $\hat{u}_K = 0.6$.}
\label{fig:aswconf}
\end{figure}

Then, we present the sandwich solution for $\lambda=20$ and $b=0.1$ associated with the choice of holonomies $n_K=0$ and $n_J=1$. We are \textit{a priori} free to set the position of the tip of the second (set of) D8-brane(s), which we will fix to $u_K/u_J= 0.6$ in the plots. The profile $\rho$ of the sandwich, as a function of the rescaled holographic coordinate $\hat{u}$, is shown in figure \ref{fig:rhoswconf}, while the associated gauge connection is shown in figure \ref{fig:aswconf}.

The sandwich defect solutions exist only if the two
D8-branes are close enough to each other. That is, for every $b$ and $\lambda$, there is a minimum value of $u$, denoted as $\hat{u}_\text{min}(b,\lambda)$, such that the sandwich solutions are found for every $\hat{u}_\text{min}(b,\lambda)<\hat{u}_K<1$. As in the baryon case, these solutions are more and more extended along $u$ and larger in $\rho$ as we consider smaller values of $\lambda$. We find numerically that $\hat{u}_\text{min}$ is a monotonously increasing function of $b$ and $\lambda$.

We conclude by mentioning that we can find a sandwich solution with $n_K$ and $n_J$ interchanged, that is, $n_K=1$ and $n_J=0$. This solution shows the opposite behavior for the profile $\rho(u)$, meaning that for all $u$, $\rho'(u)<0$.

\subsection{Deconfined phase}
\begin{figure}[H]
	\centering
	\includegraphics[width=0.5\textwidth]{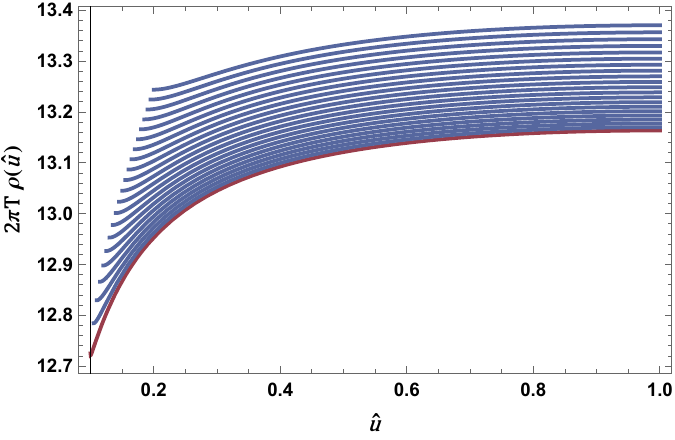}
	\caption{Family of sandwich domain walls for $u_K \to u_T$ in the deconfined phase (blue) and vorton solution (red). The black vertical line shows the position of the horizon.}
	\label{fig:vortonlim}
\end{figure}

As for the baryons, the behavior of the punctured domain walls and the sandwich vortons in the deconfined phase is very similar to their behavior in the confined phase.
We present the analysis in appendix \ref{app:deconf2}.
Here, let us just mention a connection of the sandwich solutions with the simple vortons studied in \cite{Bigazzi:2024mge}. Let us ignore for a moment the chiral symmetry restoration mechanism and consider the family of sandwich solutions as it could be defined in a larger interval of values of $u$. 
Then, if we take the limit $u_K \to u_T$ at fixed value of $\tilde{b}$ and $\lambda_T$, that is, if the tip of the lower D8-brane approaches the horizon, we observe that the limiting solution is precisely the vorton solution (discussed in detail in \cite{Bigazzi:2024mge}) at the same $\tilde{b}$ and $\lambda_T$ and with the same baryon number as the sandwich. This family of sandwich solutions transitioning to the limiting vorton solution is shown in figure \ref{fig:vortonlim}.

\section{Instabilities and decay channels}\label{sec:decays}
In this section, we will examine in detail the energy balance of all the solutions presented in this work. All the solutions are either stable (for the baryon in the confined phase) or metastable, meaning that they can decay into less energetic configurations. Even if we present the results for both the confined and the deconfined phases, we want to stress that the interpretation of the decay processes in the two phases is conceptually different. At finite temperature, the possible instabilities of the D6-brane configuration can be thought of as driven by changing the value of the temperature and, therefore, are ``dynamical'' instabilities. In the confined phase, there is no external parameter to tune and induce a decay. Therefore, in that case, we will still present the energy balance of all the configurations and discuss possible decay processes as if they were induced by a change in the parameters of the model.

In both phases, we will focus on the fate of the baryon solution, considering all the other solutions as possible decay channels 
that conserve the baryon number.

\subsection{Confined phase}
In the confined phase, it is interesting to analyze whether the baryon solution is unstable, meaning in this context that it is more energetic than another configuration with the same baryon number.

In the case of a single flavor, the $n_B=1$ baryon is completely stable in the confined phase, being the lightest charged particle. An interesting instability can arise if the model features at least two non-coincident U-shaped D8-branes, when considering sandwich solutions in the confined phase. These objects, as discussed in section \ref{sec:sandwich}, have to be interpreted from the dual field theory point of view as composite objects of two global strings charged under different (flavor) $U(1)$s living on two separated flavor branes. We can therefore think of a decay of these sandwiches in two different channels corresponding to the two different baryon solutions associated with the two flavors. Our theory only features flavor-conserving interactions, so the flavor numbers must be conserved quantities in these decays. The flavor numbers associated with D6-brane sandwich configurations are directly given by $n_K^{(W)}$ and $n_J^{(W)}$ and they count the net number of fundamental strings coming out of the flavor brane, \textit{i.e.}~the quarks. In addition to this, we associated with every actual fundamental string coming out from a D8-brane a 
charge contributing $1/N$ to $n_{K}$ or $n_J$, depending on the flavor brane. 
Let us also recall that the baryon number is given by the first Chern class $n_B$ associated with the manifold we are considering, and is also conserved. We start from a sandwich solution with baryon number $n_B$.

We consider the possibility of a decay of this sandwich vorton into two charged D6-brane configurations, one attached at $u_J$ and the other at $u_K$. These decay products have respectively baryon numbers $n_B^{(J)}$ and $n_B^{(K)}$ 
\begin{align}
    &n_B^{(J)} = n_J + p\,,& &n_B^{(K)} = n_K - p\,,&
\end{align}
compatible with the total baryon number conservation. Here $p$ is an integer divided by $N$, which will be determined by the total angular momentum conservation and flavor conservation.\footnote{Recall that a non-zero $p$ corresponds to shifting the flavor number by a global holonomy $p=-n_G$.} 

These decay products are associated in the holographic model with the splitting of the D6-brane sandwich solution into baryon-like solutions anchored respectively at $u_J$ and $u_K$. In general, they have to be complemented by fundamental strings linking the two sets of D8-branes to conserve flavor number. We first enforce the total angular momentum conservation. We study decays starting from sandwich domain walls that satisfy the anyonic relation $J = n_B^2N/2$, this choice is restrictive and enforces $n_K = 0$.\footnote{One can also consider more generally decays of sandwiches with generic $n_K$, and so they do not satisfy the anyonic relation. These configurations can decay into baryons (plus mesons) if and only if $n_B^2 - 4n_K n_B$ is a perfect square.} From this, it follows
\begin{align}
    &n_J^2 = p^2+(n_J+p)^2& &\Rightarrow& &p^2+p\,n_J = 0\,,& 
\end{align}
which is solved by $p = \{0,-n_J\}$. There are then two possibilities:
\begin{itemize}
    \item $p = 0$.
    In this case, we have 
    \begin{align}
    &n_B^{(J)} =n_J\,,& &n_B^{(K)} = 0\,,&
\end{align}
which means that in $u_J$ there is a D6-brane baryon solution with baryon number $n_B$ while in $u_K$ the D6-brane has zero baryon number and thus it is unstable under mesonic radiation. Therefore, this competing solution consists of a baryon on the D8-brane with tip in $u_J$, plus some mesonic radiation due to the decay of the unstable decay product.
    \item $p = -n_J$.
     In this case, we have 
    \begin{align}
    &n_B^{(J)} =0,& &n_B^{(K)} = n_J.&
\end{align}
As before, we can think of this as the decay of the same sandwich solution into a baryon, now hanging from $u_K$, $Nn_J$ fundamental strings linking the two sets of D8-branes oriented from $u_J$ to
$u_K$, plus some mesonic radiation due to the decay of the uncharged decay product at $u_J$.
\end{itemize}

\begin{figure}[H]
    \centering
    \includegraphics[width=0.6\textwidth]{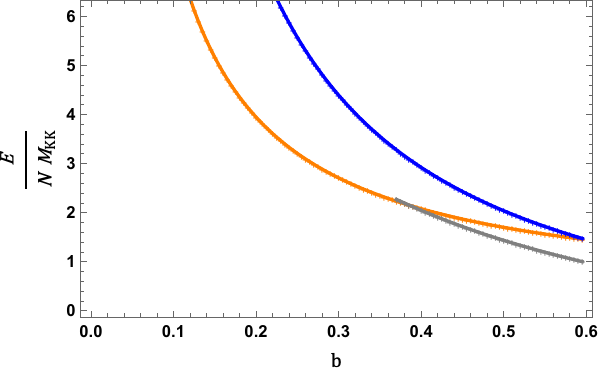}
    \caption{The plot shows the energy of the sandwich solution (with $n_K = 0$ and $n_J = 1$) in grey and the energy of its possible decay products: a baryon solution hanging from $u_K$ plus $N$ fundamental strings going from $u_J$ to $u_K$, shown in blue; or a baryon solution hanging from $u_J$, shown in orange. We present this plot for fixed $b_K = 0.6$ and $\lambda = 50$.}
\label{fig:Esandwichconf}
\end{figure}

We thus conclude that the baryon number on one of the two U-shaped D8-branes is zero, leaving only two interesting competing solutions. We also showed that, as a consequence of flavor number conservation, the allowed solutions must have a number $N|p|$ of strings, corresponding to multi-flavored mesons, stretched between the two flavor branes. Considering the example that we studied in section \ref{sec:sandwich} of a sandwich with $n_K = 0$ and $n_J=1$, the possible decays are therefore to a baryon in $u_J$ with no extra strings linking the D8 branes, or to a baryon in $u_K$, with $N$ strings going from $u_J$ to $u_K$, \textit{i.e.}~$N$ multi-flavored mesons. 
Therefore, we are interested in comparing the energies of these decay products in the large-$\lambda$ limit. As an example, their total energies are given, for fixed $b_K= \frac{u_0}{u_K} = 0.6$ and $\lambda=50$, in figure \ref{fig:Esandwichconf}. The numerical analysis shown in this figure reveals the presence of a first-order phase transition between the sandwich solution and the baryon solution (hanging from $u_J$). For the above-mentioned values of the parameters, this happens at $b\approx 0.38$. In the limit where the D8-branes are stacked together, the energy of the sandwich is the smallest, and hence the sandwich is the stable configuration. Then, increasing the distance between the D8-branes (\textit{i.e.}~decreasing $b$), the first-order transition occurs, and the baryons hanging from $u_J$ begin to be stable under this decay. 

In terms of the baryon stability, we have found that, for some range of parameters, the one-flavored baryon solution can decay into a sandwich connecting two different flavor branes. 
Nevertheless, it must be taken into account the fact that the region where the sandwich exists tends to shrink at large $\lambda$.
Moreover, the baryon solution hanging from $u_K$ plus strings is found to be always more energetic than the other configurations. 

We could also have considered a sandwich solution with $n_K = 1$ and $n_J=0$. In this case, the decay products are: a baryon solution with $n_B=1$ (plus mesonic radiation) hanging from $u_K$, or a baryon solution with $n_B=1$ hanging from $u_J$, plus $N$ fundamental strings connecting the two flavor branes with direction from $u_K$ to $u_J$. As for the previous decay, the sandwich solution is almost always less energetic than its competing solutions.

\subsection{Deconfined phase}
As discussed in the introduction of this section, the instabilities in the deconfined phase can be interpreted as ``dynamical'' instabilities driven by the temperature. The first instability to study for the baryon D6-brane solution is the melting of the baryon vertex into the horizon, which leaves $q_s$ fundamental strings attached at $u_J$ and terminating at the horizon. Therefore, focusing on the $n_B=1$ case for simplicity, we compare the energy of the D6-brane baryon solution (see equation \eqref{eq:energybaryondeconf}) with the energy of $q_s=N$ strings extended from $u=u_J$ to the horizon:
\begin{equation}
    E_\text{f-strings} = \dfrac{N}{2\pi l_s^2}(u_J-u_T) = \dfrac{4\pi}{9}\,N\lambda\dfrac{T^2}{M_{KK}}\dfrac{1-\tilde{b}}{\tilde{b}}\,.
\end{equation}

\begin{figure}[H]
    \centering
    \includegraphics[width=0.5\textwidth]{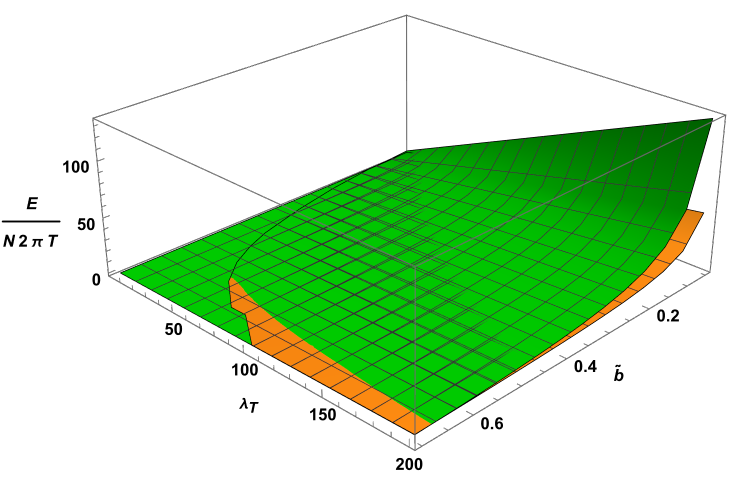}
    \caption{The plot shows the energy of baryons (in orange) and the energy of strings connecting the D8-brane to the horizon (in green) as a function of $\tilde{b}$ and $\lambda_T$.}
\label{fig:estring}
\end{figure}

We plot the two energies in figure \ref{fig:estring}. We can see that the energy of the baryons is almost always lower than the energy of a bunch of fundamental strings terminating at the horizon, apart from a small region in parameter space, when $\lambda$ and $\tilde{b}$ take large values (\textit{i.e.}~the temperature is large), for which baryons are more energetic than the strings, so that they are unstable. Let us mention that the same energy comparison for the case of a standard WSS baryon in the deconfined phase has been performed in \cite{Seki:2008mu}.

\begin{figure}[H]
    \centering
    \includegraphics[width=0.5\textwidth]{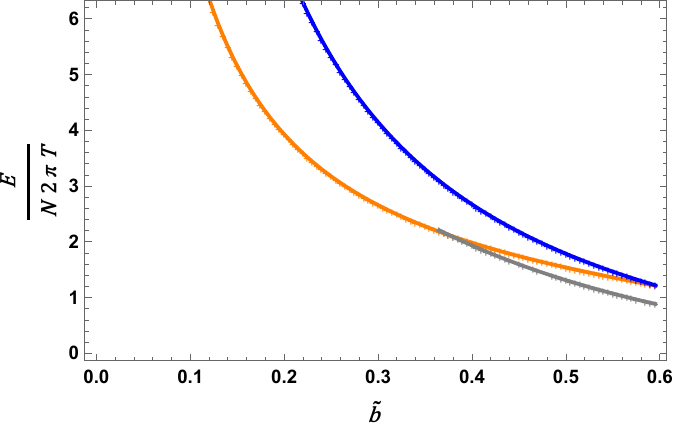}
    \caption{The plot shows the energy of the sandwich solution (with $n_K = 0$ and $n_J = 1$) in grey and the energy of its possible decay products: a baryon solution hanging from $u_K$ plus the energy of $N$ fundamental strings going from $u_J$ to $u_K$, shown in blue; or a baryon solution hanging from $u_J$, shown in orange. We present this plot for fixed ${\tilde b}_K = u_T/u_K=0.6$ and $\lambda = 50$.}
\label{fig:Esandwich}
\end{figure}

We can also study the decay of sandwich solutions into baryons in the deconfined phase. The derivation of the decay products performed in the confined phase is also applicable to the deconfined phase, and therefore we are again interested in comparing the energies of a sandwich solution extended between $u_J$ and $u_K$ with the energies of baryon solutions hanging from $u_J$ or $u_K$ (with the addition of the same number of fundamental strings to conserve flavor). We show the energy comparison in figure \ref{fig:Esandwich}.

We observe numerically the same pattern as in the confined phase, with the same first-order phase transition and energy comparison between the competing solutions.


We can also have a sandwich solution with $n_K = 1$ and $n_J=0$ (instead of $n_K = 0$, $n_J=1$), with the same decay products as in the confined phase.

Moreover, the presence of the horizon also allows for stable charged string loops, \textit{i.e.}~vortons. These objects are (meta)stable only if they carry a very large charge $n_B\sim\lambda$ (see \cite{Bigazzi:2024mge} for more details); otherwise, they are unstable. 
But we find that energetically the large-charge vortons are in fact only metastable, since they can decay into a large number (of order $\lambda$) of baryons with baryon number $n_B\sim\lambda^0$, conserving the baryon number in the process. Instead, a single large charge (of order $\lambda$) baryon could decay into a large charge vorton.

\section{Conclusions and outlook}\label{sec:conclusions}

In this work, we have provided a concrete, first-principle realization of baryons in single-flavor QCD-like theories as compact, solitonic objects holographically realized as D6-branes ending on flavor D8-branes in the Witten-Sakai-Sugimoto model. These objects manifest many of the expected features of the “baryons as quantum Hall droplets” proposal put forward in \cite{Komargodski:2018odf}, most importantly the emergence of a $U(1)_N$ Chern-Simons theory on their worldvolume. 
Such a structure emerges naturally in the holographic setup, which lends support for the idea that planar QCD with one flavor admits such pancake-like baryons, despite the absence of a traditional Skyrmion picture. The physical properties of the baryons, like the energy and the stability radius, computed within the holographic model, are collected in the tables \ref{tab:baryonstring} and \ref{tab:baryonproperties}. The spin-charge relation $J =N n_B^2 /2$, satisfied by these configurations, hints at a deeper underlying structure. It suggests the presence of collective excitations with non-trivial statistics and provides a concrete bridge between solitonic baryons and effective anyonic behavior.

From the string-theory point of view, these D6-brane configurations can be seen as formed by a D4-brane, wrapped on the background $S^4$, representing the baryon vertex in the dual field theory, and by a bunch of fundamental strings attached to it, representing the quark degrees of freedom. 
Then the strings are blown up to a D6-brane (with D4 charge) by the angular momentum, as in the standard supertube case \cite{Mateos:2001qs,Mateos:2001pi}.

A central element of our construction is the detailed analysis of the DBI-CS theory living on the worldvolume of a D6-brane, which in our setup describes \textit{effectively} a three-dimensional manifold with boundary. 
The analysis of this theory reveals a rich topological structure supported on the D6-brane, governed by the global properties of the gauge principal bundle. 

In particular, the D6-brane worldvolume theory supports a non-trivial topological structure encoded in the first Chern number, $\int f/(2\pi)$, which serves as a genuine topological invariant and directly quantifies the baryon number of the configuration. The bulk flux 
$\int f/(2\pi)$ is closely related to the boundary holonomies $\int \hat{a}/(2\pi)$, defined along the circular edges of the D6-brane, which are sensitive to large gauge transformations in contrast with the baryon number. Physically, these holonomies are associated with winding numbers $n^{(W)} =N \int \hat{a}/(2\pi)$ which precisely count the net number of fundamental strings coming out of the flavor brane. In the field theory interpretation, this is the number of quarks associated with the flavor brane.

The generality of the techniques we made use of allows us to classify and analyze a broad spectrum of solitonic objects, which we refer to as a zoo of configurations, including baryons, punctured domain walls, and sandwich vortons (besides the standard vortons studied in \cite{Bigazzi:2024mge}). The punctured domain walls correspond to D6-branes ending on the D8-brane along a circular boundary, forming metastable domain walls with holes; these holes can grow and contribute to the decay of extended domain walls, as anticipated in previous studies of defect dynamics in QCD-like theories \cite{Gabadadze:2000vw, Son:2000fh}. Sandwich vortons, on the other hand, are ring-shaped objects corresponding to D6-branes stretched between two D8-branes located at distinct holographic positions. These loops carry both baryon charge and angular momentum and provide a controlled realization of metastable string-like bound states, analogous in spirit to axionic-$\eta'$ or axionic-pionic vortons \cite{Gabadadze:2000vw} discussed in earlier holographic contexts in \cite{Bigazzi:2022luo, Bigazzi:2024mge}.

We have provided explicit numerical solutions for the D6-brane embeddings and gauge connections corresponding to each of these configurations, computing their key physical observables such as mass, size, and charge-to-spin ratio. These solutions are constructed both in the confined and deconfined phases of the WSS model.

Finally, we have explored the possible decay channels of the baryon configuration in both phases. The decay modes are constrained by charge, angular momentum, flavor conservation, and energy considerations. Our analysis suggests the existence of non-trivial phase diagrams in the space of theory parameters in which different solitonic configurations dominate. In particular, focusing on the confined phase, the sandwich solution connecting two different flavor branes can be the decay product of, or it can decay into a baryon solution with some radiated mesons and fundamental strings (\textit{i.e.}~multi-flavored mesons). In the deconfined phase, other than the aforementioned decay, the baryon solution can also melt into the plasma and decay into fundamental strings ending at the horizon, \textit{i.e.}~deconfined quarks. Finally, the baryon solution, whenever it has a large charge $n_B\sim\lambda$, can decay into large charged vortons, which themselves can decay into small charge baryons.

There are many extensions we left for the future. This work provides a description of the holographic Hall droplet baryon (and related defects) from the D6-brane perspective. We did not investigate, within the holographic model, the low-energy description of the baryon (and related defects) in terms of the mesonic degrees of freedom which live on the D8-brane worldvolume. As a matter of fact, a D6-brane attached to the D8-brane sources the flavor gauge connection, leading to non-trivial configurations. We plan to explore this direction in the near future, following the lines of \cite{Bigazzi:2022luo,Bigazzi:2022ylj,Bigazzi:2024mge}. 

A delicate issue concerns the stability of the $n_B \sim {\cal O}(1)$ baryon solutions we have found.
In fact, we found that the size of these particles is parametrically small in $\lambda$, both in the Minkowski directions and the holographic direction.
As such, the distance of (the center of) the D6-brane from the D8-brane is small in units of $\sqrt{\alpha'}$.
In flat space, the system of parallel D6 and D8 branes on top of each other has a tachyonic mode in the spectrum, with mass squared of order $1/\alpha'$, signaling the tendency of the D6-brane to melt into the D8-brane worldvolume. Even if we are considering a non-trivial background, with non-parallel, separated branes, the distance of the D6 and D8 branes could be too small, and keep this mode tachyonic. We plan to investigate this issue in the near future.

Another issue we want to clarify is the emergence of a chiral edge mode, largely studied in the context of FQHE, and its dynamics. We would like to investigate whether the edge dynamics could come from the interactions between the D6-brane and the flavor branes and hence be already present in the model. 

A further interesting question to answer is whether it is possible to realize the $N_f\geq 2$ holographic Hall droplet baryon, and how this construction is connected to the standard instantonic baryons in the WSS model, which are the holographic realizations of baryons as Skyrmions. From \cite{Komargodski:2018odf}, we expect that on the worldvolume of the $N_f\geq 2$ Hall droplet sheet should live a $U(N_f)_N$ Chern-Simons theory and this can be realized in the holographic model by considering a stack of $N_f$ D6-branes attached to the stack of $N_f$ flavor branes.

Finally, there has been interest lately \cite{Ma:2019xtx,Ma:2020nih,Rho:2024ihu} in the role played by the Hall droplet sheets and their charged versions at finite density to connect them to physical systems as neutron stars. Therefore, it would be interesting to use holographic methods to investigate these scenarios.

\section*{Acknowledgments}
We thank Riccardo Argurio, Andrea Cappelli, Carlos Hoyos, Thibaud Raymond, Angel Uranga and Riccardo Villa for suggestions, comments, and very helpful discussions. A.O. wants to thank the University of Helsinki and the King's College of London for the kind hospitality during the preparation of this work.
This project has been partially supported by the grant PRIN 20227S3M3B ``Bubble Dynamics in Cosmological Phase Transitions''.

\appendix

\section{Details on the topology of the system}\label{ap:topo}
In this appendix, we collect the details of the topological properties and theorems used in section \ref{vortex}. The results exposed here are extracted from \cite{Bott:1982xhp, Hatcher:478079, Simms_1964}.
\smallbreak
Recall that we are considering a 
	two-dimensional manifold $\Sigma$ such that $M_3 = \mathbb{R}_t \times \Sigma$, that is, topologically equivalent to a cylinder with its two bounding circles $S^1$.
\smallbreak
We now justify the statement regarding the classification of principal $U(1)$-bundles used for the baryon number. In the following, we shall work with a general smooth manifold, which we will denote as $M$. There exists a classifying space named $BU(1)$ such that principal $U(1)$-bundles over $M$ (up to gauge transformations) are isomorphic to the maps from $M$ to $BU(1)$ (up to homotopy):
\begin{equation}
\{\text{Principal }U(1)-\text{bundle on }M \}/\text{gauge} \cong \{M \to BU(1)\}/\text{homotopy}\,.
\end{equation}
The classifying space $BU(1)$ is by definition (see for instance \cite{Hatcher:478079} for a precise definition and the calculations in section 3.2 there) the infinite dimensional complex projective space $\mathbb{CP}^{\infty} \equiv S^{\infty}/U(1)$. This corresponds to the fibering:
\begin{equation}
    S^1 \hookrightarrow S^{\infty} \twoheadrightarrow \mathbb{CP}^{\infty}\,.
\end{equation}
Using the long exact homotopy sequence of this fibering (see section 4 of \cite{Hatcher:478079} or section 17 in \cite{Bott:1982xhp})  one can show that $\pi_2(\mathbb{CP}^{\infty}) = \pi_1(U(1)) = \mathbb{Z}$, and also that every other homotopy group of $\mathbb{CP}^{\infty}$ vanishes. Path-connected topological spaces that have only one non-trivial homotopy group like this one have very useful properties. They are called Eilenberg-Maclane spaces $K(G, n)$ for a group $G$ if:
\begin{equation}
    \pi_n(K(G, n)) = G\,,
\end{equation}
and all their other homotopy groups are trivial. Therefore $BU(1)$ verifies:
\begin{equation}
    BU(1) \cong K(\mathbb{Z}, 2)\,.
\end{equation}
As a consequence of a powerful theorem (Brown's representability theorem),\footnote{We are assuming here that $M$ is endowed with a \textit{cellular complex} (CW-complex) that respects nice enough properties, which can always be done for closed manifolds. Remember that a cellular complex is a topological space that is built by gluing together topological balls (so-called cells) of different dimensions in specific ways. See \cite{Simms_1964} as a useful reference.} the maps from $M$ to Eilenberg-Maclane spaces $K(G, n)$ are classified by the $n$th cohomology group with coefficients in $G$, therefore in our case, where we have $G=\mathbb{Z}$ and $n=2$, we get:
\begin{equation}
    \{M \to BU(1)\}/\text{homotopy} \cong \{M \to  K(\mathbb{Z}, 2) \}/\text{homotopy} \cong H^2(M; \mathbb{Z})\,,
\end{equation}
\textit{i.e.} the maps from $M$ to $BU(1)$ up to homotopy, or equivalently the principal $U(1)$-bundles over $M$ up to gauge transformations, are classified by the second cohomology group of $M$ with coefficients in $\mathbb{Z}$.
\smallbreak
Recall now that given that $f$ vanishes on the two bounding circles, we showed that the relevant manifold to study our setup is in fact the pinched torus $M = \mathbb{PT}$. There, we are using the one-point compactification of the cylinder. In this case, it is equivalent to performing a relative cohomology calculation of the cylinder with respect to the two bounding circles. We proceed by computing its cohomology, which can be done using the long exact sequence:
\begin{equation}
    H^*(\mathbb{PT}; \mathbb{Z}) = (\mathbb{Z}, \mathbb{Z}, \mathbb{Z}, 0\dots)\,.
\end{equation}
Then, using the definition of the first Chern class, we get that the integer in $H^2(\mathbb{PT}; \mathbb{Z}) = \mathbb{Z}$ corresponds to the first Chern number of the $U(1)$-bundle. This last definition also corresponds to our definition of the baryon number.
\smallbreak
We now wish to confront this result with the topology of the usual Yang-Mills $SU(2)$ instanton setup. To reconstruct the statements used in section \ref{vortex}, the steps are very similar to what we just did. A similar classifying space can be constructed for $SU(2)$:
\begin{equation}
\{M \to BSU(2)\}/\text{homotopy} \cong \{\text{Principal }SU(2)-\text{bundle} \}/\text{gauge}\,,
\end{equation}
which, this time, can be seen to be linked with the infinite quaternionic projective group:
\begin{equation}
   BSU(2) = \mathbb{HP}^{\infty} \equiv S^{\infty}/SU(2)\,.
\end{equation}
\begin{equation}
    SU(2) \hookrightarrow S^{\infty} \twoheadrightarrow \mathbb{HP}^{\infty}\,.
\end{equation}
This, using the homotopy exact sequence, yields that $\pi_4(\mathbb{HP}^{\infty}) = \pi_3(SU(2)) = \mathbb{Z}$, and every other homotopy group of this space vanishes. Therefore, $BSU(2) = K(\mathbb{Z}, 4)$, which yields by the same theorem the classification of principal $SU(2)$-bundles by $H^4(M; \mathbb{Z})$. Applying this to $M = S^4$ which is the manifold on which $f$ is defined in the instanton case we have (\ref{eq:insth4}):
\begin{equation}
    H^4(S^4; \mathbb{Z}) = \mathbb{Z}\,.
    \label{eq:insth4}
\end{equation}
\smallbreak
The last step is to use the properties of the second Chern class instead of the first one, which takes its values in $H^4(S^4; \mathbb{Z}) = \mathbb{Z}$, and corresponds to the instanton number.

\section{Details on the numerical analysis}\label{app:numerics}
In this appendix, we present the methods used to numerically solve the equations of motion. We first identify and exploit relations between the physical parameters in order to reduce the parameter space of the numerical study. Then we recall the boundary conditions derived in section \ref{sec:bc} and describe the methods used to solve the differential equations. We start by performing the following coordinate and field redefinition
\begin{align}
    u&\to u_J\, u  ,& \rho &\to L\,\rho, & a_\psi &\to L\, u_J \,a_\psi, & a_t &\to u_J \,a_t\,.
    \label{eq:redefinitions}
\end{align}
Let us recall that $L$ has a different expression for the confined and the deconfined phases
\begin{align*}
    &\text{Confining phase:}&  L&= \frac{R^{3/2}}{u_J^{1/2}} J(b),  &J(b)=\frac{2}{3}\int_0^1 dv \frac{ \sqrt{1-b^3} \,\,\sqrt{v}}{(1-b^3 v) \sqrt{1-b^3v - (1-b^3)v^{8/3}}}\,,\\
    &\text{Deconfined phase:}&  L&= \frac{R^{3/2}}{u_J^{1/2}} J_T(\tilde{b}),  &J_T(\tilde{b})=\frac{2}{3} \int_0^1 dv \frac{\sqrt{1-\tilde{b}^3}\,\,\sqrt{v}}{\sqrt{1-\tilde{b}^3 v}\, \sqrt{1-\tilde{b}^3v - (1-\tilde{b}^3)v^{8/3}}}\,,
\end{align*}
where we have used the standard notation of the WSS model, $b=u_0/u_J$ and $\tilde{b}=u_T/u_J$. In the following, all the fields and the coordinate $u$ have to be intended as the rescaled ones. With these transformations, the equations of motion become
\begin{itemize}
    \item Confining phase
    \begin{align}
     J^2(b)\,\partial_u &\left( \dfrac{u^4\rho'(u)\rho(u)}{D(u)} \right)  - u\, D(u)\left(1- \frac{9}{u^2}\tilde{a}_t^2 \right) = 0\,, \\
    & \partial_u\tilde{a}_\psi + \frac{3}{u} \rho(u)D(u)\,\tilde{a}_t = 0\,, \\
    & \partial_u\tilde{a}_t + \frac{3}{u} \dfrac{D(u)}{\rho(u)}\,\tilde{a}_\psi = 0\,,
\end{align}
where
\begin{align}
 &f(u)=1-\dfrac{b^3}{u^3}\,,& &D(u)= \sqrt{\frac{J^2(b)\,\rho'(u)^2\, u^3 + \dfrac{1}{f(u)}}{1+ \dfrac{9}{u^2} \left(\dfrac{\tilde{a}_\psi^2}{\rho^2}-  \tilde{a}_t^2 \right)}}\,.&
\end{align}
\item Deconfined phase
\begin{align}
    J^2(\tilde{b})\,\partial_u &\left( \dfrac{u^4\rho'(u) \rho(u) f_T(u)}{D_T(u)} \right)  - u\, D_T(u)\left(1- \dfrac{9\tilde{a}_t^2}{u^2 f_T(u)} \right) = 0\,, &&\\
    & \partial_u\tilde{a}_\psi + \frac{3}{u\,f_T(u)} \rho(u)\,D_T(u)\,\tilde{a}_t = 0\,,\\
    & \partial_u\tilde{a}_t + \dfrac{3}{u} \dfrac{D_T(u)}{\rho(u)}\,\tilde{a}_\psi= 0\,,
\end{align}
where
\begin{align}
    &f_T(u)=1-\dfrac{\tilde{b}^3}{u^3},& &  D_T(u) = \sqrt{\dfrac{1+\rho'(u)^2 \,u^3 f_T(u)  \,J^2(\tilde{b})}{1+ \dfrac{9}{u^2} \left(\dfrac{\tilde{a}_\psi^2}{\rho^2(u)}-  \dfrac{\tilde{a}_t^2}{f_T(u)}  \right)}}\,.&
\end{align}
\end{itemize}

We adapt the boundary conditions on $a$ discussed in section \ref{sec:variationalprinciple} to study configurations with generic spin. Recall that in the baryon case, $J=Nn_B^2/2$ holds and therefore the boundary conditions are those indicated in section \ref{sec:variationalprinciple}. For the other cases, we invert equation \eqref{eq:baryonnumber} and \eqref{eq:spin}:
\begin{align}
    \hat{a}_\psi(u_*) &= \frac{n_B}{2} - \frac{J}{Nn_B}\,,& \hat{a}_\psi(u_J) &= -\frac{n_B}{2} - \frac{J}{Nn_B}\,.
\end{align}
After the redefinitions in \eqref{eq:redefinitions}, using the holographic dictionary, we obtain
\begin{itemize}
    \item Confining phase:
    \begin{align}
        &\tilde{a}_\psi(1) = \frac{2 \pi l_s^2}{L u_J} \left(  -\frac{n_B}{2} - \frac{J}{Nn_B}\right) = \frac{6 \pi \sqrt{b}}{J(b)\,\lambda} \left( -\frac{n_B}{2} - \frac{J}{Nn_B} \right),\\
    & \tilde{a}_\psi(u_*/u_J) = \frac{2 \pi l_s^2}{L u_J} \left( \frac{n_B}{2} - \frac{J}{Nn_B}\right) = \frac{6 \pi \sqrt{b}}{J(b)\,\lambda} \left(\frac{n_B}{2} - \frac{J}{Nn_B} \right).
    \end{align}
    \item Deconfined phase:
    \begin{align}
        &\tilde{a}_\psi(1) =  \frac{2 \pi l_s^2 }{L \,u_J} \left( -\frac{n_B}{2} - \frac{J}{Nn_B} \right)= \frac{6\pi \sqrt{\tilde{b}} }{J_T(\tilde{b})\,\lambda_{T}}\left( -\frac{n_B}{2} - \frac{J}{Nn_B} \right),\\
    & \tilde{a}_\psi(u_*/u_J) =  \frac{2 \pi l_s^2 }{L\, u_J} \left( \frac{n_B}{2} - \frac{J}{Nn_B} \right)= \frac{6\pi\sqrt{\tilde{b}} }{ J_T(\tilde{b})\,\lambda_{T}}\left( \frac{n_B}{2} - \frac{J}{Nn_B} \right),
    \end{align}
\end{itemize}
where we have defined the effective 't Hooft coupling constant \begin{equation}
\lambda_T = \dfrac{2\pi\, T}{M_{KK}}\,\lambda = \dfrac{T}{T_c}\,\lambda\,. 
\end{equation}

Notably, the deconfined phase expressions are the same as in the confined case, up to replacing $(\lambda,b,J(b))\to(\lambda_{T},\tilde{b},J_T(\tilde{b}))$. In general, the boundary value problem we are considering depends, a priori, on four parameters: $b$ (or $\tilde{b}$), $\lambda_{(T)}$, $n_B$, and the angular momentum $J$. 

In the following, we will focus on objects with $J = N n_B^2/2$. In this case, recall that $n_B$ and $\lambda_{(T)}$ appear only in the ratio
\begin{equation}
\xi_{(T)} \equiv \dfrac{n_B}{\lambda_{(T)}},\,
\end{equation}
effectively reducing the number of parameters by one. The baryon number can be parametrically large in $\lambda_{(T)}$, $n_B = \lambda_{(T)}^\gamma$ with $0\leq\gamma\leq 1$, and we are working in large $\lambda$-limit. Therefore, the parameter $\xi$ ranges from zero when $n_B\sim\lambda_{(T)}^0$ to order 1 when $n_B\sim\lambda_{(T)}$.


We will now describe the integration method for this differential problem. The integration domain of the differential equations in the rescaled radial coordinate $u$ is $[b,1]$ for the confined phase and $[\tilde{b},1]$ for the deconfined phase. In section \ref{sec:bc}, we have defined two of our boundary conditions in $u=1$ and two in $u=u_*/u_J$. To find a numerical solution with the correct boundary conditions, we will free the two boundary conditions in $u=u_*/u_J$ and implement two more boundary conditions in $u=1$, which can be thought of as parameters, $\tilde{a}_t(1)$ and $\rho(1)$. The physical boundary conditions in $u=u_*/u_J$ will then be obtained by varying these two parameters. 

These parameter-boundary conditions are not allowed to take any value. Obviously, $\rho(1)$ can only take positive values. A more interesting inequality can be derived for $\tilde{a}_t$. Indeed, for the Lagrangian to be real, the quantity $D^2_{(T)}$ has to be positive or zero. In the confined phase, inside the integration region we are considering, this implies
\begin{equation}
       1+ \frac{9}{u^2} \left(\frac{\tilde{a}_\psi^2}{\rho^2}-  \tilde{a}_t^2 \right) >0\,.
\end{equation}

A second constraint arises from the equilibrium condition for the D6-brane, which requires that the D6-brane ends orthogonally at the D8-branes' tip. This amounts to having $\rho'(1) = 0$ and, for the lowest energy configuration, to having a concave function. Replacing these values in (\ref{eq:eomsconf}) gives the concavity inequality:
\begin{equation}
   \frac{u^2}{9} < \tilde{a}_t^2\,.
   \label{eqn:attachuJ}
\end{equation}

The value of the gauge connection component $\tilde{a}_t$ for an equilibrium configuration then has to lie in the window
\begin{equation}
    \frac{u^2}{9} < \tilde{a}_t^2 < \frac{u^2}{9} +  \frac{\tilde{a}_\psi^2}{\rho^2}\,.
\end{equation}

One can think of exotic solutions, like punctured domain walls, whose profile $\rho(u)$ is a convex function (or there is some interval in which it is convex) for which 
\begin{equation}
    0< \tilde{a}_t^2< \frac{u^2}{9}\,.
\end{equation}

Similar results can be written in the deconfined case for concave solutions
\begin{equation}
   f_T(u)\, \frac{u^2}{9}  < \tilde{a}_t^2  < f_T(u) \left(  \frac{u^2}{9} +  \frac{\tilde{a}_\psi^2}{\rho^2} \right),
\end{equation}
and convex ones
\begin{equation}
    0< \tilde{a}_t^2< f_T(u)\,\frac{u^2}{9}\,.
\end{equation}

In all the cases, we introduce a parameter $\alpha$ to span the window. Our boundary conditions for a baryon become:
\begin{itemize}[label=$\diamond$]
    \item $\tilde{a}_\psi(1) = -\dfrac{6\pi\, n_B \sqrt{b}}{\lambda J(b)}$ (Confined phase), \hspace{1.5cm} $\tilde{a}_\psi(1) = -\dfrac{6\pi \,n_B \sqrt{\tilde{b}}}{\lambda_{T} J_T(\tilde{b})}$ (Deconfined phase).
    \item $\rho'(1) = 0$, which enforces the equilibrium condition by requiring that the embedding ends orthogonally at $u=u_J$. 
    \item $\rho(1) = l$ the radius of the D6-brane at the D8-branes' tip is our first parameter-boundary condition. The value of $l$, which gives the physical boundary conditions in $u_*/u_J$ will be named $l_\text{stable}$, the stability radius. 
    \item $\tilde{a}_t(1)$, our second parameter-boundary condition, which can be given in terms of $\alpha$ as
    \begin{align}
    \text{Confined phase: }\hspace{0.2cm}\tilde{a}_t(1) &= \sqrt{\frac{1}{9} +  (1-\alpha^2)\left(\frac{(6\pi)^2 b }{\lambda^2 J^2(b)\, l^2}\right)}\,.\\
    \text{Deconfined phase: }\hspace{0.2cm}\tilde{a}_t(1) &= \sqrt{f_T(1)}\sqrt{\frac{1}{9} +  (1-\alpha^2)\left(\frac{(6\pi)^2 \tilde{b}}{\lambda_T^2 J_T^2(\tilde{b})\,l^2}\right)}\,.
\end{align}
\end{itemize}
Let us observe that the same expression for $\tilde{a}_t$ with a minus sign in front of it gives rise to nonphysical solutions, and hence, we focus on $\tilde{a}_t>0$. The function $1-\alpha^2$ was chosen heuristically and found to give faster convergence results.

We integrate the equations of motion parametrically using Mathematica. The process depends on the type of solution considered. For baryons specifically, we interrupt the integration process when $\rho$ reaches zero. For punctured domain wall solutions, we also interrupt the process if $\rho$ diverges. This allows us to consider solutions without predetermined $u_*$ like domain walls and baryons. For solutions which extend to a fixed $u_*$ like vortons or sandwich defects, the integration domain is predetermined, so there is no interruption condition.

Specifying the boundary conditions on one side of the integration domain provides the uniqueness of the solution, reducing the problem to finding the parameters $l_\text{stable}$ and $\alpha$ such that the physical boundary conditions in $u=u_*$ are satisfied. This is achieved by translating the boundary conditions in $u=u_*$ into constraints, finding a Lagrangian functional $L_0$ for these constraints, and numerically computing its minima over the parameter space:
\begin{equation}
 l_\text{stable}>0\,,\quad 0<\alpha<1\,.
\end{equation}
Since the different types of solutions we consider correspond to different boundary conditions in $u=u_*$, we need to adjust $L_0$ to each type of solution. Moreover, the Lagrange multipliers of these constraints also have to be further adjusted to ensure convergence to the desired solution. The constraint functions used for each type of solution are listed in table \ref{tab:L0}.
\begin{table}[H]
    \centering
    \begin{tabular}{|c|c|c|}
        \hline
        $L_0$ & Confined phase & Deconfined phase \\
        \hline
        Baryon & $\tilde{a}_\psi(u_E)^2 + \lambda \rho(u_E)^2$ & $\tilde{a}_\psi(u_E)^2 + \lambda_T \rho(u_E)^2$  \\
        \hline
        Vorton & (-) & $\tilde{a}_\psi(u_*)^2 + \tilde{a}_t(u_*)^2 + \frac{1}{\lambda_T }(u_T - u_*)^2$ \\
        \hline
        Punctured domain wall & $\tilde{a}_\psi(u_K)^2 +\frac{1}{\lambda} (u_K - u_*)^2$& $\tilde{a}_\psi(u_K)^2 + \frac{1}{\lambda_T }(u_K - u_*)^2$\\
        \hline
        Sandwich defect & $\tilde{a}_\psi(u_K)^2 + \frac{1}{\lambda}(u_K - u_*)^2 + \lambda\rho'(u_K)^2$ & $\tilde{a}_\psi(u_K)^2 + \frac{1}{\lambda_T}(u_K - u_*)^2 + \lambda_T\rho'(u_K)^2$ \\
        \hline
    \end{tabular}
    \caption{Constraint functions used for each configuration in both phases of the WSS model.}
    \label{tab:L0}
\end{table}
It is easy to verify that every solution to the physical boundary problem is at a minimum of this function, and that they are realized for $L_0 = 0$. We will say that the solution is found when the value of $L_0$ at the minimum is lower than
\begin{equation}
    \text{Min } L_0 < \varepsilon\equiv10^{-10}
\end{equation}
For baryon solutions, the only choice lies in the Lagrange multiplier for each configuration. For the explored range of $\lambda_{(T)}$, this was found to give the best results. For vortons and punctured domain walls, we add a $(u_* - u_K)^2$ term, which controls the coordinate $u$ at which the solution ends, either by entering the horizon (then $u_K = u_T$) or by diverging ($\rho \to \infty$). For the sandwich vortons, we also add a $\rho'(u_*)^2$ term to control the stability at the second D8-brane setup.

\section{Hall droplet baryons in the deconfined phase}\label{app:deconf1}
\begin{figure}[H]
	\centering
	\begin{subfigure}{0.45\textwidth}
		\centering
		\includegraphics[width=\textwidth]{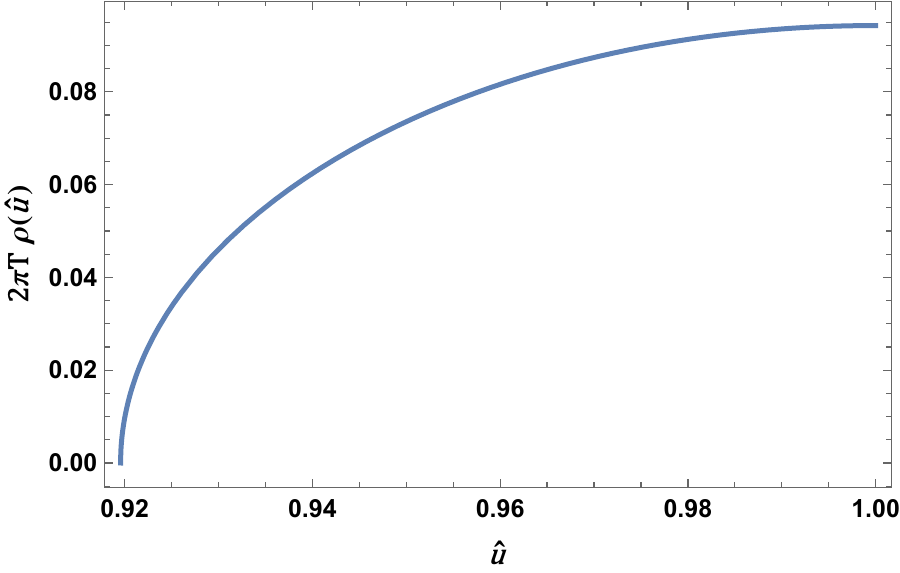}
	\end{subfigure}
	\hfill
	\begin{subfigure}{0.45\textwidth}
		\centering
		\includegraphics[width=\textwidth]{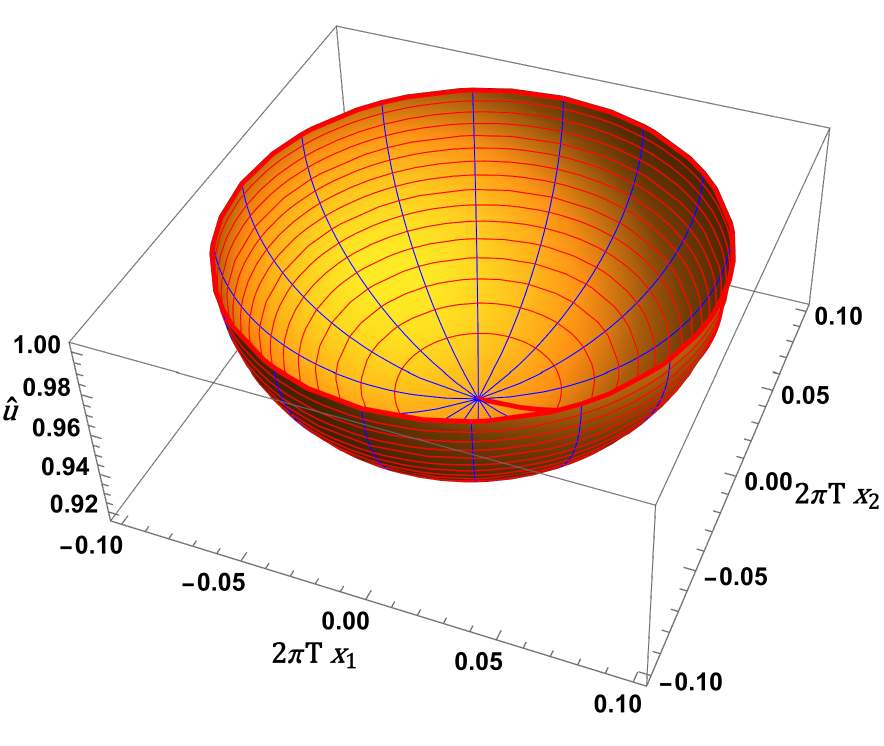}
	\end{subfigure}
	\caption{Numerical solution of $\rho$ (times $2\pi T$) for $\tilde{b}=0.1$ and $\lambda_T=100$ as a function of the rescaled holographic coordinate $\hat{u}=u/u_J \in [0.92,1]$,
		describing a (meta)stable baryon with baryon number $n_B=1$. The profile's derivative at $\hat{u} = 1$ is zero, indicating the (meta)stability of this embedding, while it goes to $\infty$ at $\hat{u}=u_E/u_J$.}
	\label{fig:rhobaryondeconf}
\end{figure}

\begin{figure}[H]
	\centering
	\begin{subfigure}{0.45\textwidth}
		\centering
		\includegraphics[width=\textwidth]{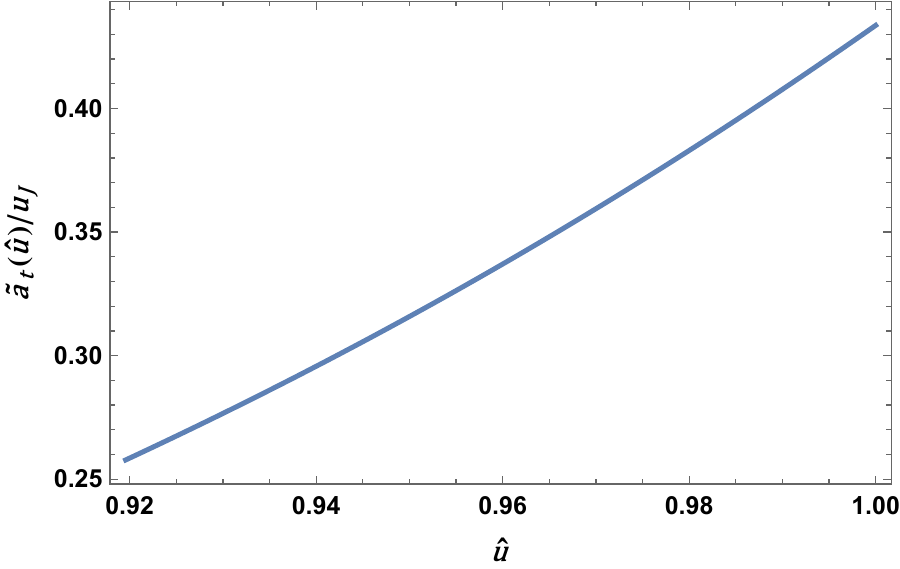}
	\end{subfigure}
	\hfill
	\begin{subfigure}{0.45\textwidth}
		\centering
		\includegraphics[width=\textwidth]{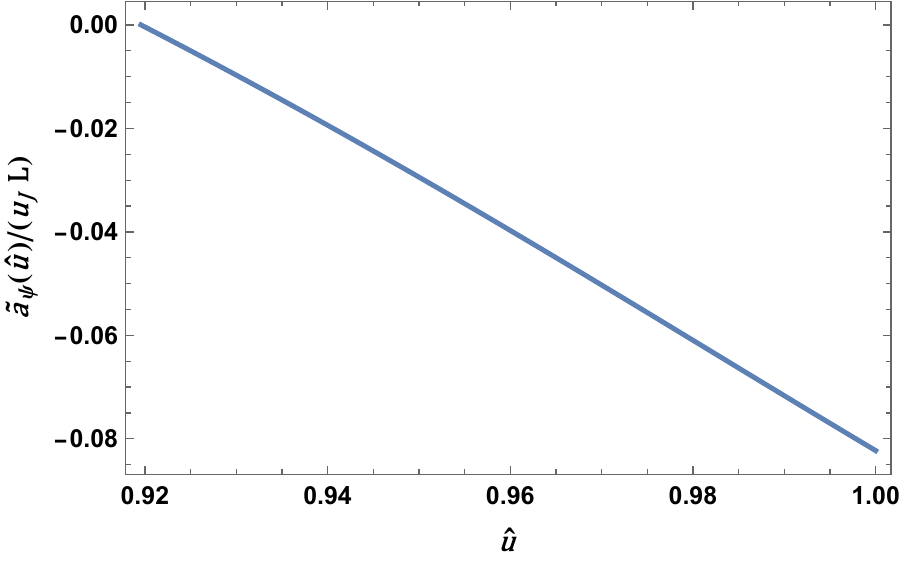}
	\end{subfigure}
	\caption{Numerical solutions for $u_J^{-1}\tilde{a}_t$ (left panel) and $(u_J L)^{-1}\tilde{a}_\psi$ (right panel) as a function of the rescaled holographic coordinate $\hat{u}$ for $\tilde{b}=0.1$ and $\lambda_T=100$.}
	\label{fig:abaryondeconf}
\end{figure}

In this appendix, we present the baryon solution with $n_B=1$ in the deconfined phase for $\lambda_T=100$ and $\tilde{b} =0.1$, in figures \ref{fig:rhobaryondeconf} and \ref{fig:abaryondeconf}. For these values of the parameters, the integration stops at $u_E\approx 0.92$. The plots show functions of the rescaled holographic coordinate $\hat{u}=u/u_J$.

\begin{figure}[H]
	\centering
	\begin{subfigure}{0.45\textwidth}
		\centering
		\includegraphics[width=\textwidth]{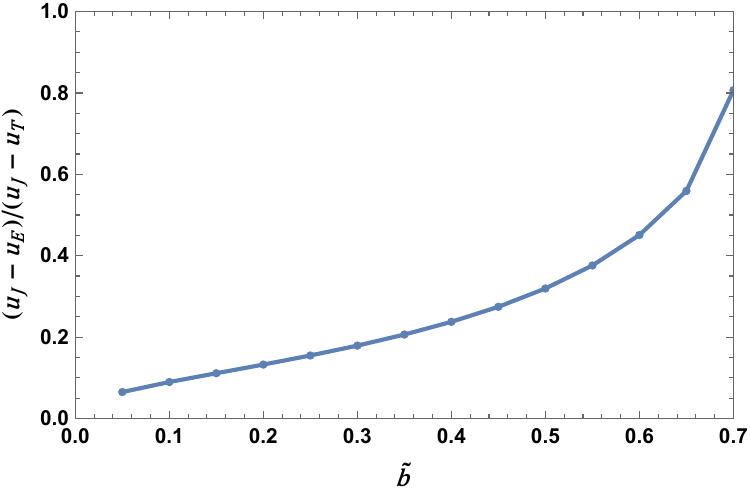}
	\end{subfigure}
	\hfill
	\begin{subfigure}{0.45\textwidth}
		\centering
		\includegraphics[width=\textwidth]{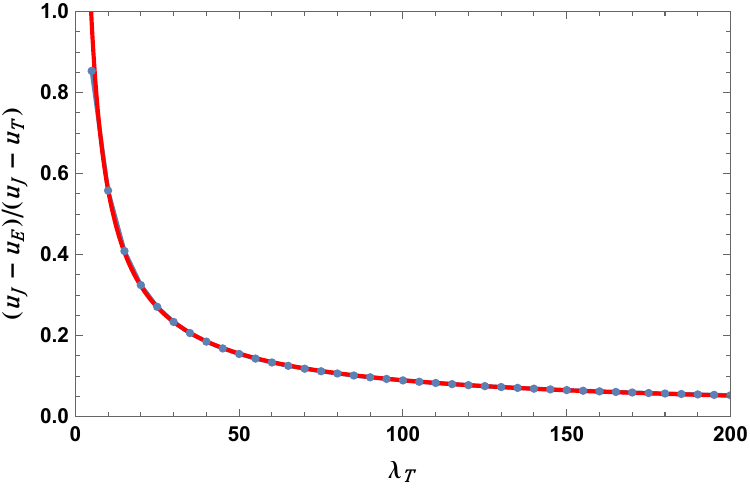}
	\end{subfigure}
	\caption{In these plots, we show the behavior of the extension of the D6-brane along the $u$ coordinate $(u_J-u_E)/(u_J-u_T)$. In the left panel, we fix $\lambda_T=100$ and we show the behavior as a function of $\tilde{b}$. In the right panel, we fix $\tilde{b} = 0.1$ and we show the behavior of $(u_J-u_E)/(u_J-u_T)$ as a function of $\lambda_T$. A fit with a function of $\lambda_T$ is indicated in red and goes like $\sim\lambda_T^{-0.8}$.}
	\label{fig:udeconf}
\end{figure}

We can compute the extension along $u$ of the D6-brane embedding with baryon number $n_B=1$. The left panel of figure \ref{fig:udeconf} shows the plot of $(u_J-u_E)/(u_J-u_T)$ as a function of $\tilde{b}$ for fixed $\lambda_T=100$. In the right panel of figure \ref{fig:udeconf}, we show the plot of the same quantity for fixed $\tilde{b}=0.1$ as a function of $\lambda_T$ (in blue). We also show a fit (in red) with a power function of $\lambda_T$
\begin{align}
	&\dfrac{u_J-u_E}{u_J-u_T}\bigg|_{\tilde{b}\,\text{fixed}}\sim\,\lambda_T^{-\gamma}& &\text{with}& &\gamma\approx 0.8\,.&
\end{align}

In all the plots for the deconfined phase, we remind that $\tilde{b}\simeq 0.75$ is the value for which we have chiral symmetry restoration, after which the D6-brane configurations do not make sense anymore. Therefore, we stop all the deconfined phase plots at $\tilde{b}=0.7$.

\begin{figure}[H]
	\centering
	\begin{subfigure}{0.45\textwidth}
		\centering
		\includegraphics[width=\textwidth]{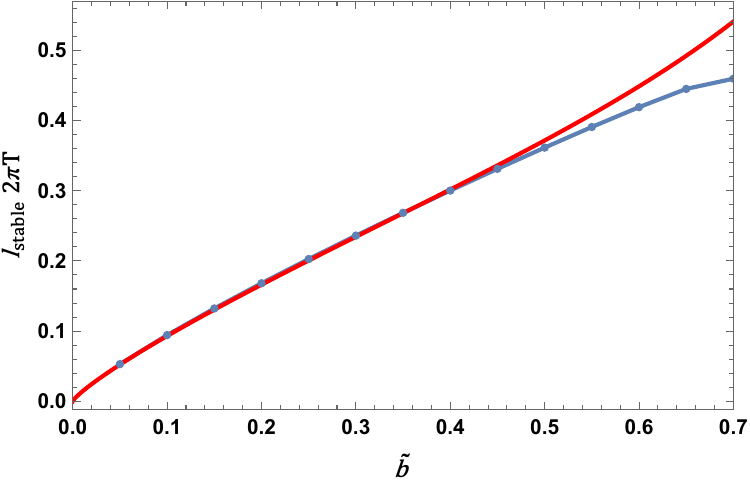}
	\end{subfigure}
	\hfill
	\begin{subfigure}{0.45\textwidth}
		\centering
		\includegraphics[width=\textwidth]{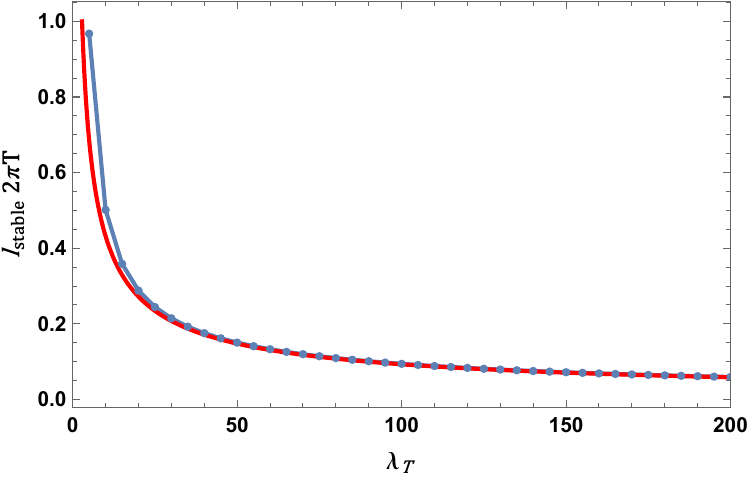}
	\end{subfigure}
	\caption{The plot of $l_\text{stable}2\pi T$ is shown (in blue) for fixed $\lambda_T=100$ as a function of $\tilde{b}$ (left panel) and for fixed $\tilde{b}=0.1$ as a function of $\lambda_T$ (right panel). The red lines are the analytical estimates coming from the rigid rotor.}
	\label{fig:lstabledeconf}
\end{figure}

Also in the deconfined phase, the rigid rotor approximation perfectly fits the behavior of the stability radius $l_\text{stable}$ as a function of 
$\lambda$. In the left panel of figure \ref{fig:lstabledeconf}, it is shown the plot of $l_\text{stable}$ as a function of $\tilde{b}$ for fixed $\lambda_T=100$ in blue, together with the analytical result in red coming from the rigid rotor approximation \eqref{eq:lstablerigiddeconf}. In the right panel of figure \ref{fig:lstabledeconf}, it is shown the plot of $l_\text{stable}$ for $\tilde{b}=0.1$ as a function of $\lambda_T$ (in blue) together with the analytic prediction at fixed $\tilde{b}=0.1$ coming from the rigid rotor. The analytic prediction is very reliable for large $\lambda_T$, while it starts to fail at very small $\lambda_T$ where, anyway, the holographic results are not reliable.

Therefore, also in the deconfined phase, the approximation of the D6-brane as a rotating disk looks reasonable since the extension along $u$ of the D6-brane ($\sim\lambda^{-0.8}$) is parametrically smaller than the radius of the D6-brane ($\sim\lambda^{-2/3}$).

\begin{figure}[H]
	\center
	\includegraphics[height = 5cm]{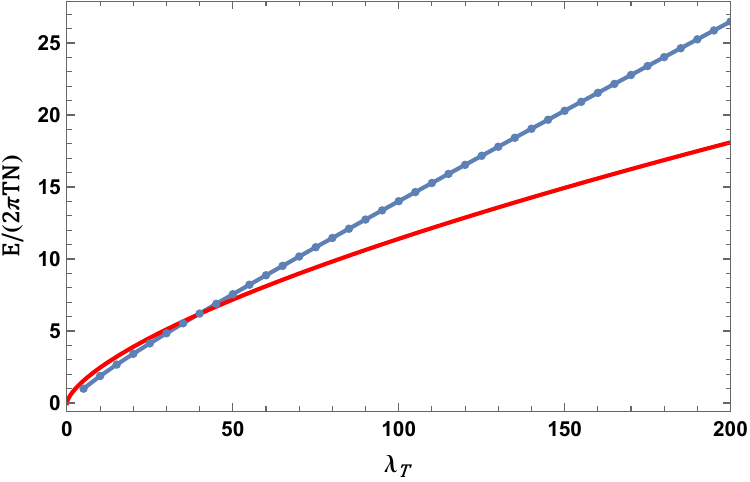}
	\caption{The plot shows the energy of the baryon solution with $n_B=1$ in blue for $\tilde{b}=0.1$ as a function of $\lambda_T\in[5,200]$. The red curve is the corresponding analytical prediction from the rigid rotor.}
	\label{fig:energyfitdeconf}
\end{figure}

The energy of the solution in the deconfined phase can be computed numerically starting from the following formula (see also equation (\ref{energy2}))
\begin{equation}\label{eq:energybaryondeconf}
	E = N\,\lambda^2\dfrac{2^2\pi\,J_T(\tilde{b})}{3^4\,\tilde{b}^{3/2}}\dfrac{T^3}{M_{KK}^2}\int^{1}_{u_E/u_J}d\hat{u}\,\hat{u}\,\rho(\hat{u})\left(D_T(\hat{u})+\dfrac{(\partial_u\tilde{a}_t)^2}{D_T(\hat{u})}\right),
\end{equation}
where $\rho(\hat{u})$ and $D_T(\hat{u})$ are computed for the rescaled fields. In figure \ref{fig:energyfitdeconf} we plot the energy for $\tilde{b} =0.1$ as a function of $\lambda_T$ and show that for this observable, the rigid rotor approximation \eqref{eq:energyrigiddeconf} is not reliable. Again, we suspect that dropping the gauge connection in the mass term of the rigid rotor energy might be too drastic, and it spoils the $\lambda_T$-dependence of the energy.

\begin{figure}[H]
	\center
	\includegraphics[height = 5cm]{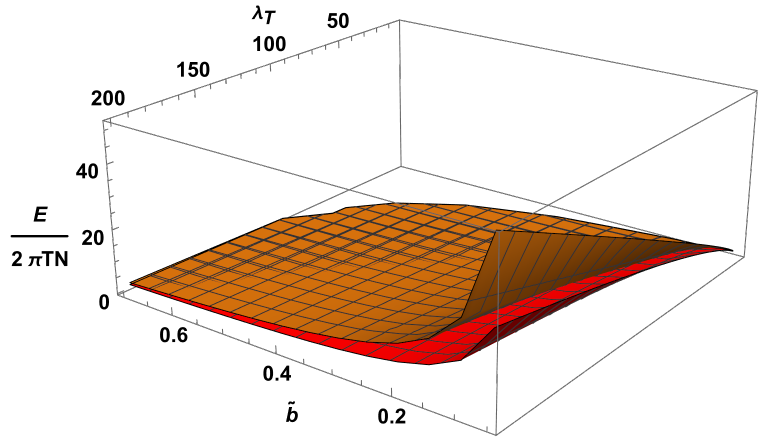}
	\caption{The figure shows the plot the energy for a D6-brane solution with $n_B=1$ in orange as a function of $\tilde{b}$ and $\lambda_T\in[5,200]$. We plot it together with the associated Bogomol'nyi bound \eqref{BogoboundE} in red.}
	\label{fig:energybogdeconf}
\end{figure}

In figure \ref{fig:energybogdeconf}, we plot the energy for a D6-brane solution with $n_B=1$ in orange as a function of $\tilde{b}$ and $\lambda_T$. We plot it together with the associated Bogomol'nyi bound \eqref{BogoboundE} in red.

\section{Punctured domain walls and sandwiches in the deconfined phase}\label{app:deconf2}
\begin{figure}[H]
	\centering
	\begin{subfigure}{0.4\textwidth}
		\centering
		\includegraphics[width=\textwidth]{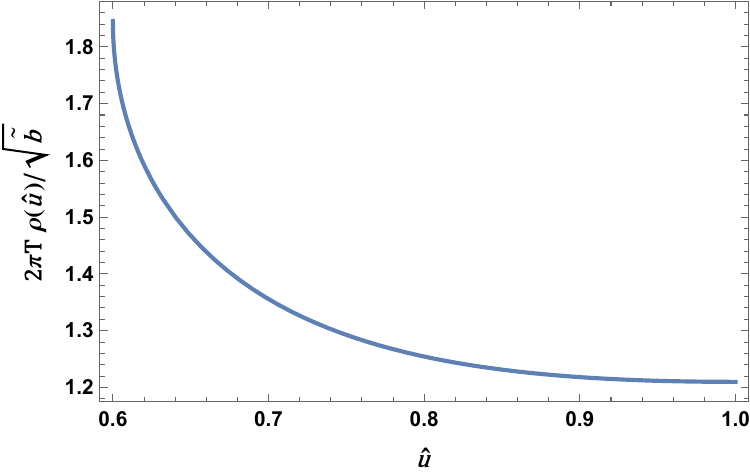}
	\end{subfigure}
	\hfill
	\begin{subfigure}{0.55\textwidth}
		\centering
		\includegraphics[width=\textwidth]{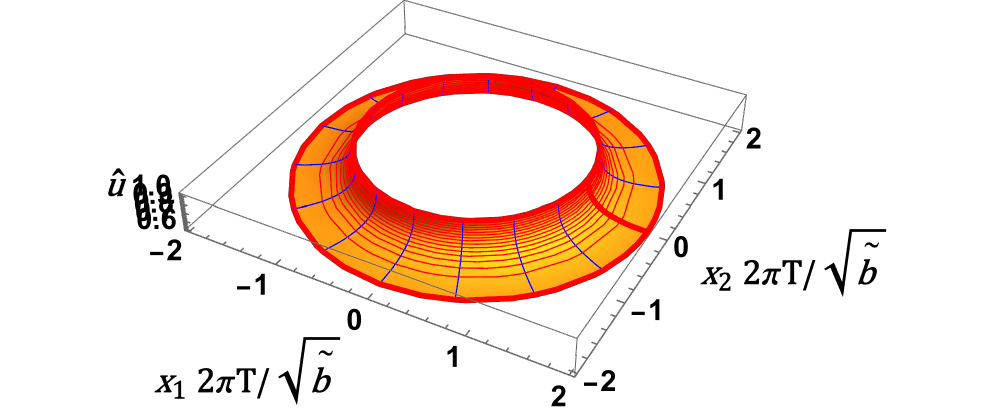}
	\end{subfigure}
	\caption{Numerical solution of $\rho$ (times $2\pi T/\sqrt{\tilde{b}}$) for $\lambda_T=20$, $\tilde{b}=0.1$ and $\hat{u}_* = 0.6$, as a function of the rescaled holographic coordinate $\hat{u}=u/u_J \in [0.6,1]$, describing a stable punctured domain wall with baryon number $n_B=1$. The profile's derivative $\hat{u} = 1$ is zero, indicating the (meta)stability of this embedding. At $\hat{u}_*=u_*/u_J = 0.6$ the profile's radius goes to infinity while in the plot we cut it for clarity at some finite value.}
	\label{fig:rhodwdeconf}
\end{figure}

\begin{figure}[H]
	\centering
	\begin{subfigure}{0.45\textwidth}
		\centering
		\includegraphics[width=\textwidth]{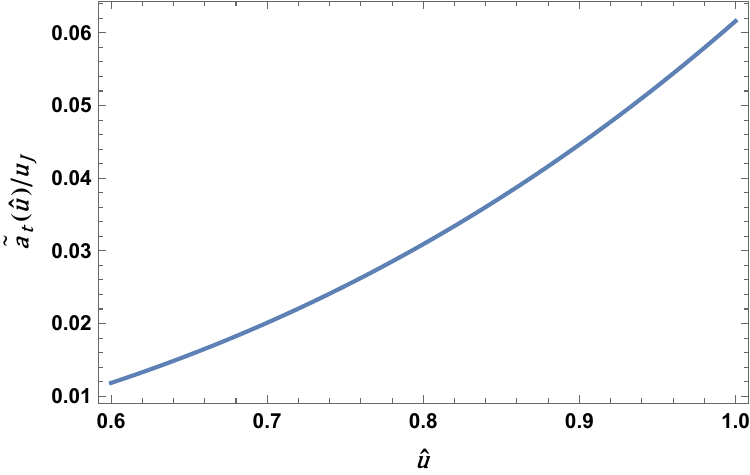}
	\end{subfigure}
	\hfill
	\begin{subfigure}{0.45\textwidth}
		\centering
		\includegraphics[width=\textwidth]{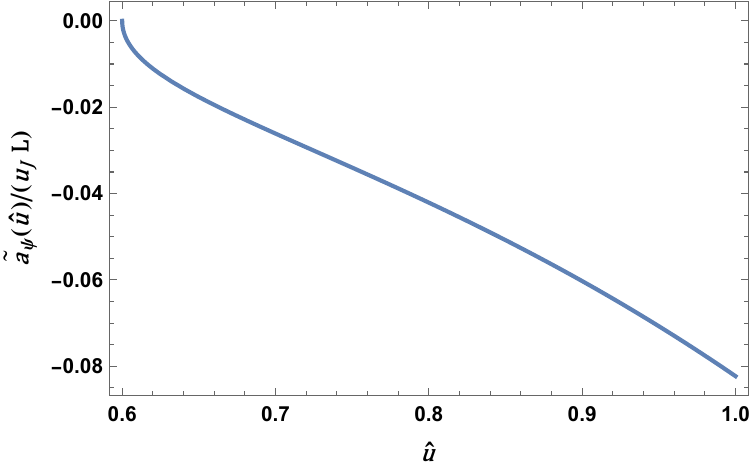}
	\end{subfigure}
	\caption{Numerical solutions for $u_J^{-1}\tilde{a}_t$ (left panel) and $(u_J L)^{-1}\tilde{a}_\psi$ (right panel) as a function of the rescaled holographic coordinate $\hat{u}$ for $\tilde{b}=0.1$ and $\lambda=100$, with the singularity in $\hat{u}_* = 0.6$.}
	\label{fig:adwdeconf}
\end{figure}

We start by presenting the punctured domain wall solutions in the deconfined phase. In figures \ref{fig:rhodwdeconf} and \ref{fig:adwdeconf} we show the solutions to the equations of motion for $\tilde{b}=0.1$, $\lambda_T=100$, and for which we have chosen to end the integration at $u_*/u_J=0.6$.

\begin{figure}[H]
	\centering
	\begin{subfigure}{0.4\textwidth}
		\centering
		\includegraphics[width=\textwidth]{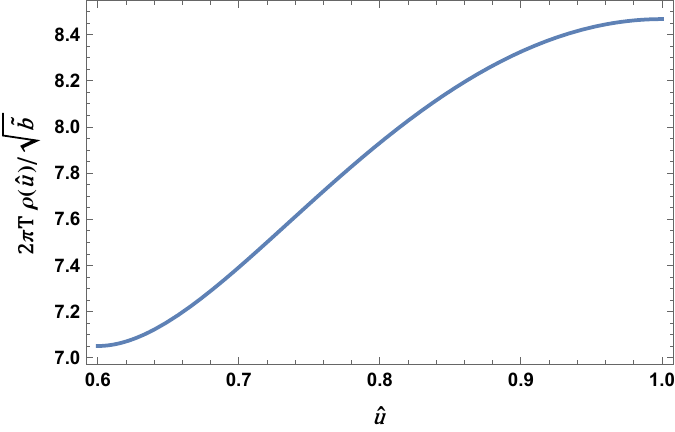}
	\end{subfigure}
	\hfill
	\begin{subfigure}{0.55\textwidth}
		\centering
		\includegraphics[width=\textwidth]{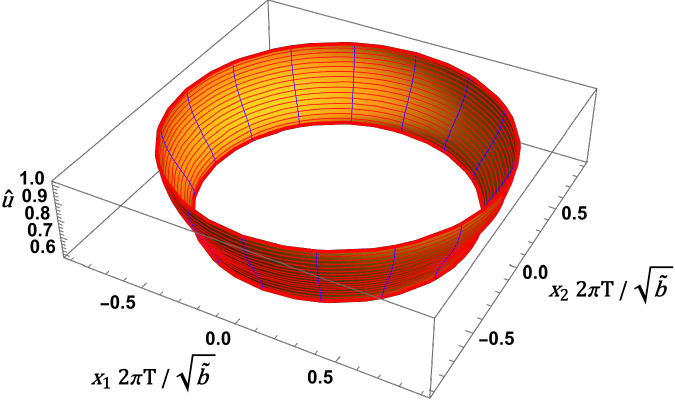}
	\end{subfigure}
	\caption{Numerical solution of $\rho$ (times $2\pi T/\sqrt{\tilde{b}}$) for $\tilde{b}=0.1$ and $\lambda_T=100$ as a function of the rescaled holographic coordinate $\hat{u}=u/u_J \in [0.6,1]$, describing a stable baryon with baryon number $n_B=1$. The profile's derivative at both $\hat{u}_K$ and $\hat{u} = 1$ is zero, indicating the (meta)stability of this embedding. At $\hat{u}=u_*/u_J$, the profile's radius goes to infinity while in the plot we cut it for clarity at some finite value.}
	\label{fig:rhoswdeconf}
\end{figure}

\begin{figure}[H]
	\centering
	\begin{subfigure}{0.45\textwidth}
		\centering
		\includegraphics[width=\textwidth]{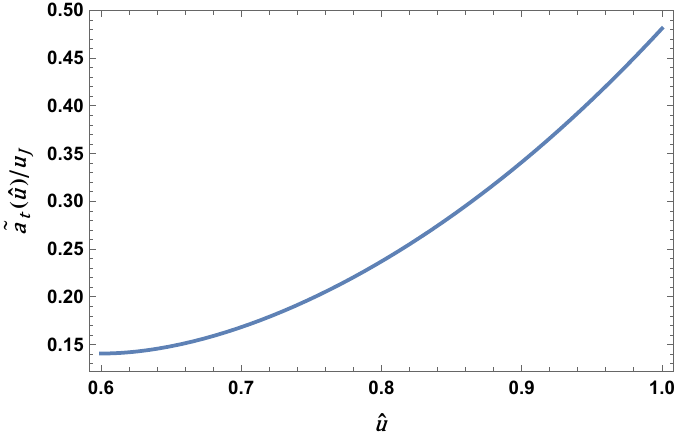}
	\end{subfigure}
	\hfill
	\begin{subfigure}{0.45\textwidth}
		\centering
		\includegraphics[width=\textwidth]{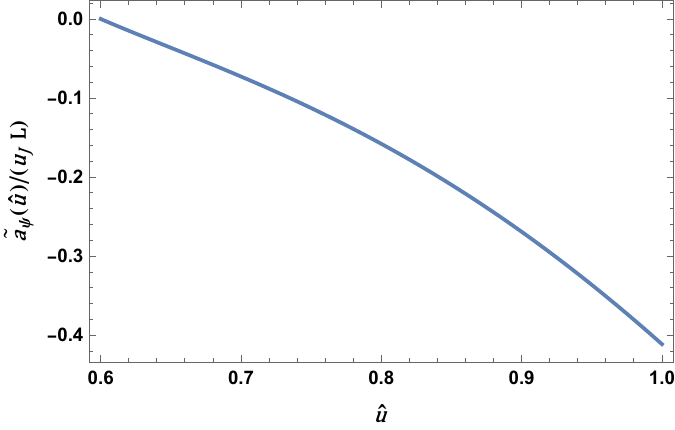}
	\end{subfigure}
	\caption{Numerical solutions for $u_J^{-1}\tilde{a}_t$ (left panel) and $(u_J L)^{-1}\tilde{a}_\psi$ (right panel) as a function of the rescaled holographic coordinate $\hat{u}$ for $\tilde{b}=0.1$ and $\lambda_T=100$.}
	\label{fig:aswdeconf}
\end{figure}

In figures \ref{fig:rhoswdeconf} and \ref{fig:aswdeconf}, we present sandwich solutions in the deconfined phase.

\begin{figure}[H]
	\centering
	\includegraphics[width=0.4\textwidth]{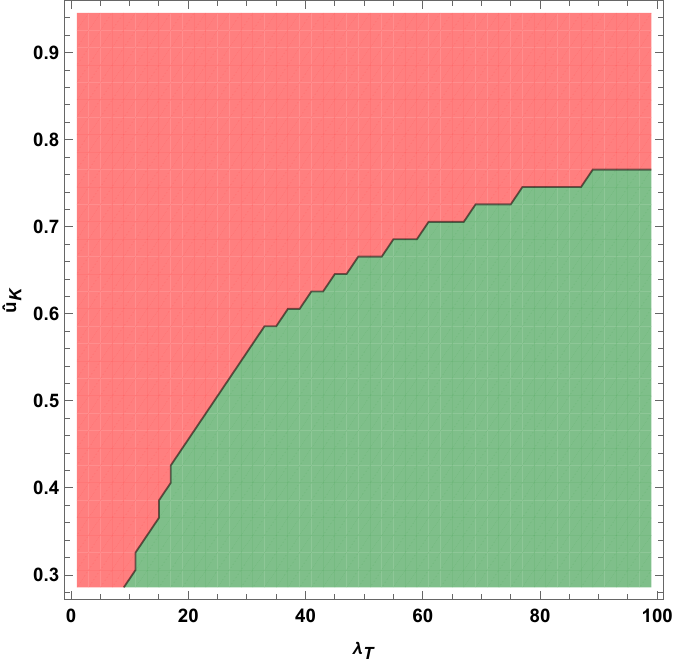}
	\caption{Region of existence (red) of sandwich domain walls solution in the deconfined phase between $\hat{u}_K = 0.27$ and $\hat{u}_K=1$, for values of $\lambda_T \in [1, 100]$.}
	\label{fig:regionSWdec}
\end{figure}

Much like in the confined phase, we obtain these solutions for every value of $\hat{u}_K$ above a minimal value $\hat{u}_\text{min}$. This function is once again an increasing function of both $\tilde{b}$ and $\lambda_T$. As mentioned previously, both $\tilde{b}=u_T/u_J$ and $\tilde{b}_K=u_T/u_K$ are bounded from above by a fixed value ($\sim 0.75$) if we want the D8-branes to be in the U-shaped configuration corresponding to the chirally broken phase.
The region of existence of physical sandwich domain walls with $n_B=1$ for $\tilde{b} = 0.2$ is shown in red in figure \ref{fig:regionSWdec}.

We conclude by saying that also in the deconfined phase we can find sandwich solutions with $n_K=1$ and $n_J=0$ with an associated profile $\rho(u)$ for which for all $u$, $\rho'(u)<0$.

\bibliographystyle{JHEP}
\bibliography{Ref.bib}
\end{document}